\documentclass[aps,pre,onecolumn,preprint,showpacs,superscriptaddress,groupedaddress
]{revtex4-1}
\usepackage{amsmath,amssymb}
\usepackage{amsmath,bm}
\usepackage{graphicx}
\usepackage{dcolumn}
\usepackage{bm}
\usepackage{xcolor}

\newcommand{\sech}{\mbox{sech}}

\newcommand{\bea}{\begin{eqnarray}}
\newcommand{\eea}{\end{eqnarray}}
\newcommand{\bes}{\begin{subequations}}
\newcommand{\ees}{\end{subequations}}

\begin{document}

\title{ Vector soliton molecules and their collisions }
\author{S. Stalin\footnote{Corresponding author}}
\email{stalin.cnld@gmail.com}
\author{M. Lakshmanan}
\email{lakshman.cnld@gmail.com}
\affiliation{Department of Nonlinear Dynamics, Bharathidasan University, Tiruchirapalli--620 024, India}

\begin{abstract}
In recent times, bound soliton states have often been referred to as soliton molecules in the nonlinear optics literature. The striking analogies between photonic bound states and matter molecular structures in chemistry and physics have intensified studies on optical soliton molecules in both conservative and dissipative systems. In this paper, we demonstrate the existence of vector soliton molecules and their related isomer structures in a conservative optical fiber system by considering the integrable Manakov equation. We show their existence by applying the velocity resonance condition and appropriate choice of temporal separations to the degenerate $N=(\bar{N}+\bar{M})$-soliton solution. Then, we classify the obtained molecular states as either dissociated or synthesized molecular states based on the temporal locations of the constituent solitons. Furthermore, we analyze the collision properties of vector soliton molecules in the present conservative system. The collision scenarios reveal that the soliton molecules undergo intriguing energy-sharing collisions through energy redistribution among the modes. To characterize these collisions, we have carried out an appropriate asymptotic analysis and found that elastic collisions arise as a special case of energy-sharing collisions under specific choices of polarization constants. Finally, we numerically verify the robustness of vector soliton molecules. We believe that the results presented in this paper show potential for soliton molecule-based applications such as optical computation and multi-level encoding for communications. 

\end{abstract}

\maketitle


\section{Introduction}
Temporal optical solitons are highly localized nonlinear wave packets in time, resulting from an exact balance between linear and nonlinear physical effects. A natural tendency of these solitons is that they can propagate over large distances without diminishing their sizes and energies, and preserve their integrity under collisions \cite{kiv-book}.  Due to these facts,  optical solitons have been considered as valuable assets for long-haul optical communication applications \cite{gp}. Recently, multi-soliton structures, namely soliton molecules (SMs), which are multi-soliton bound states composed of two or more fundamental optical solitons co-propagate with equal or degenerate group velocities, have been proposed to improve the bit-rates in multi-level optical communication applications \cite{fedor1,fedor2,fedor3}. The need for finding alternative new scheme arises because the data-carrying capacity of the fiber is approaching its maximum limit, as restricted by the Shannon's theorem \cite{shanon},  using the conventional coding scheme.  Therefore,  the demand for advancing telecommunication technology on the one hand and perceiving knowledge on the characteristics of SMs on the other hand necessitates intense study of SMs both theoretically and experimentally starting from the basic level.  

Mathematically, bound soliton states are described by the multi-soliton solutions of the nonlinear evolution equations in certain conservative and dissipative systems under a special condition on the velocities of the solitons. For example, to explain such bound soliton state (BSS) solutions one can consider the multi-soliton solution of the focusing NLS equation or the NLS equation with anomalous dispersion \cite{zakharov}. The bound bright soliton states of the NLS equation are described by $N$-soliton solution with velocity degeneracy condition: $v_i=v_j=v_{mol}$, $1\le i\neq j \le N$, where $v_i$ and $v_j$'s are the fundamental bright soliton velocities in which all or some of them are degenerate. Such soliton states can also be brought out by assuming the complete degeneracy condition: $v_1=v_2=...=v_n=v_{mol}$, $n=1,2,...,N$, in which all the solitons co-propagate with an identical velocity denoted by the  molecular velocity ($v_{mol}$). These velocity resonance conditions are very much essential for the formation of bound soliton states in the NLS equation. The soliton solutions obeying the aforementioned velocity resonance conditions have been referred in the literature as doubly degenerate and completely (or fully) degenerate soliton solutions \cite{biondini}, respectively, which we refer in the present paper as bound soliton state solutions. This situation is similar to studying simple-pole solutions with  two or more discrete eigenvalues having the same real parts through the Inverse Scattering Transform formalism \cite{ablowitz-ist}. In the Hirota's bilinear framework \cite{hirota-book}, it is equivalent to analysing $N$-soliton solution with the imaginary part of $N$-propagation constants ($k_j$'s) being equal. The simplest bound soliton state of the NLS equation can be obtained by considering the corresponding two-soliton solution with     $v_1(\equiv 2k_{1I})=v_2(\equiv 2k_{2I})$. The higher-order bound soliton states can be obtained by considering higher-order degeneracy. For instance,  multi-fold degeneracy on velocities of solitons yields a triplet SM (constituted by three soliton atoms), a quadruplet SM (made by four-soliton atoms), and a more complex macro molecular state (composed of arbitrary $N$ soliton atoms) from the respective multi-soliton solutions.

The temporal separation is another essential degree of freedom to form a stable SM. If the temporal separation is relatively larger than the width of the constituents, then the corresponding SM is a dissociated molecule \cite{gelash-breather-molecule,kirane}. When this value is relatively small, then the nonlinear interaction between the soliton atoms becomes significant. Consequently,  they experience force on each other. At a particular equilibrium separation, where the net force is zero between the basic soliton pulses, the soliton atoms constitute a stable SM. This phenomenon is referred as synthesis of molecule \citep{gelash-breather-molecule,kirane}. Further, BSS solutions or  soliton molecular structures exhibit periodic oscillations, characterized by molecular frequency, when the soliton parameters are commensurate with each other.

After the advent of real-time ultra-fast measurements \cite{krupa,solli}, the multi-soliton bound structures in dissipative systems have become one of the main areas of research in fiber lasers. Because they display internal motion akin to diatomic molecules in chemistry, including synthesis, vibration \cite{grelu2,grelu1}, phase sliding \cite{krupa}, and step-wise evolution \cite{solli}, their study has gained much interest. The analogy between optical SMs and matter molecules, though they are fundamentally different entities,  further elevates research on  complex molecular structures, made by either soliton pulses or breather pulses, in many directions. For instance, it is interesting to point out that the existence of multi-color SMs \cite{willms,melchert,boris-malomed}, breather molecules in passively mode-locked lasers \cite{peng}, harnessing of photonic SMs by tuning the temporal separation through the dispersion losses \citep{boris-malomed1}, Hopf bifurcation induced intrinsic oscillations in SMs \cite{grelu1,sakaguchi}, isomers of higher-order soliton molecular states \cite{huang-apl}, and the  formation of supra-molecular structure by considering the long-range interactions of solitons \cite{menyuk} have been discussed in the recent literature in the context of photonic molecular states. These molecular states are constituted by the individual units, the so called dissipative solitons or dissipative breathers, which are formed through the balance of energy exchange with the surroundings and in the presence of dispersion and nonlinearity \cite{DS-book}. However, these studies on SMs have gained interest from the earlier studies on soliton compound of conservative and dissipative systems and from the works that have shown the existence of force (either attractive or repulsive) between the constituents of a bound state \cite{karpmann,gordon,fedor4,anderson,boris-1,usama,afanasjec,soto,
boris-2,fedor1,yulin,weng,kirane,soliton-complex,gelash-breather-molecule}. 

Further, we wish to point out that the BSSs were analyzed in detail in the following systems: The NLS equation with pumping and dissipation \cite{boris-1}, the complex scalar and coupled Ginzburg-Landau equations  \cite{boris-1,afanasjec,soto,boris-2}, and the Gross-Pitaevskii equation \cite{usama}, dispersion-managed fibers \cite{fedor1}, twin-core fibers \cite{yulin}, and optical microresonators \cite{weng}.   These studies have revealed that in-phase solitons always attract and lead to the formation of coalescence phenomenon where constructive interference takes place in the overlapping region. Otherwise, a destructive interference will occur when two out-of-phase solitons repel against each other. We wish to note that soliton molecules having remarkable properties in dissipative systems can be treated as the extended concept of BSSs in integrable systems and their properties are different from the flat-top SMs, as recently reported for the Fermi-Pasta-Ulam-Tsingou lattice with quadratic-cubic nonlinear interactions \cite{kirane}. It is also interesting to note that the concept of multi-soliton complexes was introduced in the integrable multi-mode theory \cite{soliton-complex} and it has been demonstrated that the conservative fiber system supports the formation of breather molecules \cite{gelash-breather-molecule}.    

  From the above mentioned studies, we infer that the SMs or in other words multi-bound soliton structures are investigated prominently in the dissipative systems since they possess non-zero binding energy. In the case of integrable Hamiltonian systems, they were less studied/neglected  or mostly their existence was pointed out simply by the graphical demonstration without any further analysis. This is because bound states are neutrally stable since the binding potential is absent (or zero binding energy) in between the binding partner soliton atoms.  A small perturbation can cause instability and hence breaks the degeneracy/velocity resonance condition. Consequently, the constituents propagate independently with their own identities. However, by considering all the above facts, we feel that  it is still interesting to analyse bound soliton state solutions or multi-soliton solutions with degenerate velocities, the various soliton molecular structures admitted by such solutions, underlying their formation mechanism, and unveiling their collisional properties in the conservative fiber systems modelled by the integrable nonlinear Schr\"{o}dinger family of equations. In particular, to the best of our knowledge, characterization of the bound soliton solutions in the conservative two mode fiber system, governed by the Manakov equation \cite{manakov}, is missing in the literature.  Then, analysing the various isomer molecular structures formed by tuning the amplitude dependent temporal separations (or bond lengths), and investigating the collision properties of SMs under various physical situations have not been clearly addressed so far in the literature except in Ref. \cite{sun}, where the basic stationary bound soliton state, constituted by a pair of scalar bright solitons, is displayed for the Manakov system by assuming the complex phase constants as unity ($\alpha_1^{(j)}=\alpha_2^{(j)}=1$, $j=1,2$, see Sec. II for further details). 

Also, till date, bound states of two solitons have been the most studied soliton molecular structures in the literature. The molecular structure composed of three or more than three soliton atoms and their collisional properties under different physical situations, including the collision between two SMs, each made by a pair of vector bright solitons have been given less attention in the literature for both integrable and non-integrable systems \cite{demircan}. By taking into account all the above mentioned facts, we restrict ourselves to the integrable case in the present study and it can be extended, in principle, to dissipative case as well by following the equivalent theory developed in the literature on dissipative solitons \cite{DS-book}.             

To address these issues, we consider a basic system of two-coupled nonlinear Schr\"odinger (2-CNLS) equations, which was introduced a long time ago by Manakov \cite{manakov} to describe the propagation of  electromagnetic optical pulses in a two-mode optical fiber, of the form 
\begin{eqnarray}
iq_{j,z}+q_{j,tt}+2(|q_1|^2+|q_2|^2)q_j=0,~j=1,2.
\label{manakov}
\end{eqnarray} 
In the above equations, the dependent variables $q_j\equiv q_j(z,t)$'s represent the complex field envelope functions, and the independent variables $z$ and $t$ denote the partial derivatives with respect to those dimensionless variables. In nonlinear fiber optics,  these independent variables generally represent the normalized distance along the fiber and retarded time, respectively. The model (\ref{manakov}) is a simple generalization of the scalar NLS equation and is considered as a basic coupled field model since it widely appears in several branches of physics most notably in nonlinear optics \cite{kiv-book}, Bose-Einstein condensates \cite{kevrekidis}, hydrodynamics \cite{osborne}, and plasma physics \cite{shukla}. It is well known fact that, Eq. (\ref{manakov}) admits a Lax pair  \cite{manakov}, infinite number of symmetries and conserved quantities and $N$-soliton solution. It is interesting to note that the vector two bright solitons, each characterized by identical wave number in both the modes, of Eq. (\ref{manakov}) undergo energy sharing collision via intensity redistribution among the modes \cite{radha}. Besides this, as we have shown in our recent works \cite{ss,ramakrishnan}, one can also obtain  a more general form of fundamental nondegenerate vector bright soliton solution, which contains  two distinct propagation constants, for the Manakov system. However, in the present study, we wish to show that the bound states are essentially constituted by the energy sharing collision exhibiting vector bright solitons of degenerate type \cite{radha,ss}. In principle, one can also extend this study by considering non-degenerate vector bright solitons, which  we leave for future study. We wish to note that  the existence of bound soliton states of dark-bright vector solitons was demonstrated in the literature for the defocusing Manakov model \cite{sheppard, radha1}, and the three-component CNLS equations \cite{jiang}. The bound soliton states were also further explored in other physical systems such as Bose-Einstein condensates \cite{usama}, optical systems \cite{khawaja2, crasovan}, hydrodynamics \cite{pelinovsky} and other closely related systems \cite{matre}.   

 In order to investigate our main objective, we wish to present the general $N=(\bar{N}+\bar{M})$-vector bright soliton solution of degenerate type, which contains two sets of $\bar{N}$ and $\bar{M}$-bright solitons with arbitrary velocities, for the Manakov system (\ref{manakov}). Introducing velocity resonance conditions to such multi-soliton solution, we deduce the bound soliton state solutions describing vector soliton molecular structures. Based on these bound soliton solutions, we obtain doublet, triplet, quadruplet SMs, and a macro molecule (or soliton molecular complex) by restricting the number of constituents appropriately. Then, by tuning the amplitude dependent relative temporal separations, we bring out the various possible isomer structures associated with these molecules. Further, we analyse the robustness of these molecules (in particular we restrict ourselves to the doublet SM or the fundamental molecular state) by considering three distinct physical situations. We first investigate the collision between the fundamental molecule and a single vector bright soliton. Then, we analyze the collision between a stationary doublet SM and two basic vector solitons. Finally, we also consider the head-on collision between two doublet SMs by making them to propagate in opposite directions. By doing so, we find that the molecular structures in each of these collision scenarios undergo interesting energy sharing collisions with its interacting partners. Besides these, we also bring out an elastic collision in each of these cases by suppressing the energy sharing nature of SMs. We achieve this by choosing the complex phase constants or the polarization constants appropriately. In addition to these, we also verify the stability of vector SMs numerically through the Split-Step Fourier (SSF) method.       

The rest of the paper is organized as  follows. In section II, we present the $(\bar{N}+\bar{M})$-bright soliton solution of the Manakov system. It is deduced by rewriting the $N$-soliton solution of degenerate type appropriately. In this section, we also point out the partial, complete,  and double degeneracy velocity resonance conditions in order to get the mixed bound-non-bound soliton solutions, pure bound soliton solutions, and two distinct bound soliton solutions, respectively. The various vector molecular structures, their formation mechanism (especially for the doublet SM), their isomer structures, and their corresponding solutions are presented in Sec III. Further, we analyze the interesting energy sharing collisions exhibited by the doublet molecular structures in Section IV. 
The energy sharing nature of the molecular states is confirmed in each of these collision scenarios through appropriate asymptotic analysis. In this section, we also point out the possibility of observing elastic collision in all of these situations. Section V presents the numerical stability analysis of vector doublet SM while Section VI concludes the present work. In Appendix A, we present the various constants which arise during the asymptotic analysis in Sections III A and III B.     

\section{($\bar{N}+\bar{M}$)-soliton solution and velocity degeneracy conditions}
To begin with, we consider the arbitrary degenerate $N$-soliton solution of the Manakov system (\ref{manakov}) using Gram determinants \cite{ohta,viji-epjplus} (For nondegenerate $N$-soliton solution, see for example Refs. \cite{ss, ramakrishnan}).The explicit form of this solution is   
\begin{eqnarray}
q_j(z,t)=\frac{g^{(j)}(z,t)}{f(z,t)},~j=1,2,~
g^{(j)}=
\begin{vmatrix}
A & I & \varphi\\
-I & B & {\bf 0}^T\\
{\bf 0} & C_j & 0
\end{vmatrix},~f=\begin{vmatrix}
A & I \\
-I & B
\end{vmatrix}, \label{2}
\end{eqnarray}
where the elements of ($N\times N$) matrices $A$ and $B$ are defined as $A_{il}=\frac{e^{\eta_i+\eta_l^*}}{k_i+k_l^*}$, $B_{il}=\frac{\psi_i^\dagger\sigma\psi_l}{k_i^*+k_l}$, $\psi_l=\begin{pmatrix}
\alpha_l^{(1)}\\\alpha_l^{(2)}
\end{pmatrix} $, $i,l=1,2,..,N$. Then, the matrices $\varphi$,  ${\bf 0}$, and $C_j$ are defined as follows: $(N\times 1)$ column  matrix $\varphi=\begin{pmatrix}
e^{\eta_1},&e^{\eta_2},&\cdot &\cdot &, &e^{\eta_N}
\end{pmatrix}^T$, $(1\times N)$ row matrix $C_j=-\begin{pmatrix}
\alpha_1^{(j)},&\alpha_2^{(j)}&\cdot &\cdot &\cdot &\alpha_N^{(j)}
\end{pmatrix}$, ($1\times N$) row matrix ${\bf 0}=\begin{pmatrix}
0,&0,&\cdot &\cdot &\cdot &0
\end{pmatrix}$, and $\sigma$ and $I$ are ($N\times N$) identity matrices. The above multi-soliton solution  (\ref{2}) is governed by $3N$-complex parameters, namely $\alpha_n^{(j)}$ and $k_j$, where $j=1,2$ and $n=1,2,...,N$. 

We wish to point out that one can also obtain bound soliton states by using the solution (\ref{2}). However, we modify it in a convenient way and refer the resultant multi-soliton solution as $(N=\bar{N}+\bar{M})$-soliton solution, where $\bar{N}$ and $\bar{M}$ represent the two distinct sets of arbitrary number of solitons. We make this classification essentially to bring out the coexistence of soliton molecules and vector solitons as well as two different SMs, apart from constructing a macro molecular state.  To obtain such a solution, we rewrite all the matrices involved in the Gram determinant forms of $g^{(j)}$'s and $f$ in the following way: The matrices $A=\begin{pmatrix}
A_{nn'}& a_{nm}\\
\tilde{a}_{mn} & \hat{A}_{mm'}
\end{pmatrix}$ and $B=\begin{pmatrix}
B_{nn'}& b_{nm}\\
\tilde{b}_{mn} & \hat{B}_{mm'}
\end{pmatrix}$ now become $(\bar{N}+\bar{M})\times (\bar{N}+\bar{M})$ matrices and their corresponding elements are defined as 
\begin{eqnarray}
&&\hspace{-0.2cm}A_{nn'}=\frac{\text{exp}(\eta_n+\eta_{n'}^*)}{(k_n+k_{n'}^*)},~a_{nm}=\frac{\text{exp}(\eta_n+\xi_{m}^*)}{(k_n+l_{m}^*)},~\tilde{a}_{mn}=\frac{\text{exp}(\xi_m+\eta_{n}^*)}{(l_m+k_{n}^*)},~\hat{A}_{mm'}=\frac{\text{exp}(\xi_m+\xi_{m'}^*)}{(l_m+l_{m'}^*)},~~~~\nonumber\\
&&\hspace{-0.2cm}B_{nn'}=\frac{\psi_n^{\dagger}\sigma\psi_{n'}}{(k_n^*+k_{n'})},~  b_{nm}=\frac{\psi_n^{\dagger}\sigma\psi_{m}'}{(k_n^*+l_{m})}, ~\tilde{b}_{mn}=\frac{\psi_m'^{\dagger}\sigma\psi_{n}}{(l_{m}^*+k_n)}, ~\hat{B}_{mm'}=\frac{\psi_m'^{\dagger}\sigma\psi'_{m'}}{(l_m^*+l_{m'})},\nonumber\\
&&\hspace{-0.2cm} \psi_{n/n'}=\begin{pmatrix}
\alpha_{n/n'}^{(1)}\\
\alpha_{n/n'}^{(2)}
\end{pmatrix}, ~\psi_{m/m'}'=\begin{pmatrix}
\beta_{m/m'}^{(1)}\\
\beta_{m/m'}^{(2)}
\end{pmatrix},~\eta_n=k_nt+ik_n^2z, ~\xi_m=l_mt+il_m^2z, \label{2a}
\end{eqnarray} 
where $n,n'=1,2,..,\bar{N}$, and $m,m'=1,2,..,\bar{M}$.
The matrices $\sigma$ and $I$ are $2\bar{N}\times 2\bar{M}$ identity matrices and the remaining matrices are defined as follows: 
\begin{eqnarray}
&&\varphi=\begin{pmatrix}
e^{\eta_1} & e^{\eta_2} & \cdot &\cdot & e^{\eta_{\bar{N}}} &e^{\xi_1} & e^{\xi_2}  & \cdot &\cdot & e^{\xi_{\bar{M}}}
\end{pmatrix}^T, ~{\bf 0}=\begin{pmatrix}
0, & 0, &\cdot & \cdot &\cdot & 0
\end{pmatrix},   \nonumber\\
&&C_j=-\begin{pmatrix}
\alpha_1^{(j)} &\alpha_2^{(j)} &\cdot  &\cdot & \alpha_{\bar{N}}^{(j)}&\beta_1^{(j)} &\beta_2^{(j)}& \cdot &\cdot &\beta_{\bar{M}}^{(j)}
\end{pmatrix}.\label{2b}
\end{eqnarray}
In the above, the order of the column matrix $\varphi$ is $(\bar{N}+\bar{M})\times 1$ and the order of the row matrices ${\bf 0}$ and $C_j$  is $1\times (\bar{N}+\bar{M})$. We note that one can derive the $N$-soliton solution (\ref{2}) as well as $(\bar{N}+\bar{M})$-soliton solution (Eq. (\ref{2}) along with Eqs. (\ref{2a}) and (\ref{2b})) of Eq. (\ref{manakov}) through the standard Hirota's bilinear method by solving its corresponding bilinear equations: $(iD_z+D_t^2)g^{(j)}\cdot f=0$, ~$D_t^2f\cdot f=2(|g^{(1)}|^2+|g^{(2)}|^2)$,~$j=1,2$, with appropriate forms of seed solutions. However, to derive the $(\bar{N}+\bar{M})$-soliton solution, we have assumed the general form of seed solutions, $g_1^{(j)}=\sum_{n=1}^{\bar{N}}\alpha_n^{(j)}e^{\eta_n}+\sum_{m=1}^{\bar{M}}\beta_m^{(j)}e^{\xi_m}$ during the solution construction process. The resultant $(\bar{N}+\bar{M})$-soliton solution (\ref{2}) contains $(\bar{N}+\bar{M})$ number of vector bright solitons that propagate with arbitrary velocities $v_n=2k_{nI}$ and $v'_m=2l_{mI}$, $n=1,2,...,N$, $m=1,2,...,M$ (here $k_{nI}$ and $l_{mI}$ are the imaginary parts of the wave numbers $k_n$ and $l_m$, respectively). The solution (\ref{2}) is characterized by $3(\bar{N}+\bar{M})$ complex free parameters: ($\alpha_n^{(j)},k_n$) and ($\beta_m^{(j)},l_m$), $j=1,2$, $n=1,2,...,\bar{N}$, $m=1,2,...,\bar{M}$. The interesting fact about the ($\bar{N}+\bar{M}$)-soliton solution ((\ref{2}) with (\ref{2a}) and (\ref{2b})) is that one can deduce the possible degenerate and non-degenerate vector bright soliton solutions of the Manakov system \cite{radha,ss,ramakrishnan} by appropriately fixing the complex phase constants $\alpha_n^{(j)}$ and $\beta_m^{(j)}$'s as zero. 
\subsection{Bound soliton solutions and velocity resonance conditions:}
To deduce the bound soliton state solution, we impose the velocity resonance condition or degeneracy condition on the $(\bar{N}+\bar{M})$-soliton solution (\ref{2}) with Eqs. (\ref{2a}) and (\ref{2b}). As we pointed out earlier, this can be achieved  by restricting the imaginary parts of the wave numbers $k_n$ and $l_m$, $n=1,2,...,\bar{N}$, $m=1,2,...,\bar{M}$. Based on these restrictions, we classify the $(\bar{N}+\bar{M})$-soliton solution (\ref{2}) as three types: (i) partially degenerate BSS solution, (ii) completely degenerate BSS solution, and (iii) doubly degenerate BSS solution. 

(i) To get the partially degenerate BSS solution, we apply the degeneracy condition \bes \begin{eqnarray}
&&\text{\it either}~~v_1=v_2=...=v_n=v_{mol}=2k_{1I}, ~n=1,2,...,\bar{N}, ~~\text{for} ~\text{set} \{\bar{N}\}, \label{3a} \\
&&\text{\it or}~~ v'_1=v'_2=...=v'_m=v'_{mol}=2l_{1I}, ~m=1,2,...,\bar{M}, ~~\text{for}~ \text{set} \{\bar{M}\},\label{3b}
\end{eqnarray} \ees
to the $(\bar{N}+\bar{M})$-soliton solution (\ref{2}) with Eqs. (\ref{2a}) and (\ref{2b}). This yields a macro bound soliton molecular state constituted by either the vector bright solitons from the set $\{\bar{N}\}$ or the solitons from the set $\{\bar{M}\}$ (Note: $6\bar{N}$ or $6\bar{M}$ soliton real parameters should be non-zero, if the corresponding solitons from a particular set do not contribute to the formation of bound states so that they move with arbitrary velocities). Under this situation, the entire molecular structure propagates with  a single molecular velocity $v_{mol}=2k_{1I}$ (or $v'_{mol}=2l_{1I}$) and it is characterized by $5\bar{N}+1$ (or $5\bar{M}+1$)-real parameters: $\alpha_{nR}^{(j)}$,  $\alpha_{nI}^{(j)}$, $j=1,2$, $k_{nR}$, and $k_{1I}$ (or $\beta_{mR}^{(j)}$, $\beta_{mI}^{(j)}$, $l_{mR}$, and $l_{1I}$). Here, we note that the suffices $R$ and $I$ appearing above and in the following denote the real and imaginary parts of that particular complex parameter.  Depending on the sign of $k_{1I}$ and $l_{1I}$, the molecular structure moves either to the left or to the right directions and it becomes stationary when the values of these parameters are equal to zero (that is, $k_{1I}=0$ and $l_{1I}=0$). Consequently, the stationary molecular structure is governed by $5\bar{N}$-real parameters.  Therefore, the solution (\ref{2}) obeying the partial degeneracy condition (\ref{3a}) or (\ref{3b}) provides avenue to investigate the collision between SMs and vector solitons with arbitrary velocities. 

(ii) Next, the $(\bar{N}+\bar{M})$-soliton solution (\ref{2}), along with Eqs. (\ref{2a}) and (\ref{2b}), yields a completely degenerate BSS solution when it obeys the following velocity resonance condition:
\begin{equation}
v_1=v_2=...=v_n=v
_1'=v'_2=...=v'_m=v_{mol}, \label{4}
\end{equation} 
where $v_n=2k_{nI}$, $v'_{m}=2l_{mI}$, and $k_{nI}=l_{mI}$, with $n,=1,2,...,\bar{N}$ and $m=1,2,...,\bar{M}$.
In this situation, every one of the soliton atoms in each set equally contributes to the formation of a non-stationary soliton molecular structure, which is governed by a set of $5(\bar{N}+\bar{M})+1$-real parameters. A static molecular structure results from this case by losing a degree of freedom, $k_{1I}=0$. 

(iii) Finally, the double degenerate BSS solution emerges from the solution (\ref{2}) with Eqs. (\ref{2a}) and (\ref{2b}) by considering the velocity degeneracy conditions (\ref{3a}) and (\ref{3b}) simultaneously. Consequently, the solitons from each set assemble themselves parallelly and generate their own molecular structures. Overall, the bound soliton structure emerging from the sets $\{\bar{N}\}$ and $\{\bar{M}\}$ is governed by $5(\bar{N}+\bar{M})+2$-real parameters. From this case also one can bring out the two distinct set of static soliton molecular structures when their velocities are assumed to be zero. This consideration of the double degeneracy condition in the soliton solution (\ref{2}) facilitates us to investigate the interaction dynamics between two distinct SMs. 

Apart from the above, we wish to remark the following important points related to the BSS solutions mentioned above: (i) The soliton molecular structures can also be generated by assuming some or all of the soliton velocities as degenerate either in the set $\{\bar{N}\}$ or in the set $\{\bar{M}\}$ as per the following velocity resonance condition:   
\begin{eqnarray}
&&v_n=v_j=v_{mol}, ~~1\le n\neq j \le \bar{N}, ~ ~\text{for} ~\text{set} \{\bar{N}\} \nonumber\\
\text{or} &&v'_m=v'_k=v'_{mol}, ~~1\le m\neq k \le \bar{M},~~\text{for} ~\text{set} \{\bar{M}\}.\label{5}
\end{eqnarray}
(ii). Existence of periodicity is an important property of the BSS solution of the Manakov system. In general, the $(\bar{N}+\bar{M})$-soliton solution is not periodic. However, it becomes periodic in $z$ when it obeys the velocity resonance conditions given in  Eqs. (\ref{3a}), (\ref{3b}), (\ref{4}) and (\ref{5}),  and appropriate commensurate choices of the real parts of the wave numbers. The periodic nature of the BSS solution is characterized by oscillation frequencies ($\omega_{i,j}$'s) and oscillation periods $T_{i,j}$'s.  For example, as we have illustrated later in Section III, the periodicity of the doublet SM is characterized by a single recurrence frequency $\omega_{12}$ and the period of oscillation $T_{12}=\frac{2\pi}{\omega_{12}}$. \\(iii) In the above, all types of BSS solution, the bound soliton states  in the two modes are constituted by the fundamental degenerate vector bright solitons of the following form \cite{radha},          
\begin{eqnarray}
q_j=k_{1R}A_je^{i\eta_{1I}}\sech(\eta_{1R}+\phi),~j=1,2,~\eta_{1R}=k_{1R}(t-2k_{1I}z), ~\eta_{1I}=k_{1I}t+(k_{1R}^2-k_{1I}^2)z.
\end{eqnarray}
Here, $A_j=\frac{\alpha_1^{(j)}}{\sqrt{|\alpha_1^{(1)}|^2+|\alpha_1^{(2)}|^2}}$, $j=1,2$, represents the unit polarization vector and $k_{1R}A_j$ describes the amplitude of the above vector bright soliton.  The amplitude dependent central position is given by $\phi=\frac{1}{2}\ln\frac{|\alpha_1^{(1)}|^2+|\alpha_1^{(2)}|^2}{(k_1+k_1^*)^2}(=\frac{1}{2}\ln\frac{|\alpha_1^{(1)}|^2+|\alpha_1^{(2)}|^2}{4k_{1R}^2})$. It implies that the polarization vectors and the amplitude dependent temporal positions corresponding to two vector solitons determine the various bound soliton structures in the Manakov system through intensity distribution between the modes. Note that the nondegenerate vector one-soliton solution \cite{ss,ramakrishnan} is defined by
\bes\begin{eqnarray}
q_1&=&\frac{\alpha_1^{(1)}e^{\eta_1}+e^{\eta_1+\xi_1+\xi_1^*+\delta_{11}}}{1+e^{\eta_1+\eta_1^*+\delta_1}+e^{\xi_1+\xi_1^*+\delta_2}+e^{\eta_1+\eta_1^*+\xi_1+\xi_1^*+\delta_3}},\\
q_2&=&\frac{\alpha_1^{(2)}e^{\xi_1}+e^{\xi_1+\eta_1+\eta_1^*+\delta_{12}}}{1+e^{\eta_1+\eta_1^*+\delta_1}+e^{\xi_1+\xi_1^*+\delta_2}+e^{\eta_1+\eta_1^*+\xi_1+\xi_1^*+\delta_3}},
\end{eqnarray}\ees
where $e^{\delta_{11}}=\frac{(k_1-l_1)\alpha_1^{(1)}|\alpha_1^{(2)}|^2|}{(l_1+l_1^*)^2(k_1+l_1^*)}$, $e^{\delta_{12}}=-\frac{(k_1-l_1)\alpha_1^{(2)}|\alpha_1^{(1)}|^2|}{(k_1+k_1^*)^2(k_1^*+l_1)}$,  $e^{\delta_1}=\frac{|\alpha_1^{(1)}|^2}{(k_1+k_1^*)^2}$, $e^{\delta_2}=\frac{|\alpha_1^{(2)}|^2}{(l_1+l_1^*)^2}$, $e^{\delta_3}=\frac{|k_1-l_1|^2|\alpha_1^{(1)}|^2|\alpha_1^{(2)}|^2}{(k_1+k_1^*)^2(l_1+l_1^*)^2|k_1+l_1^*|^2}$, $\xi_1=l_1t-il_1^2z$, and $\eta_1=k_1t-ik_1^2z$. The bound state solitons corresponding to this type of soliton will be analyzed separately as their analysis require more complicated calculations.
\\(iv) Bound soliton structures of the Manakov system (\ref{manakov}) obtained using three types of bound soliton solutions are further classified as either dissociated or synthesized molecule based on the bond length or temporal separation between the binding partners. If the relative separation between the sub-vector soliton pulses is sufficiently small, synthesis of molecule will occur due to their strong interactions, thereby leading to the formation of breathing pattern characterized by oscillation frequencies.  The dissociated molecule forms when the interaction between the sub-soliton pulses is weak and it is specified by a large temporal separation. We wish to note that in the present case, the temporal separation is calculated by the separation between the intensity peaks in a SM structure.


\section{Soliton molecular structures}
In this section, we intend to analyze the explicit structure of SMs admitted by the Manakov system (\ref{manakov}). In particular, we will display the doublet, triplet and quadruplet soliton molecular structures and  analyse their molecular properties. In addition to these, we will also display the possible isomer structures related to triplet and quadruplet SMs through the temporal distribution of soliton atoms. Then, we will also explain the mechanism behind the formation of all these structures in the present integrable case. In order to unveil all these aspects, we wish to analyse the associated bound soliton solutions deduced from the $(\bar{N}+\bar{M})$-soliton solution (\ref{2}) under the complete degeneracy condition (\ref{4}) by restricting the soliton numbers appropriately. 
\subsection{Fundamental soliton molecular structure}
To start with, we consider the basic molecular structure, namely the doublet SM and deduce its analytical formula from the following ($1+1$)-soliton solution:
\begin{eqnarray}
q_j=\frac{g^{(j)}}{f},~
g^{(j)}=
\begin{vmatrix}
\frac{e^{\eta_1+\eta_1^*}}{k_1+k_1^*}&\frac{e^{\eta_1+\xi_1^*}}{k_1+l_1^*}&1&0&e^{\eta_1}\\
\frac{e^{\eta_1^*+\xi_1}}{l_1+k_1^*}&\frac{e^{\xi_1+\xi_1^*}}{l_1+l_1^*}&0&1&e^{\xi_1}\\
-1&0&B_{11}&b_{11}&0\\
0&-1 &\tilde{b}_{11}&\hat{B}_{11}&0\\
0&0&-\alpha_1^{(j)}&-\beta_1^{(j)}&0
\end{vmatrix},~
f=\begin{vmatrix}
\frac{e^{\eta_1+\eta_1^*}}{k_1+k_1^*}&\frac{e^{\eta_1+\xi_1^*}}{k_1+l_1^*}&1&0\\
\frac{e^{\eta_1^*+\xi_1}}{l_1+k_1^*}&\frac{e^{\xi_1+\xi_1^*}}{l_1+l_1^*}&0&1\\
-1&0&B_{11}&b_{11}\\
0&-1 &\tilde{b}_{11}&\hat{B}_{11}
\end{vmatrix},j=1,2,~~ \label{7}
\end{eqnarray}
where $\eta_1=k_1t+ik_1^2z$, $\xi_1=l_1t+il_1^2z$, $B_{11}=\frac{\psi_1^{\dagger}\sigma\psi_1}{(k_1^*+k_1)}$, $b_{11}=\frac{\psi_1^{\dagger}\sigma\psi_1'}{(k_1^*+l_1)}$,
$\tilde{b}_{11}=\frac{\psi_1'^{\dagger}\sigma\psi_1}{(l_1^*+k_1)}$,
$\hat{B}_{11}=\frac{\psi_1'^{\dagger}\sigma\psi_1'}{(l_1^*+l_1)}$, $\sigma=\begin{pmatrix}
1 &0\\
0&1
\end{pmatrix}$, 
$\psi_1=\begin{pmatrix}
\alpha_1^{(1)}\\ \alpha_1^{(2)}
\end{pmatrix}$, and $\psi'_1=\begin{pmatrix}
\beta_1^{(1)}\\ \beta_1^{(2)}
\end{pmatrix}$. The six complex parameters, $\alpha_1^{(j)}$, $\beta_1^{(j)}$, $j=1,2$, $k_1$, and $l_1$ determine the properties of two vector bright solitons in the above soliton solution. The corresponding soliton velocities are defined by $v_1=2k_{1I}$ and $v_1'=2l_{1I}$. These velocities are always preserved  asymptotically  when two individual incoherent solitons undergo shape changing collision through intensity redistribution among the modes \cite{radha1}.  However, to form the basic soliton molecular state these unequal velocities have to follow the velocity degeneracy condition, $v_1=v_1'=v_{mol}$ (or $k_{1I}=l_{1I}$). Under this situation, the two solitons always stay close to each other asymptotically and they evolve with a common molecular velocity, $v_{mol}=2k_{1I}$.    By applying the latter resonance condition on the soliton solution (\ref{7}), we deduce the basic bound soliton solution in the form
\bes\begin{eqnarray}
q_j=\frac{g^{(j)}}{f},~
g^{(j)}&=&
\begin{vmatrix}
\frac{\text{exp}{2\eta_{1R}}}{2k_{1R}}&\frac{\text{exp}{(\eta_{1R}+\xi_{1R}+i(\eta_{1I}-\xi_{1I}))}}{k_{1R}+l_{1R}}&1&0&e^{\eta_1}\\
\frac{\text{exp}{(\eta_{1R}+\xi_{1R}-i(\eta_{1I}-\xi_{1I}))}}{k_{1R}+l_{1R}}&\frac{\text{exp}{2\xi_{1R}}}{2l_{1R}}&0&1&e^{\xi_1}\\
-1&0&B_{11}&b_{11}&0\\
0&-1 &\tilde{b}_{11}&\hat{B}_{11}&0\\
0&0&-\alpha_1^{(j)}&-\beta_1^{(j)}&0
\end{vmatrix},~j=1,2,~~~~~~~\label{8a}\\
f&=&\begin{vmatrix}
\frac{\text{exp}{2\eta_{1R}}}{2k_{1R}}&\frac{\text{exp}{(\eta_{1R}+\xi_{1R}+i(\eta_{1I}-\xi_{1I}))}}{k_{1R}+l_{1R}}&1&0\\
\frac{\text{exp}{(\eta_{1R}+\xi_{1R}-i(\eta_{1I}-\xi_{1I}))}}{k_{1R}+l_{1R}}&\frac{\text{exp}{2\xi_{1R}}}{2l_{1R}}&0&1\\
-1&0&B_{11}&b_{11}\\
0&-1 &\tilde{b}_{11}&\hat{B}_{11}
\end{vmatrix}.\label{8b}
\end{eqnarray}\ees
 Here, $\eta_1=\eta_{1R}+i\eta_{1I}=[k_{1R}(t-2k_{1I}z)]+i[k_{1I}t+(k_{1R}^2-k_{1I}^2)z]$,  $\xi_1=\xi_{1R}+i\xi_{1I}=[l_{1R}(t-2k_{1I}z)]+i[k_{1I}t+(l_{1R}^2-k_{1I}^2)z]$, $B_{11}=\frac{(|\alpha_1^{(1)}|^2+|\alpha_1^{(2)}|^2)}{2k_{1R}}$, $b_{11}=\frac{(\alpha_1^{(1)}\beta_1^{(1)*}+\alpha_1^{(2)}\beta_1^{(2)*})}{k_{1R}+l_{1R}}$, $\tilde{b}_{11}=\frac{(\beta_1^{(1)}\alpha_1^{(1)*}+\beta_1^{(2)}\alpha_1^{(2)*})}{k_{1R}+l_{1R}}$, and $\hat{B}_{11}=\frac{(|\beta_1^{(1)}|^2+|\beta_1^{(2)}|^2)}{2l_{1R}}$. The eleven real parameters, $\alpha_{1R}^{(j)}$,  $\alpha_{1I}^{(j)}$, $\beta_{1R}^{(j)}$,  $\beta_{1I}^{(j)}$, $j=1,2$, $k_{1R}$, $l_{1R}$, and $k_{1I}$, appearing in the above simplest BSS solution describe the basic molecular structure. It is essentially built by the contribution of a single soliton from each set $\{\bar{N}\}$ and $\{\bar{M}\}$.  
 
 

In order to understand the properties of the fundamental soliton molecular state, constituted by a pair of incoherent vector soliton atoms, we rewrite the Gram determinant forms of basic bound soliton state solution (\ref{8a})-(\ref{8b}) as  
\begin{subequations}\begin{eqnarray}
q_1&=&\frac{1}{D}\bigg(e^{i\xi_{1I}}c_{11}\cosh(\eta_{1R}+\phi_1^{(1)})+e^{i\eta_{1I}}c_{21}\cosh(\xi_{1R}+\phi_2^{(1)})\bigg),\label{9a}\\
q_2&=&\frac{1}{D}\bigg(e^{i\xi_{1I}}c_{12}\cosh(\eta_{1R}+\phi_1^{(2)})+e^{i\eta_{1I}}c_{22}\cosh(\xi_{1R}+\phi_2^{(2)})\bigg),\label{9b}\\
D&=&c_{3}\cosh(\eta_{1R}+\xi_{1R}+\phi_3)+c_{4}\cosh(\eta_{1R}-\xi_{1R}+\phi_4)\nonumber\\
&&+c_{5}\big(\cosh\phi_5\cos(\eta_{1I}-\xi_{1I})+i\sinh\phi_5\sin(\eta_{1I}-\xi_{1I})\big),\label{9c}
\end{eqnarray}
where 
\begin{eqnarray}
&&\hspace{-0.5cm}c_{1j}=\frac{(k_{1R}-l_{1R})^{1/2}\sqrt{\beta_1^{(j)}}[\alpha_1^{(j)}\tilde{b}_{11}-\beta_1^{(j)}B_{11}]^{1/2}}{\sqrt{2k_{1R}}(k_{1R}+l_{1R})^{1/2}},~c_{4}=\frac{(B_{11}\hat{B}_{11})^{1/2}}{2\sqrt{k_{1R}l_{1R}}}, c_{5}=\frac{(b_{11}\tilde{b}_{11})^{1/2}}{(k_{1R}+l_{1R})},  \nonumber\\
&&\hspace{-0.5cm}c_{2j}=\frac{(k_{1R}-l_{1R})^{1/2}\sqrt{\alpha_1^{(j)}}[\alpha_1^{(j)}\hat{B}_{11}-\beta_1^{(j)}b_{11}]^{1/2}}{\sqrt{2l_{1R}}(k_{1R}+l_{1R})^{1/2}}, ~c_{3}=\frac{(k_{1R}-l_{1R})(B_{11}\hat{B}_{11}-\tilde{b}_{11}b_{11})}{2\sqrt{k_{1R}l_{1R}}(k_{1R}+l_{1R})}, \nonumber\\
&&\hspace{-0.5cm}\phi_1^{(j)}=\frac{1}{2}\ln \frac{(k_{1R}-l_{1R})[\alpha_1^{(j)}\tilde{b}_{11}-\beta_1^{(j)}B_{11}]}{2\beta_1^{(j)}k_{1R}(k_{1R}+l_{1R})},~\phi_2^{(j)}=\frac{1}{2}\ln \frac{(k_{1R}-l_{1R})[\alpha_1^{(j)}\hat{B}_{11}-\beta_1^{(j)}b_{11}]}{2\alpha_1^{(j)}l_{1R}(k_{1R}+l_{1R})},\nonumber\\
&&\hspace{-0.5cm}\phi_3=\frac{1}{2}\ln \frac{(k_{1R}-l_{1R})^2(B_{11}\hat{B}_{11}-b_{11}\tilde{b}_{11})}{4k_{1R}l_{1R}(k_{1R}+l_{1R})^2}, ~\phi_4=\frac{1}{2}\ln \frac{B_{11}l_{1R}}{\hat{B}_{11}k_{1R}}, ~\phi_5=\frac{1}{2}\ln \frac{b_{11}}{\tilde{b}_{11}}, ~j=1,2.~~\label{9d}
\end{eqnarray}\end{subequations}
From the above fundamental BSS solution, one can identify that the amplitude parameters $k_{1R}$, $l_{1R}$, $\alpha_1^{(j)}$'s, and $\beta_1^{(j)}$'s explicitly appear in the phase terms $\phi_1^{(j)}$, $\phi_2^{(j)}$, $j=1,2$, $\phi_3$, $\phi_4$, and $\phi_5$. Consequently, the amplitude parameters and the temporal positions of the soliton atoms get related. Therefore, one can tune the location of the soliton atoms through these amplitude parameters. 

Then, an interesting feature of the BSS solution (\ref{9a})-(\ref{9d}) is that it exhibits periodic oscillations along the fiber length ($z$). It can be confirmed by the appearance of periodic functions $\cos(\eta_{1I}-\xi_{1I})$ and $\sin(\eta_{1I}-\xi_{1I})$, $\eta_{1I}-\xi_{1I}=(k_{1R}^2-l_{1R}^2)z$, in the BSS solution (\ref{9a})-(\ref{9d}). Such periodic behaviour is characterized by a single recurrence frequency $\omega_{12}=|k_{1R}^2-l_{1R}^2|$ and its period of oscillation is given by $T_{12}=\frac{2\pi}{\omega_{12}}=\frac{2\pi}{|k_{1R}^2-l_{1R}^2|}$. These formulae show that the breathing nature of basic SM is predominantly influenced by the real parameters $k_{1R}$ and $l_{1R}$. Further, the molecular frequency and periodic oscillation formulae imply that, one can have a molecular state  with zero oscillation or weak oscillations in its intensity profile, if the difference between the real parts of the wave numbers $k_1$ and $l_1$ approaches zero. We also wish  to point out here that the temporal separation ($\Delta t_{21}=t_2-t_1$ or $\Delta t_{12}=t_1-t_2$, where $t_1$ and $t_2$ are the temporal locations of the solitons) gets affected not only  by $k_{1R}$ and $l_{1R}$ but also by the polarization vectors:  $A_j^1=\frac{\alpha_1^{(j)}}{\sqrt{|\alpha_1^{(1)}|^2+|\alpha_1^{(2)}|^2}}$,   and $A_j^2=\frac{\beta_1^{(j)}}{\sqrt{|\beta_1^{(1)}|^2+|\beta_1^{(2)}|^2}}$, $j=1,2$, of the two vector bright solitons. Therefore, in the present coupled system, the structure of the fundamental SM is decided by the intensity distribution among the modes through polarization vectors and amplitude dependent temporal separation involving the parameters $k_{1R}$, $l_{1R}$, $\alpha_1^{(j)}$'s, and $\beta_1^{(j)}$'s. This fact is different for the bound soliton states of the scalar NLS equation, where the temporal separation is determined only by the real parts of wave numbers. We also wish to note here that the explicit forms of $t_1$ and $t_2$ can be obtained by finding the extremum points employing  the first and second derivative tests to Eqs. (\ref{9a})-(\ref{9d}) of the basic bound soliton state.  In the present case, we have numerically calculated the positions of solitons by fixing the aforementioned arbitrary parameter values in Eqs. (\ref{9a})-(\ref{9d}) with a fixed value of $z$.  The mechanism behind the building of SMs in the present vector case  is analysed below in detail. 
\begin{figure*}
	\centering
	\includegraphics[width=0.75\linewidth]{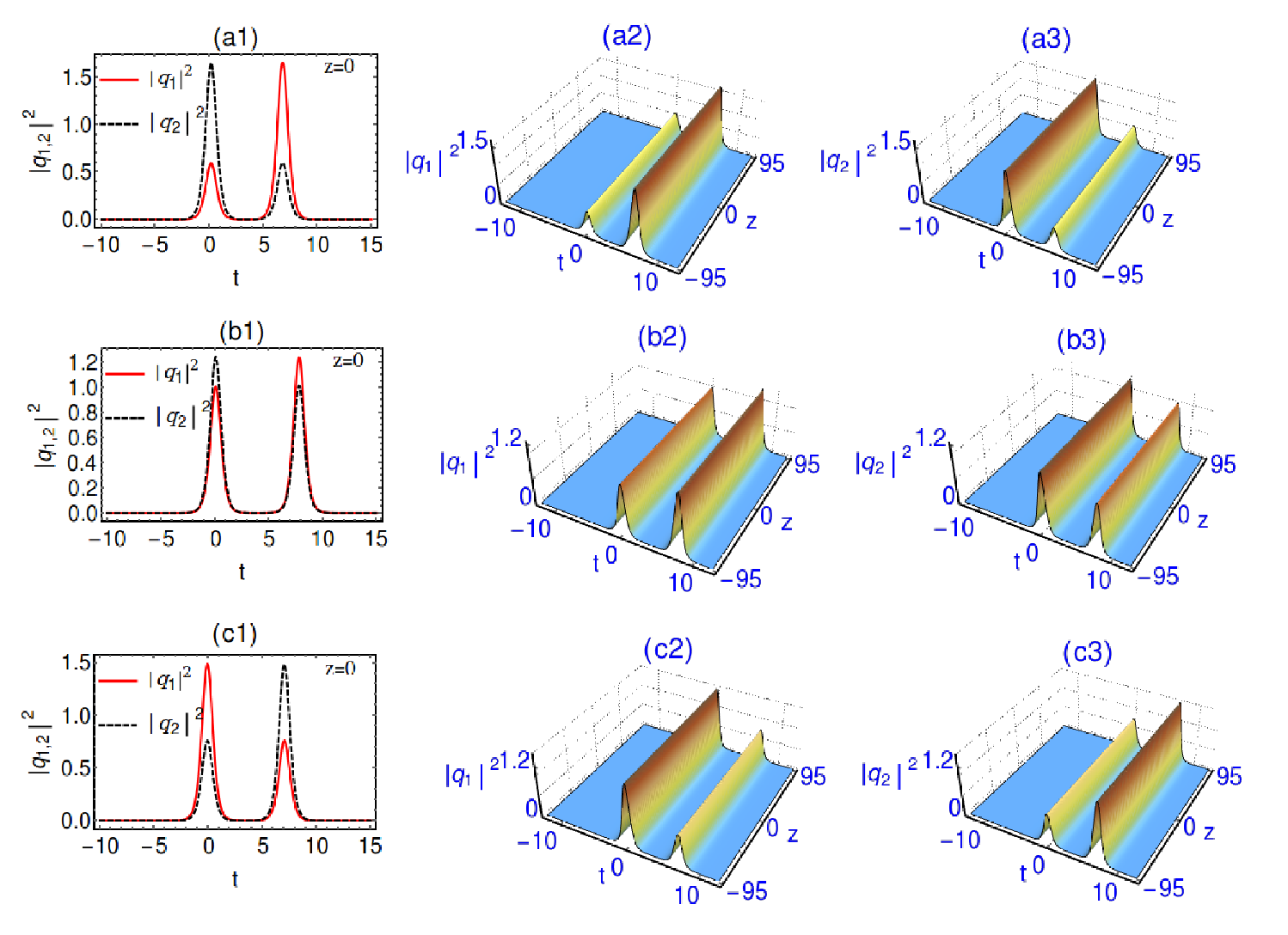}
	\caption{A dissociated SM is formed due to an asymmetric intensity distribution between the modes. Top-panels (a1)-(a3): The parameters are set as $k_1=1.5$, $l_1=1.499$, $\alpha_1^{(1)}=0.2$, $\alpha_1^{(2)}=\beta_1^{(j)}=1$, $j=1,2$, and $\Delta t_{21}=6.6175$. Middle Panels (b1)-(b3): Increasing  $\alpha_1^{(1)}$ to $0.8$ while keeping all the other parameters unchanged leads to a temporal separation of $\Delta t_{21}=7.7326$. Bottom Panels (c1)-(c3): Further increasing $\alpha_1^{(1)}$ to $1.8$ results in a temporal separation, $\Delta t_{21}=7.1364$. }
	\label{f1}
\end{figure*}

\begin{figure*}
	\centering
	\includegraphics[width=1.0\linewidth]{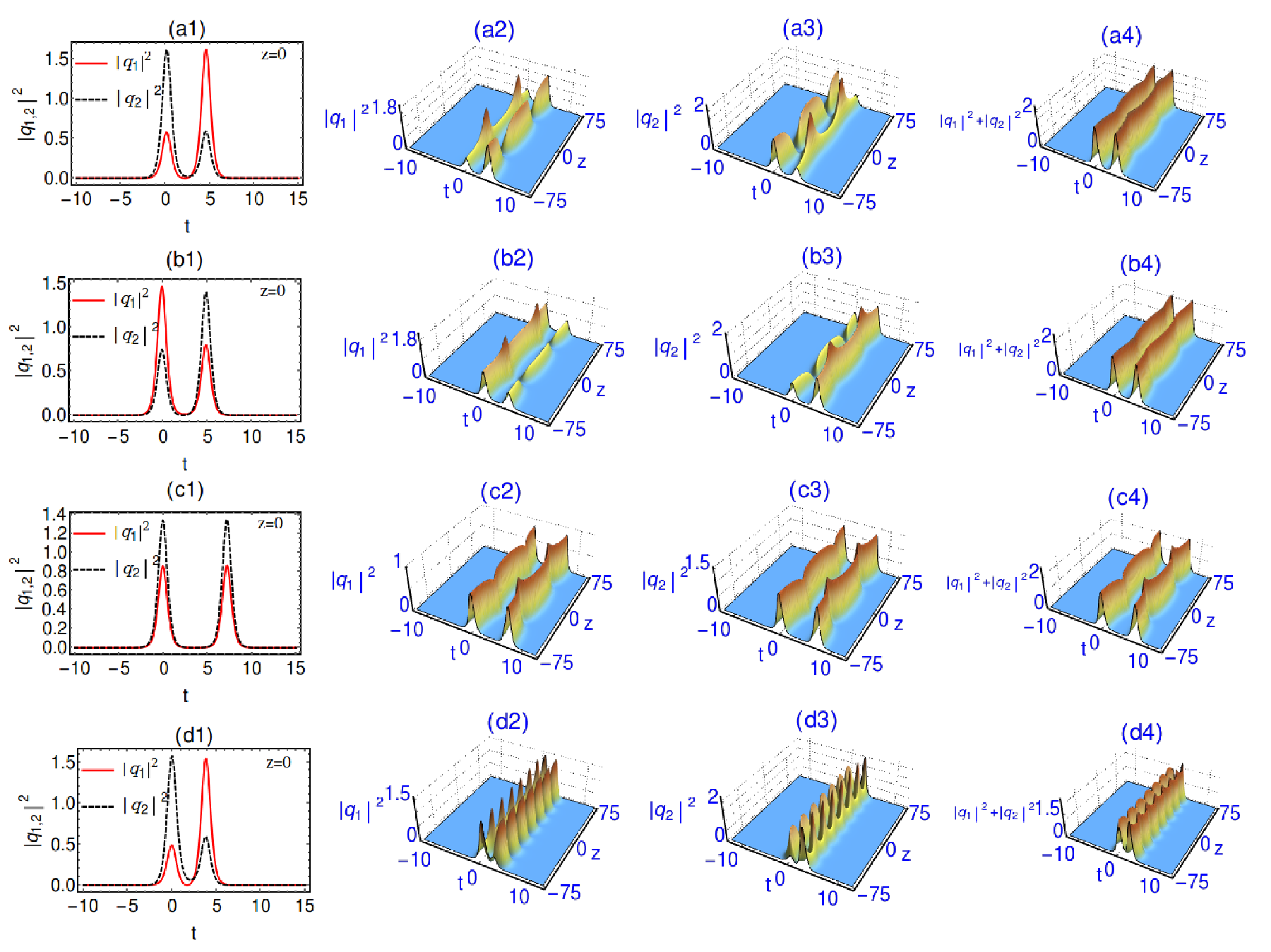}
	\caption{The first- and second-row panels (Figs. \ref{fig1}(a1)-\ref{fig1}(a4) and \ref{fig1}(b1)-\ref{fig1}(b4)) demonstrate the formation of a partially dissociated doublet SM state through an asymmetric intensity distribution among the modes, with temporal separations of $\Delta t_{21} = 4.4328$ and $\Delta t_{21} = 4.9558$, respectively. The third-row panels (Figs. \ref{fig1}(c1)-\ref{fig1}(c4)) illustrate a symmetrically intensity-distributed partially dissociated doublet SM with $\Delta t_{21} = 7.2303$, whereas the fourth-row panels (Figs. \ref{fig1}(d1)-\ref{fig1}(d4)) display the synthesis of a doublet molecule as two soliton atoms experience a strong interaction force. }
	\label{fig1}
\end{figure*}
\subsubsection{Mechanism for the formation of vector soliton molecule: Role of amplitude parameters } 
To unravel the mechanism behind the formation of the doublet SM and its related structures, we wish to analyse the solution (\ref{9a})-(\ref{9d}) for fixed values of $k_{1R}$ and  $l_{1R}$ while varying the value of any one of the complex constants $\alpha_1^{(j)}$ and $\beta_1^{(j)}$, $j=1,2$. For example, we start the analysis by varying the value of $\alpha_1^{(1)}$ in the solution (\ref{9a})-(\ref{9d}) with fixed values $k_{1R}=1.5$, $l_{1R}=1.499$, $\alpha_1^{(2)}=\beta_1^{(1)}=\beta_1^{(2)}=1$, see Figs. \ref{f1}(a1), \ref{f1}(a2) and \ref{f1}(a3). This case corresponds to analysing the dissociated molecular state ($k_{1R}-l_{1R}=0.001$), as we have classified SM below in Table I. By doing so we observe that the soliton intensities are distributed asymmetrically between the two modes for $\alpha_1^{(1)}=0.2$. Such intensity distribution is demonstrated in Figs. \ref{f1}(a1)-\ref{f1}(a3). In this case, the two solitons (left and right) are located at $t_1=0.1674$ and $t_2=6.7849$, respectively, resulting in a temporal separation of $\Delta t_{21}=6.6175$. Next, we increase the value of $\alpha_1^{(1)}$ from $0.2$ to $0.8$ while keeping the other parameters unchanged to examine the corresponding changes in the SM profile. This situation is illustrated in Figs. \ref{f1}(b1)-\ref{f1}(b3), which clearly demonstrate that the intensity distributions of the solitons remain slightly asymmetric, with further shifts in the positions of solitons. The computed temporal separation in this case is  $\Delta t_{21}=7.8049-0.0723=7.7326$, which is slightly larger than in the previous case. Further increasing  the value of $\alpha_1^{(1)}$  to $1.8$ results in another asymmetric distribution of soliton intensities, but in a different manner. Such asymmetric intensity profiles are shown in Figs. \ref{f1}(c1)-\ref{f1}(c3). The corresponding temporal separation between the soliton constituents is calculated as $\Delta t_{21}=7.0448-(-0.0916)=7.1364$. The latter numerical value, along with Figs. \ref{f1}(c1)-\ref{f1}(c3), confirms that the temporal separation decreases slightly in this third situation. From these observations, we conclude that tuning $\alpha_1^{(1)}$ predominantly influences the intensity distribution between the modes and it alters the temporal positions of the soliton atoms slightly. The slight temporal shifts acquired by the soliton atoms do not induce the breathing pattern in the soliton molecule since in the present case soliton atoms repel each other. Therefore, no molecular synthesis (or oscillation) can occur as evidenced by the intensity profiles of SM from Fig. \ref{f1}.  Therefore, the complex parameters, $\alpha_1^{(j)}$ and $\beta_1^{(j)}$, $j=1,2$, mainly determine the intensity distributions between the modes $q_1$ and $q_2$.  

We repeat the same analysis by varying $\alpha_1^{(1)}$  while fixing the difference: $k_{1R}-l_{1R}=1.5=1.472=0.028$. The other parameter values are fixed as $\alpha_1^{(2)}=\beta_1^{(1)}=\beta_1^{(2)}=1$. We observe an asymmetric intensity distribution occurs in between the modes for $\alpha_1^{(1)}=0.2$. Such intensity distribution is depicted in Figs. \ref{fig1}(a1)-\ref{fig1}(a3), where two bright solitons having unequal intensities along with temporal separation, $\Delta t_{21}=t_2-t_1=4.4328$, set up the partially dissociated doublet SM. Here, $t_1$ and $t_2$ are the central positions of left and right solitons located at $t=0.1561$ and  $t=4.5889$, respectively. If we increase the value of $\alpha_1^{(1)}$ from $0.2$ to $1.8$, the intensities of the two incoherent solitons again get distributed asymmetrically among the modes but in a different way as it is displayed in Figs. \ref{fig1}(b1)-\ref{fig1}(b3). In this case, the temporal separation between the constituent gets increased only slightly. That is, $\Delta t_{21}=t_2-t_1=4.8598-(-0.0959)=4.9558$. Note that the periodicity does not change in the intensity profiles of SM. In addition, it is also possible to construct a doublet molecular structure through a symmetric intensity distribution when we choose the parameter values: $\alpha_1^{(1)}=0.8$, $\alpha_1^{(2)}=1$, $\beta_1^{(1)}=1.2$, and $\beta_1^{(2)}=1.5$, which correspond to the parameter ratio $\alpha_1^{(1)}:\beta_1^{(1)}=\alpha_1^{(2)}:\beta_1^{(2)}=0.6666$. The corresponding symmetric intensity profiles of  the fundamental SM are depicted in the third-panels of Fig. \ref{fig1} (Figs. \ref{fig1}(c1)-(\ref{fig1}(c3)). In this situation, the soliton atoms are well separated from each other, specified by the large temporal separation $\Delta t_{21}=7.1789-(-0.0514)=7.2303$. From the first three panels of Fig. \ref{fig1}, we observe that while the periodicity remains preserved, the intensity gets distributed either asymmetrically or symmetrically among the modes. Another important observation here is that, although the temporal positions of the solitons vary, the weakly periodic nature remains unchanged throughout the evolution. This suggests that the complex parameters $\alpha_1^{(j)}$'s and $\beta_1^{(j)}$'s do not influence the partially dissociated nature (or breathing pattern) of the SM.  In all these three cases the total energy, $|q_1|^2+|q_2|^2$, is always constant. The figures explaining this fact are displayed in the right most figures of Fig. \ref{fig1}. 

An important observation from all the first three-panels of Fig. \ref{fig1} mentioned above is that, the intensity profiles (symmetric/asymmetric) of the doublet molecular structure are non-identical in the two modes ($|q_1|^2\neq |q_2|^2$) and they are also weakly periodic, meaning that the periodicity occurs on a much larger scale in the $z$-direction. This implies that the soliton atoms experience a weak force imparted by the self- and cross-Kerr effects. However, they start to experience a strong force if the temporal separation is close to the width of a single vector soliton. Such a possibility is demonstrated in the bottom-panel of Fig. \ref{fig1} (Figs. \ref{fig1}(d1)-\ref{fig1}(d4)) when we fix the separation $\Delta t_{21}=3.7843$ (this value is close to the width of a single soliton). To fix the latter temporal separation we consider the parameter values as $k_{1R}=1.5$, $l_{1R}=1.4$, $\alpha_1^{(1)}=0.2$, $\alpha_1^{(2)}=\beta_1^{(1)}=\beta_1^{(2)}=1$ (note that the value of $l_{1R}$ is different in this fourth case). From the Figs. \ref{fig1}(d1)-\ref{fig1}(d4), we observe that a rapid periodic oscillations emerging in the structure of basic molecular state due to the attractive and repulsive forces acting periodically on the constituents.  

\begin{figure}
\includegraphics[scale=0.5]{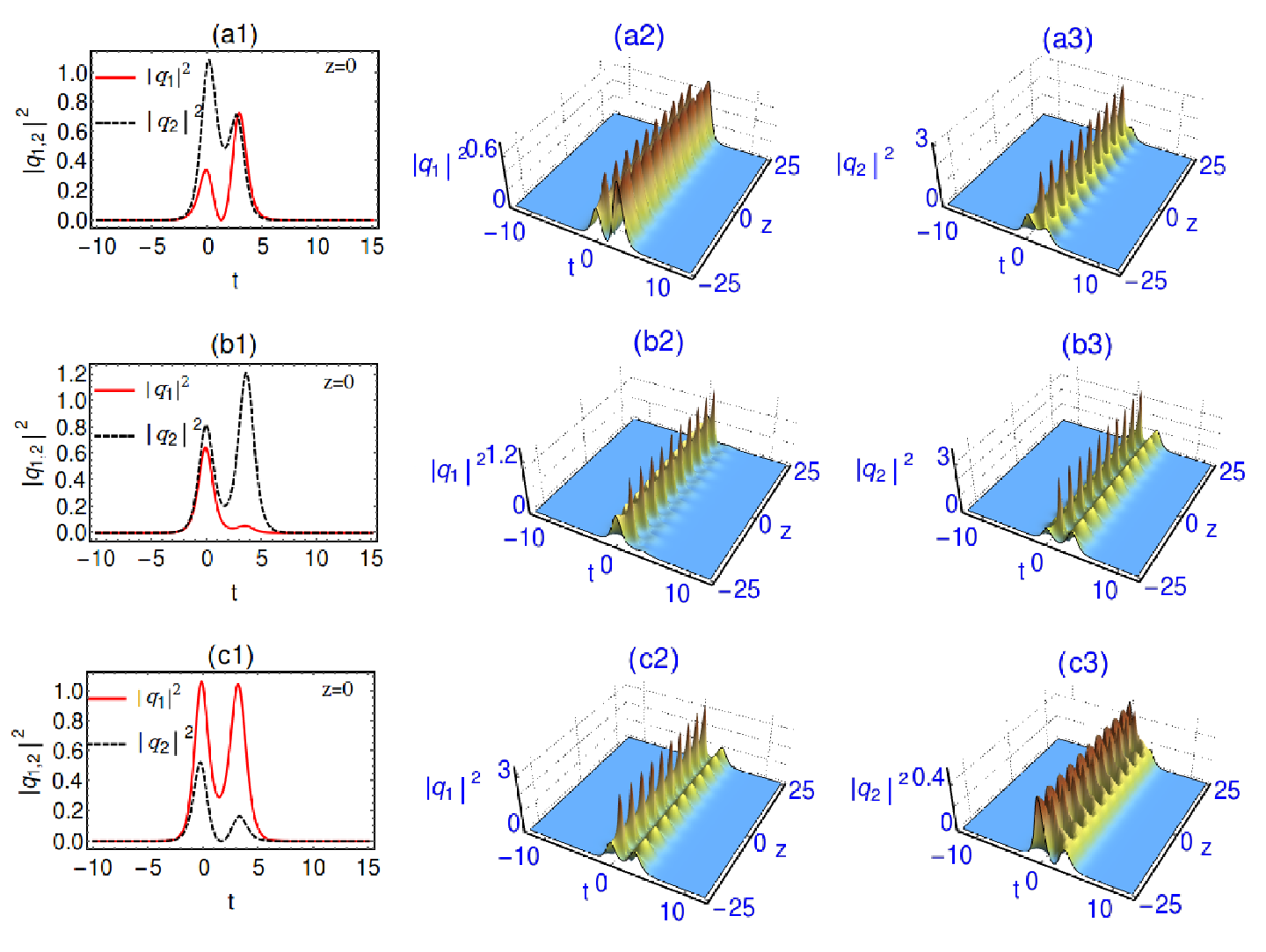}
\caption{Top-panels (a1)-(a3): A synthesized SM is formed with an asymmetric intensity distribution in the $q_1$ mode and a symmetric intensity distribution in the $q_2$ mode. These figures are obtained by setting $k_1=1.5$, $l_1=1.0$, $\alpha_1^{(1)}=0.2$, $\alpha_1^{(2)}=\beta_1^{(j)}=1$, $j=1,2$, and $\Delta t_{21}=3.0074$. Middle Panels (b1)-(b3): A synthesized SM is formed with an asymmetric intensity distribution between the modes. Here, $\alpha_1^{(1)}=0.8$ (while the other parameter values remain the same as in the previous case) and $\Delta t_{21}=3.5604$. Bottom Panels (c1)-(c3): A synthesized SM is displayed for $\alpha_1^{(1)}=1.8$ and $\Delta t_{21}=3.32$.}
\label{f3}
\end{figure}
Besides the above, we also consider the condition $k_{1R}-l_{1R}>0$ and repeat the same analysis as we have done for the previous cases. For this purpose, we set $\alpha_1^{(1)}=0.2$  and $k_{1R}-l_{1R}=1.5-1.0=0.5$. Under this parametric choices, a synthesized molecular state emerges with strong periodic oscillations. This molecular state is illustrated in Figs. \ref{f3}(a1)-\ref{f3}(a3). By determining the soliton positions, we calculate the temporal separation as $\Delta t_{21}=2.898-(-0.1094)=3.0074$ (this value is close to the width of a single soliton). Similarly, we examine the synthesized molecular state by setting $\alpha_1^{(1)}=0.8$ and $\alpha_1^{(1)}=1.8$, as shown in the middle (Figs. \ref{f3}(b1)-\ref{f3}(b3)) and bottom (Figs. \ref{f3}(c1)-\ref{f3}(c3)) panels of Fig. \ref{f3}. These figures confirm that the strong periodic nature of the synthesized molecular state is preserved, with changes occurring only in the intensity distribution (either asymmetric or symmetric). These simple changes in the intensity distribution do not affect the breathing nature of soliton molecules. Therefore, the complex parameters $\alpha_1^{(j)}$'s and $\beta_1^{(j)}$'s do not influence the breathing pattern of the synthesized SM but they determine the intensity distribution in the synthesized soliton molecular structure. On the other hand the real parts of the wave numbers $k_1$ and $l_1$ predominantly influence the breathing pattern of bound molecular state, which is evident from the values of $|k_{1R}-l_{1R}|$ in Figs. \ref{f1}, \ref{fig1} and \ref{f3}, which are respectively $0.001$, $0.028$ and $0.5$.

\begin{figure*}
	\centering
	\includegraphics[width=0.27\linewidth]{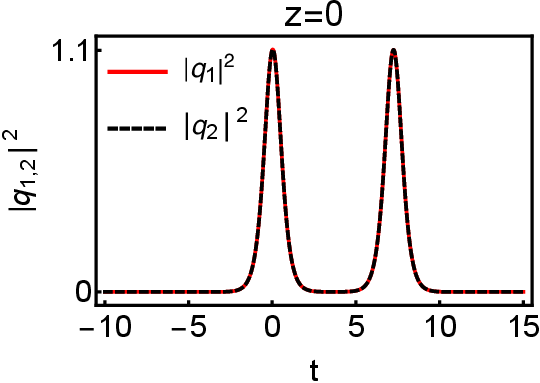}
	~\includegraphics[width=0.63\linewidth]{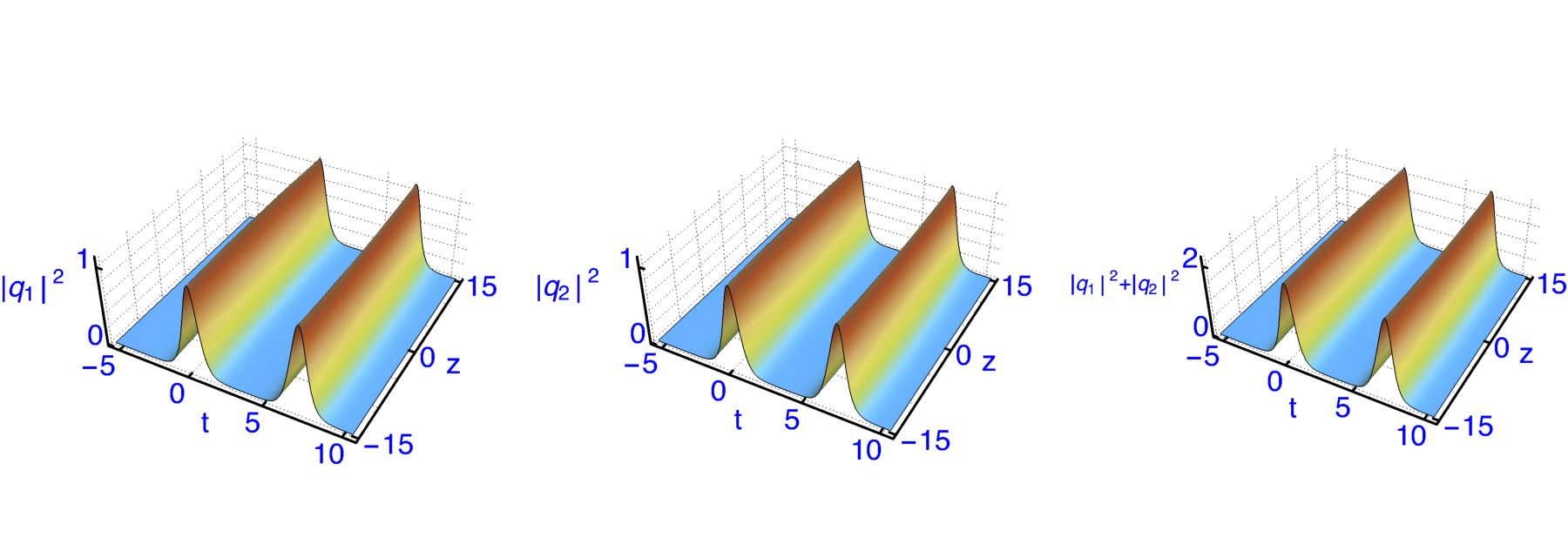}
	\caption{A dissociated SM made by an identical intensity profiles in both the modes is displayed for $k_1=1.5$, $l_{1}=1.492$, and $\alpha_{1}^{(j)}=\beta_{1}^{(j)}=1$, $j=1,2$. }
	\label{fig2}
\end{figure*}

In addition, for completeness, we also display the fundamental molecular structure having identical intensity profiles in both the modes in Fig. \ref{fig2}, for the temporal separation $\Delta t_{21}=7.2472-0.3333=7.2139$. This molecular state can be treated as the basic bound soliton state of the scalar NLS equation. It is because the solution (\ref{9a})-(\ref{9d}) of the Manakov system (\ref{manakov}) can be reduced to the basic BSS solution of the NLS equation when the wave parameters are chosen as $\alpha_1^{(1)}=\alpha_1^{(2)}=\beta_1^{(1)}=\beta_1^{(2)}=1$. This choice makes $q_1=q_2$ \cite{sun}.  
   
\begin{figure*}
	\centering
	\includegraphics[width=0.65\linewidth]{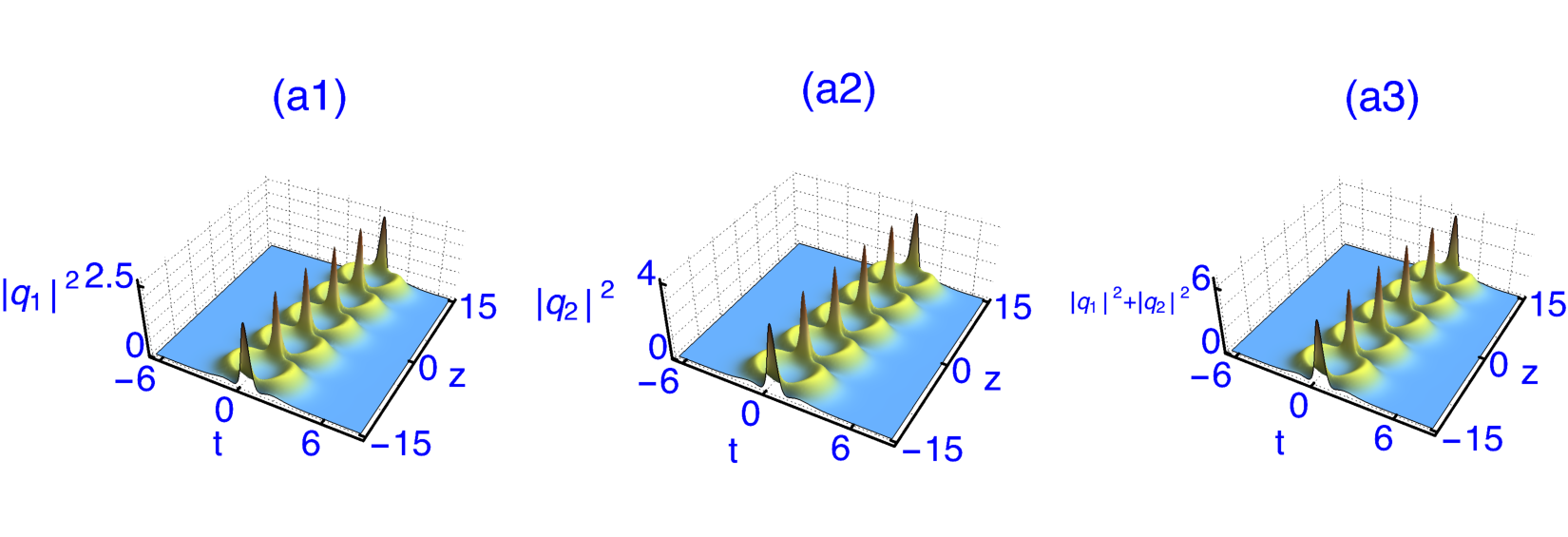}\\
	\includegraphics[width=0.65\linewidth]{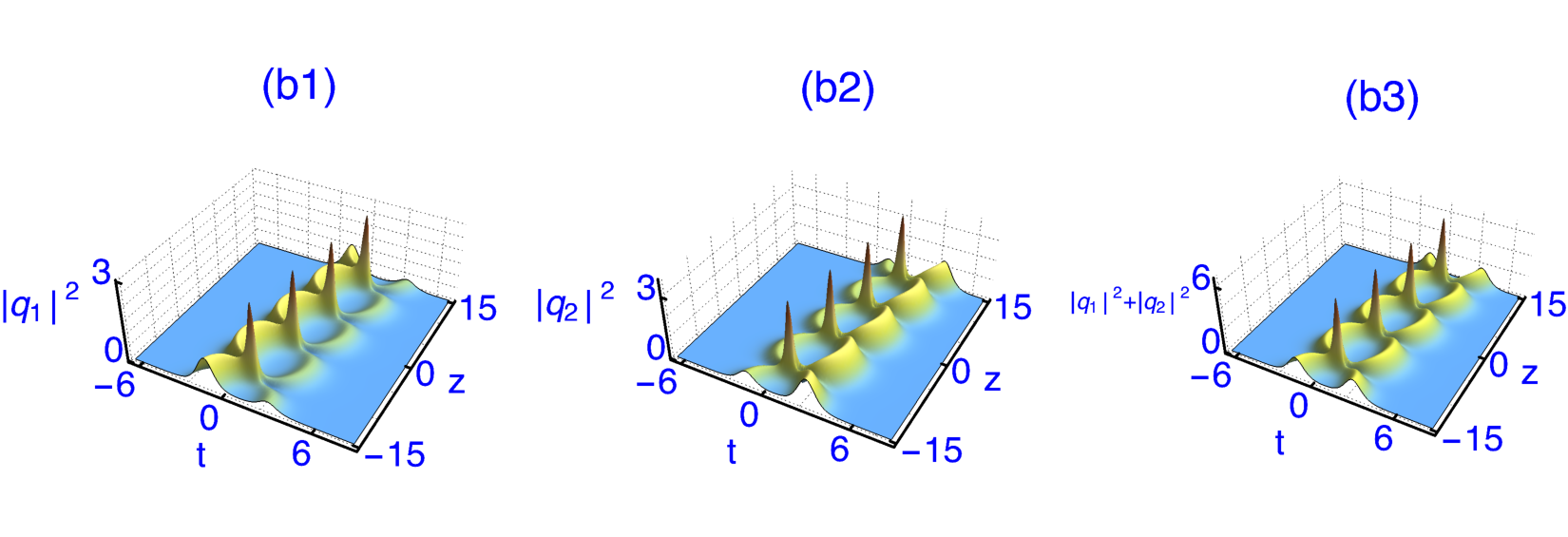}\\
	\includegraphics[width=0.65\linewidth]{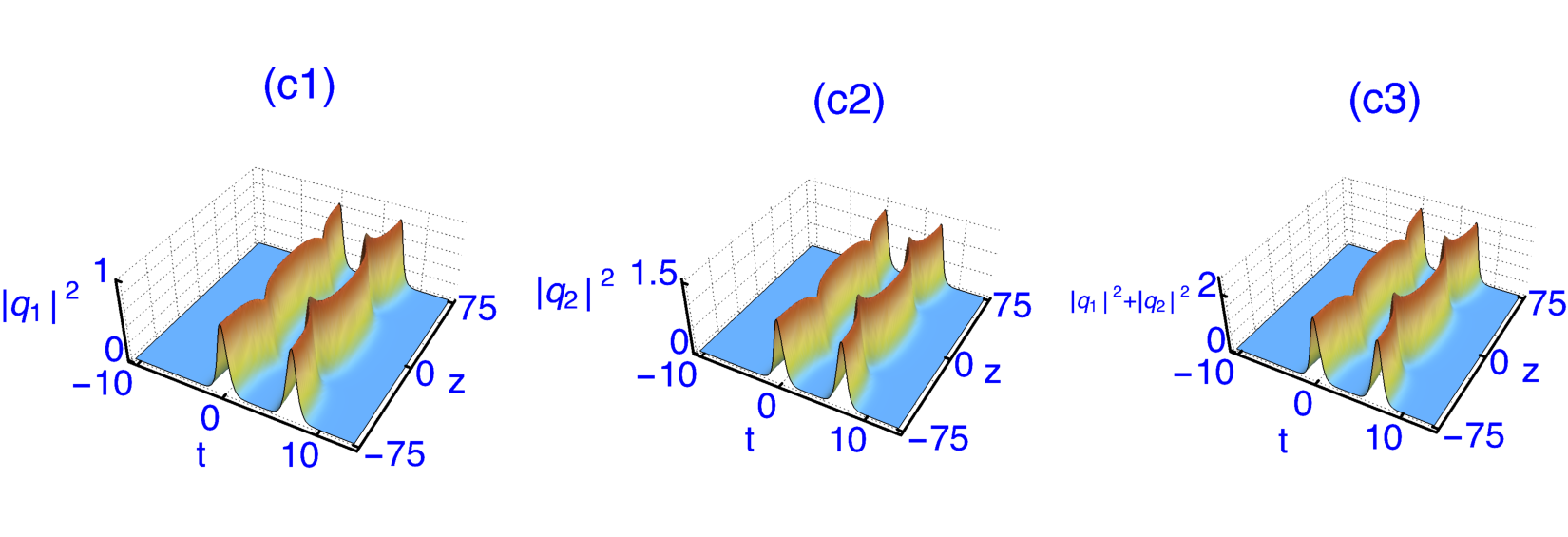}\\
	\includegraphics[width=0.65\linewidth]{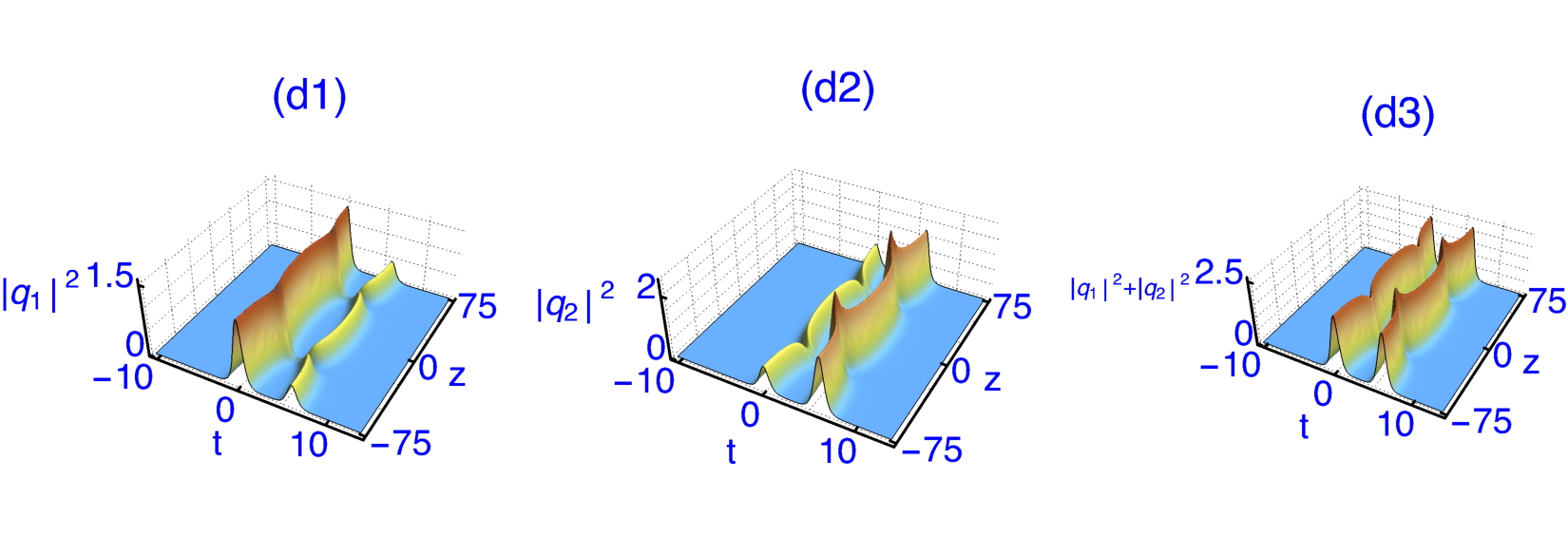}\\
	\caption{The various possible structures admitted by the fundamental SM or doublet SM are demonstrated by tuning the amplitude dependent temporal separation. To draw these figures we fix the parameter values as follows: A symmetric doublet SM synthesis is displayed in (a1)-(a3) for $k_{1}=1.5+0.04i$, $l_{1}=1.1+0.04i$, $\alpha_{1}^{(1)}=0.8$, $\alpha_{1}^{(2)}=1$, $\beta_{1}^{(1)}=1.2$ and $\beta_{1}^{(2)}=1.5$. An asymmetric fundamental SM is brought out in Figs. (b1)-(b3) for $k_{1}=1.5+0.04i$, $l_{1}=1.2+0.04i$, $\alpha_{1}^{(1)}=1$, $\alpha_{1}^{(2)}=0.8$, $\beta_{1}^{(1)}=1.4$ and $\beta_{1}^{(2)}=1$. A partially dissociated SM with symmetric intensity distribution is depicted in Figs. (c1)-(c3) for $k_1=1.5$, $l_{1}=1.472$, $\alpha_{1}^{(1)}=0.8$, $\alpha_{1}^{(2)}=1$, $\beta_{1}^{(1)}=1.2$ and $\beta_{1}^{(2)}=1.5$. A partially dissociated doublet SM with asymmetric intensity distribution is displayed in Figs. (d1)-(d3) for $k_1=1.5$, $l_{1}=1.472$, $\alpha_{1}^{(1)}=1$, $\alpha_{1}^{(2)}=0.8$, $\beta_{1}^{(1)}=1.4$ and $\beta_{1}^{(2)}=1$. }
	\label{fig3}
\end{figure*}
\subsubsection{Classification of fundamental soliton molecules}
\begin{figure*}
	\centering
	\includegraphics[width=0.7\linewidth]{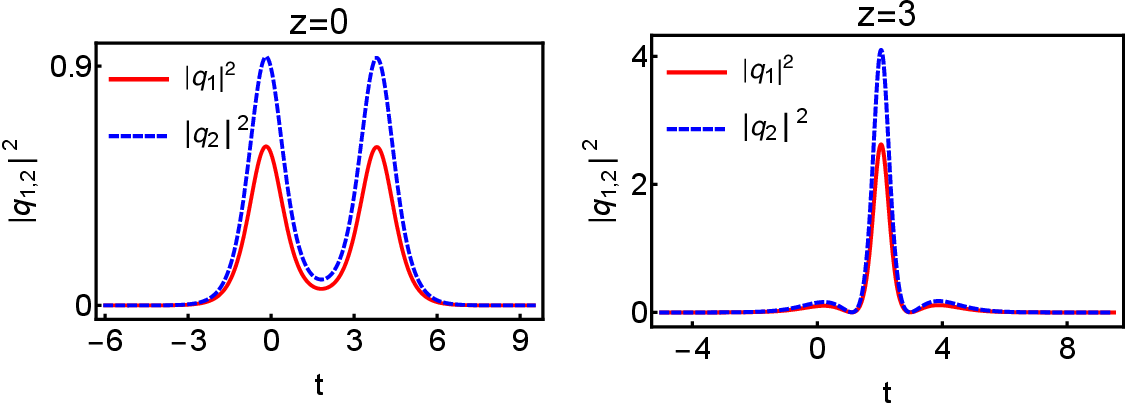}\\
	\includegraphics[width=0.35\linewidth]{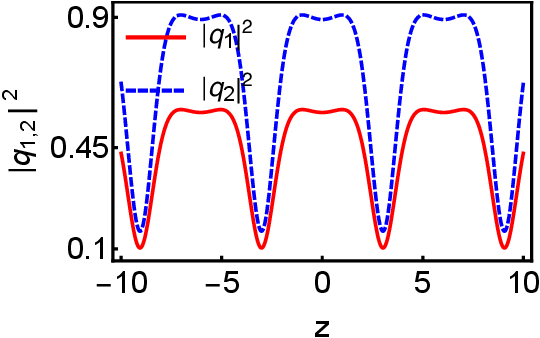}~
	\caption{During the repulsive process, the maximum temporal separation covered by the two soliton atoms is displayed in the top left panel. The formation of intensity peak during the attractive process is depicted in the top right panel for $z=3$. The bottom panel illustrates the periodic nature of a doublet SM and is drawn for $t=1$.   }
	\label{fig4}
\end{figure*}

Next, based on the temporal separation and periodicity, we classify the basic SM admitted by the BSS solution (\ref{9a})-(\ref{9d}) as either dissociated or synthesized molecular state. This classification can be made simply by the appearance of breathing pattern in the overlapping region. We note here that the breathing pattern is predominantly governed by the real parameters $k_{1R}$ and $l_{1R}$, so it is more relevant to classify the SMs based on these amplitude parameters only, as we have pointed out in Table I below.  If the breathing pattern is not visible in the structure of SM, then such a molecule is categorized as dissociated type. To bring out this type of molecular state the real parameters $k_{1R}$ and $l_{1R}$ should be very close to each other.  On the other hand, the emergence of periodic oscillation signifies the synthesis of a novel molecule state, which can be brought by assuming either $k_{1R}>l_{1R}$ or $k_{1R}<l_{1R}$.  To demonstrate this difference, we first consider the doublet molecular states which are displayed in Figs. \ref{fig3}(a1)-(a3) and \ref{fig3}(b1)-(b3). These molecular states  are essentially formed by the symmetric and asymmetric intensity distributions and also due to the fact that the temporal separation between the constituents is of the order of width of a single soliton. In both the cases, the two soliton atoms get overlapped and they exhibit periodic oscillations in the coalescence region due to the continuous action of attractive and repulsive forces alternately. Therefore, these states are categorized as synthesized molecular states. During the periodic attraction and repulsion, merging and splitting of two solitons occur at a regular interval of fiber length $z$. For example, during the repulsion, the maximum temporal separation covered by the soliton atoms in Fig. 3(a1)-(a3) is $\Delta t_{21}=3.8227-(-0.1799)=4.0026$. Such a maximum separation repeatedly occurs at every full cycle $nT_{12}$, $n=0,\pm 1, \pm 2,...$, where $T_{12}$ is the period of oscillation. We illustrate this separation for $z=0$ in the top-panel (left) of Fig. \ref{fig4}. Then, during the periodic attraction, the intensity peaks are formed periodically along the fiber length at every half cycle $(n+\frac{1}{2})T_{12}$, $n=0,\pm 1, \pm 2,...$. A typical intensity profile in the top-panel (right) of Fig. \ref{fig4} demonstrates the merging of two solitons at  $z=3$. To further confirm the periodic nature of molecular state demonstrated in Fig. \ref{fig3}(a1)-(a3) we illustrate the corresponding BSS solution (\ref{9a})-(\ref{9d}) in the bottom-panel of Fig. \ref{fig4} for a fixed time $t=1$ along the propagation direction $z$. This figure also confirms that the breathing pattern always appears once the soliton molecule is synthesized.      
\begin{table}[h!]
\centering
\begin{tabular}{|p{4.0cm}|p{4.0cm}|p{4.0cm}|p{4.0cm}|}
\hline
\multicolumn{4}{|c|}{Classification of Fundamental Soliton Molecule} \\
\hline
Condition & Temporal separation $\Delta t_{12}$ & Periodic Nature & Types of SMs \\
\hline
$k_{1R} \approx l_{1R}$  &  It is greater than the width of a single soliton & almost zero or zero oscillation & completely  dissociated SM \\
\hline
$k_{1R}$ is slightly greater than $l_{1R}$ & It is slightly greater than the width of a single soliton & weakly periodic & partially dissociated SM \\ \hline
$k_{1R} > l_{1R}$ or $k_{1R} < l_{1R}$ & It is equal to the width of a single soliton & strongly periodic & synthesized SM \\ 
\hline
$k_{1R}\gg l_{1R}$ or $k_{1R}\ll l_{1R}$ & It is less than the width of a single soliton & oscillation with single-humped profile & composite SM \\
\hline
\end{tabular}
\caption{Classification of fundamental soliton molecule.}
\label{table:soliton_classification}
\end{table}

\begin{figure*}
	\centering
	\includegraphics[width=0.4\linewidth]{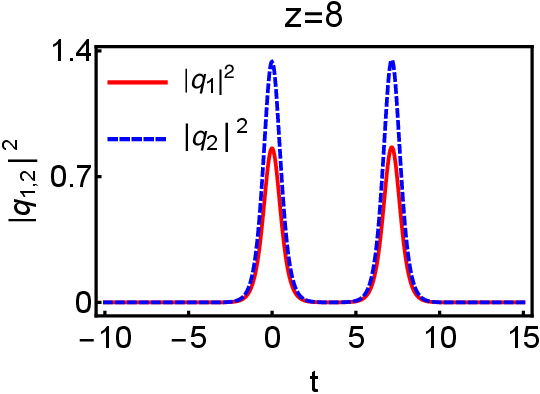}
	\includegraphics[width=0.4\linewidth]{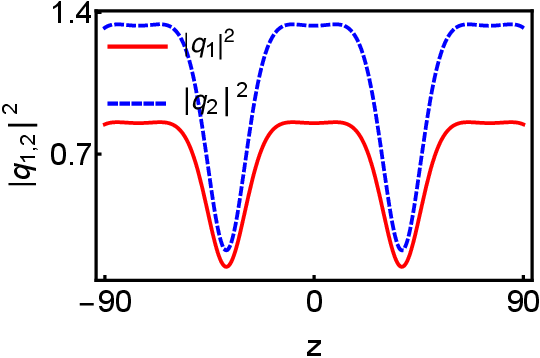}
	\caption{The intensity profile of a slightly dissociated molecular state is displayed in the left panel to demonstrate the maximum relative temporal separation between the constituents. The corresponding periodic nature of doublet SM is illustrated in the right panel for $t=1$.  }
\label{fig5}
\end{figure*}

Then, to distinguish a dissociated molecule from the synthesized  molecular state the basic SM states are depicted in Figs. \ref{fig3}(c1)-(c3) and \ref{fig3}(d1)-(d3) with large temporal separation. These doublet molecular states are made by symmetric and asymmetric intensity distributions respectively. We designate these two molecular states as dissociated molecules since their binding soliton atoms are well separated temporally and they exhibit weak periodic oscillations along the $z$-direction. To confirm this, we have considered a molecular state that is displayed in Figs. \ref{fig3}(c1)-(c3) and calculated the temporal separation value between the binding soliton atoms as $\Delta t_{21}=7.1415-(-0.0151)=7.1566$. This value shows that the two solitons are well separated with each other as it is demonstrated in Fig. \ref{fig5} for a particular value of fiber length $z=8$. Then, the difference $k_{1R}^2-l_{1R}^2=2.25-2.1668=0.0832$ implies that the period of oscillation is large. Because of this, the periodic attraction and repulsion do not occur rapidly as it is shown in the right-side panel of Fig. \ref{fig5}. This fact is also true in the case of asymmetric doublet molecular state demonstrated in Figs. \ref{fig3}(d1)-(d3), where the two solitons are well separated and the  weak oscillations appear in their structure for large values of $z$.  

\begin{figure*}
	\centering
	\includegraphics[width=0.8\linewidth]{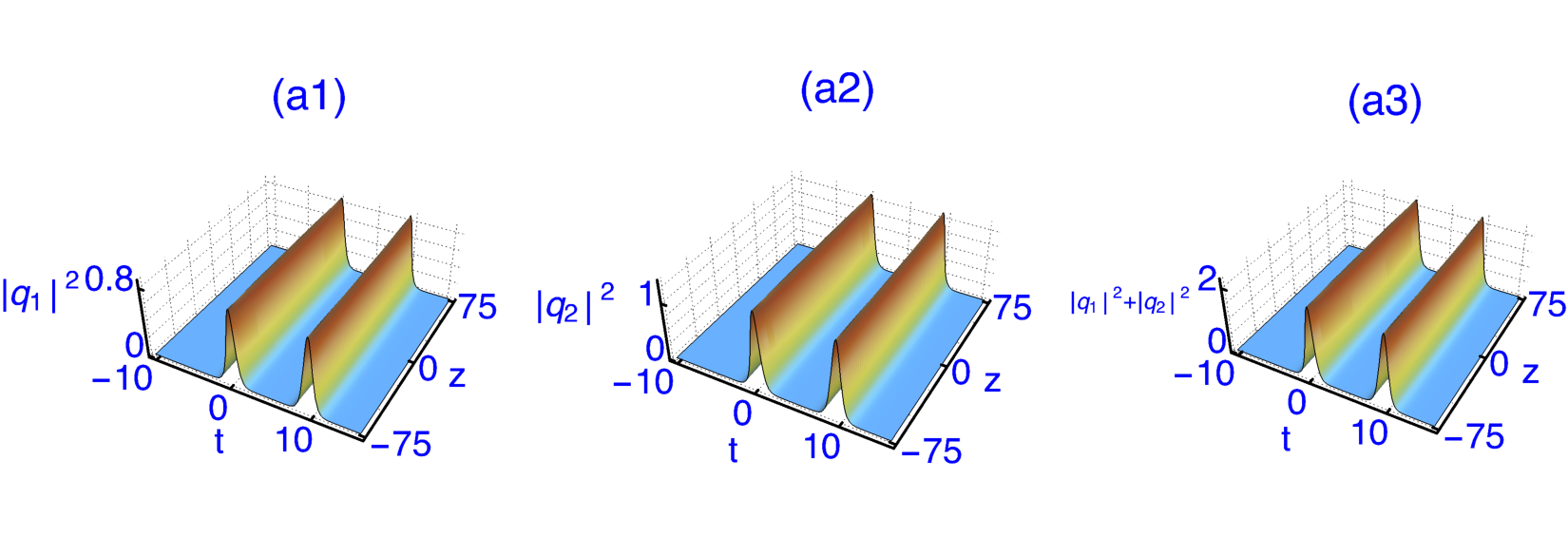}\\
	\includegraphics[width=0.8\linewidth]{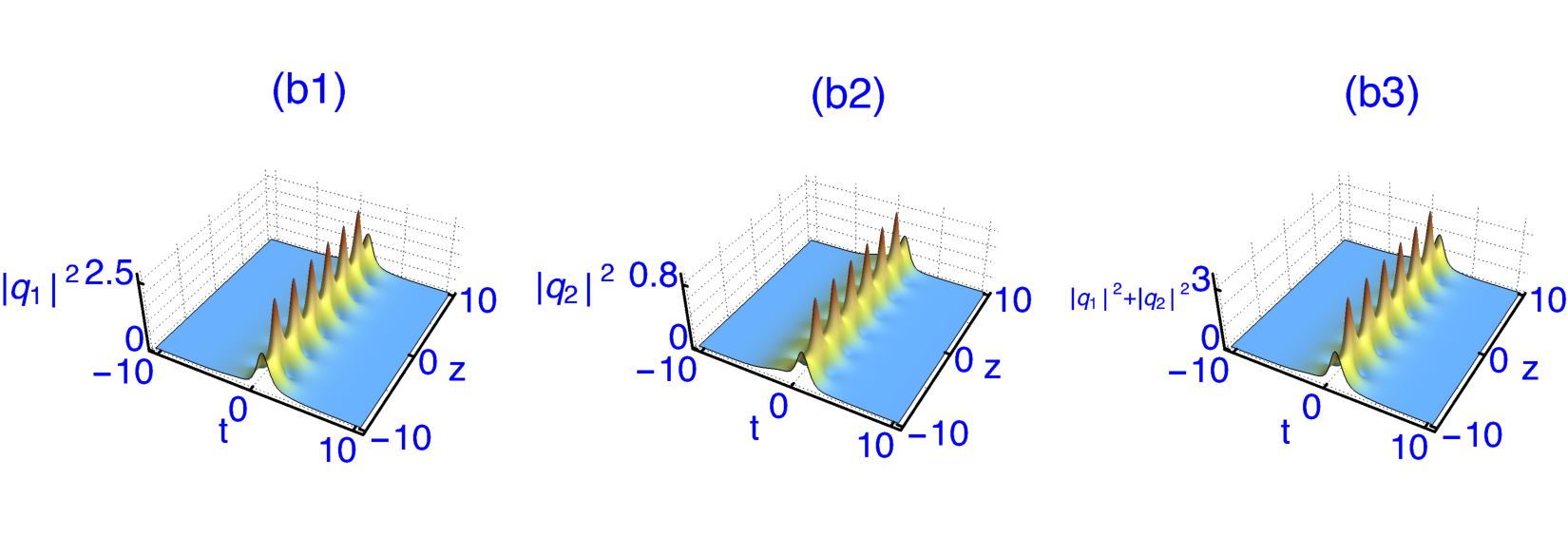}
	\caption{A completely dissociated doublet SM structure is demonstrated in the top-row panels by fixing $k_{1}=1.5$, $l_{1}=1.495$, $\alpha_{1}^{(1)}=0.8$, $\alpha_{1}^{(2)}=1$, $\beta_{1}^{(1)}=1.2$ and $\beta_{1}^{(2)}=1.5$. In the bottom panels, a composite doublet molecular state is illustrated for $k_{1}=1.5$, $l_{1}=0.45$, $\alpha_{1}^{(1)}=1$, $\alpha_{1}^{(2)}=0.7$, $\beta_{1}^{(1)}=1$, and $\beta_{1}^{(2)}=1$. }
\label{fig6}
\end{figure*}

Further, one can also have a completely dissociated doublet SM having zero oscillations in its structure. Such a molecular state is displayed in Figs. \ref{fig6}(a1)-(a3),  where the constituents experience a net zero force. In this case, the temporal separation value is $\Delta t_{21}=9.4469-(-0.0447)=9.4916$ and the period of oscillation is, $T_{12}=\frac{2\pi}{2.25-2.235}=\frac{2\pi}{0.015}$. Besides this, we also depict the existence of a composite doublet SM, which is essentially formed by merging the two solitons. That is, the soliton having a small amplitude coincides with its binding partner which is having a large amplitude. This type of SM is depicted in Fig. \ref{fig6}(b1)-(b3). From this figure, one can notice that the periodic oscillation persists in this case and is dictated by the time period $T_{12}=\frac{2\pi}{2.25-0.2025}=\frac{2\pi}{2.0475}$.    

\begin{figure*}
	\centering
	\includegraphics[width=0.7\linewidth]{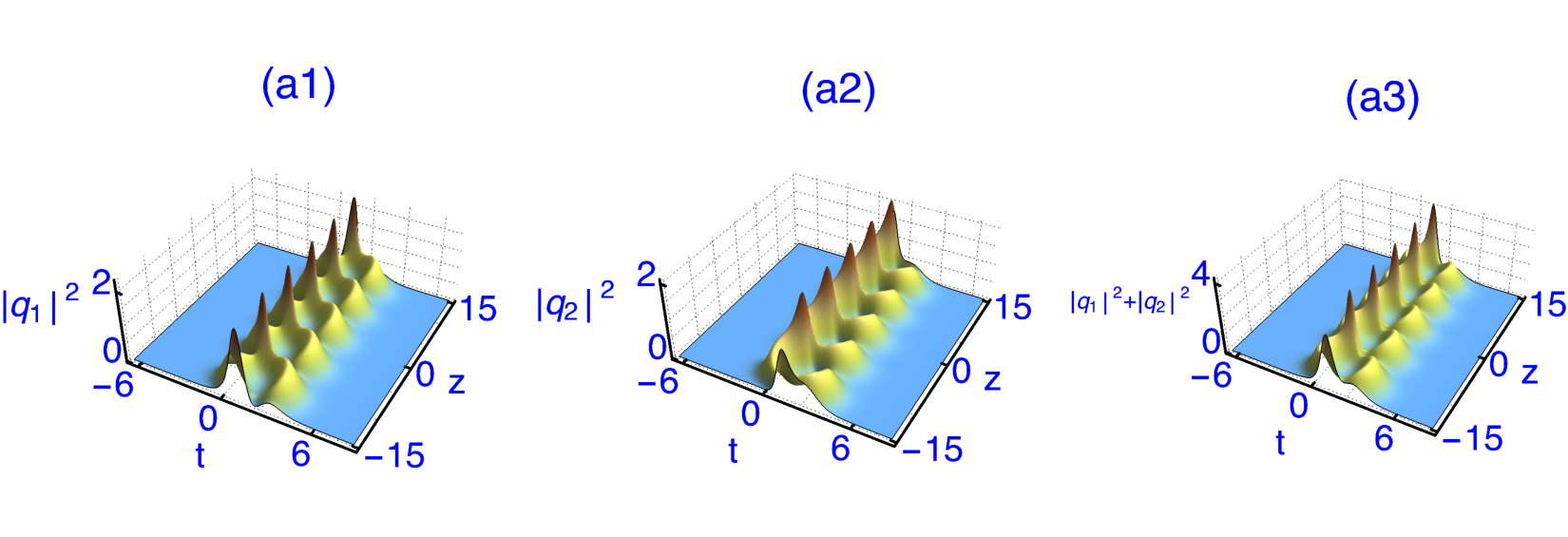}\\
	\includegraphics[width=0.7\linewidth]{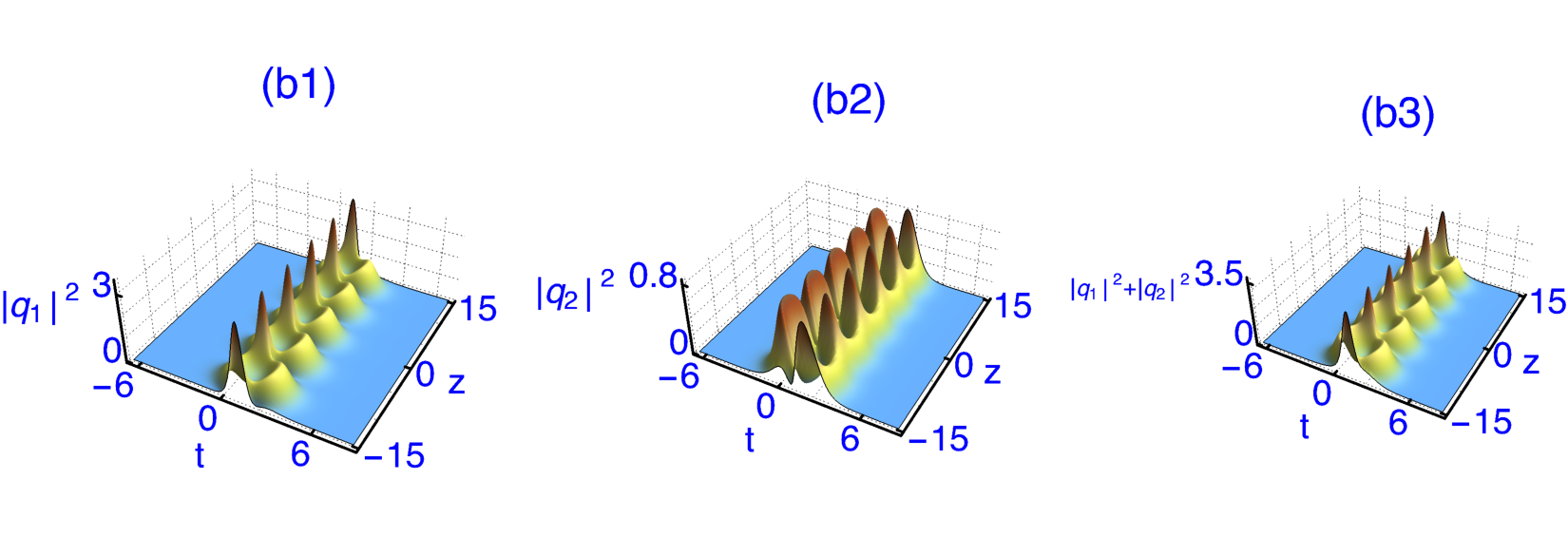}\\
	\includegraphics[width=0.7\linewidth]{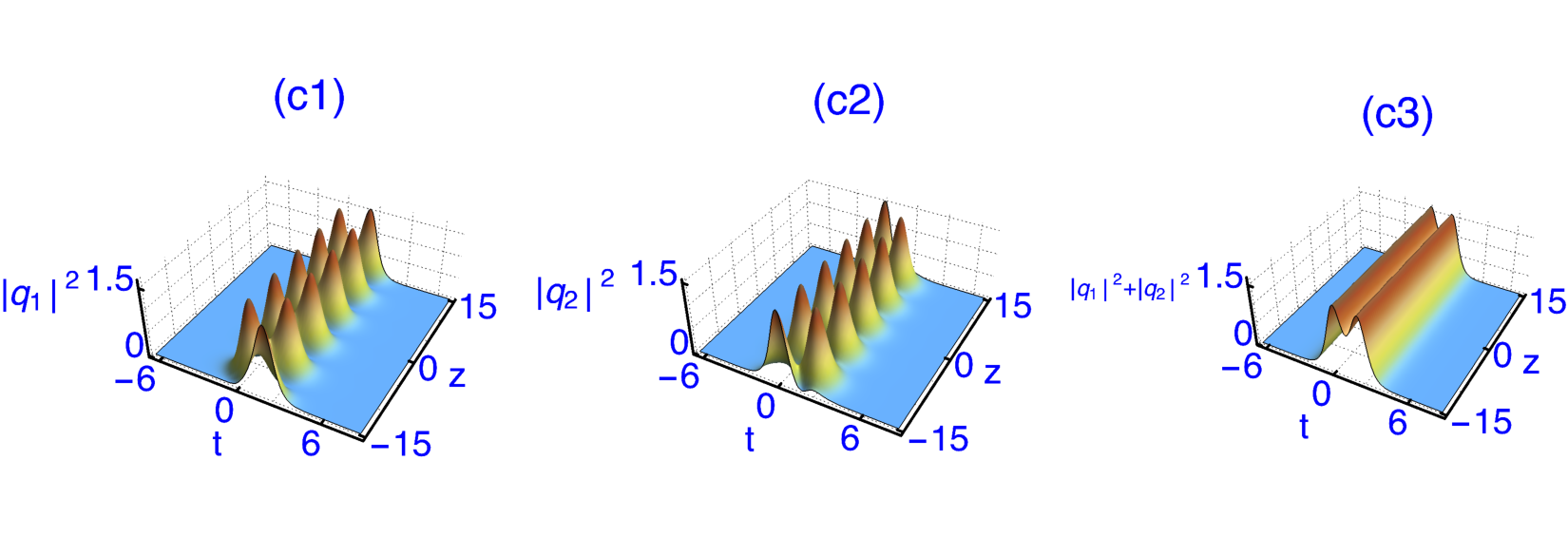}\\
	\includegraphics[width=0.7\linewidth]{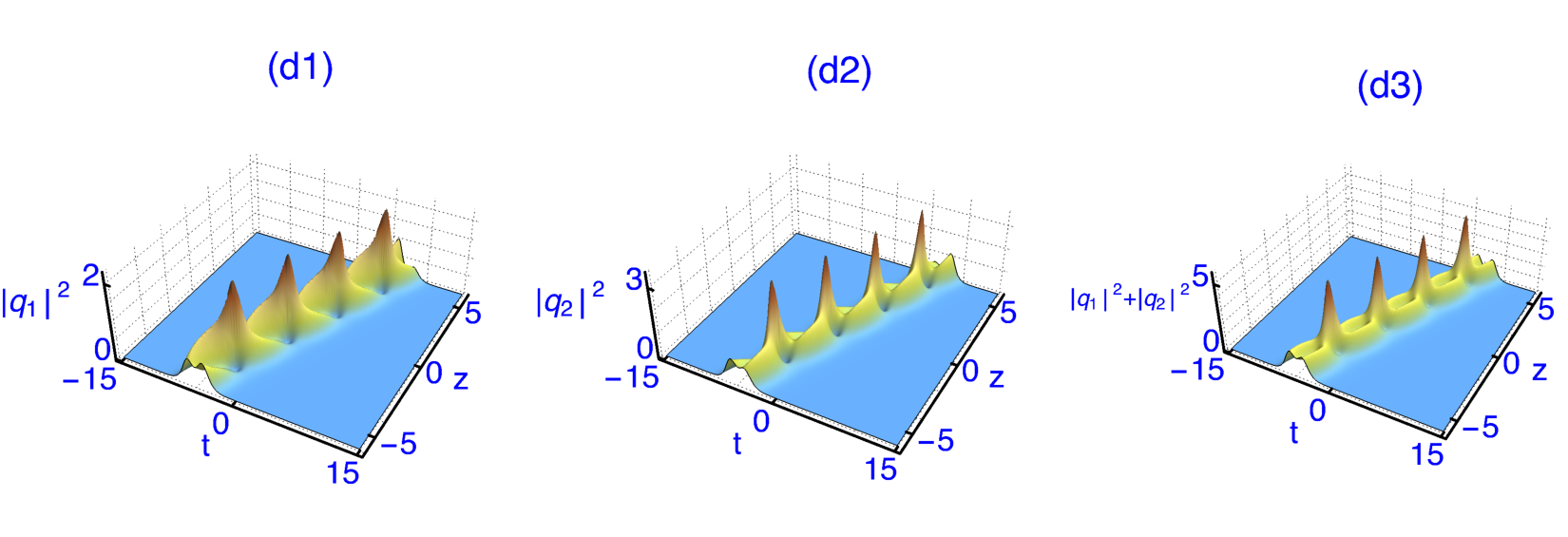}
	\caption{The various possible synthesized fundamental molecular structures formed by asymmetric intensity distribution are displayed. Here, we fix the parameter values as follows: For \ref{fig7}(a1)-(a3): $k_{1}=1.5$, $l_{1}=1.1$, $\alpha_{1}^{(1)}=0.9+0.5i$, $\alpha_{1}^{(2)}=1$, $\beta_{1}^{(1)}=1$, and $\beta_{1}^{(2)}=1.2+i$. For Figs. \ref{fig7}(b1)-(b3): $k_{1}=1.5$, $l_{1}=1.1$, $\alpha_{1}^{(1)}=1+i$, $\alpha_{1}^{(2)}=1$, $\beta_{1}^{(1)}=0.7+0.7i$, and $\beta_{1}^{(2)}1.5$. For Figs. \ref{fig7}(c1)-(c3): $k_{1}=1.5$, $l_{1}=1.1$, $\alpha_{1}^{(1)}=1+i$, $\alpha_{1}^{(2)}=-1$, $\beta_{1}^{(1)}=0.7+0.7i$, and $\beta_{1}^{(2)}=1.5$. In Figs. \ref{fig7}(d1)-(d3), we display a non-stationary right-direction moving asymmetric doublet SM for $k_{1}=1.5+0.5i$, $l_{1}=1+0.5i$, $\alpha_{1}^{(1)}=1+i$, $\alpha_{1}^{(2)}=1$, $\beta_{1}^{(1)}=1$, and $\beta_{1}^{(2)}=1$.}
	\label{fig7}
\end{figure*}

 We note that all the fundamental molecular structures described above are obtained by considering the values of the phase constants $\alpha_1^{(j)}$ and  $\beta_1^{(j)}$, $j=1,2$ as real. However, one can also bring out the basic molecular state by treating these constants as complex as well. For example, a twisted doublet SM is illustrated in Figs. \ref{fig7}(a1)-(a3) with asymmetric intensity distribution in the two modes. Besides this, one can also have a molecular state with symmetric intensity distribution in one mode and an asymmetric distribution in the other mode. Such a possibility is illustrated in Figs. \ref{fig7}(b1)-(b3). Then, in Figs. \ref{fig7}(c1)-(c3), a molecular state formed by an asymmetric intensity distribution among the modes, is illustrated by fixing a negative value to one of phase constants ($\alpha_1^{(2)}=e^{i\theta}=-1$, $\theta=\pi$). Finally, a stable propagation of doublet SM along the $z$ and $t$-directions is displayed in Figs. \ref{fig7}(d1)-(d3). We remark that, these four molecular states are periodic in nature. Thus, they are all synthesized molecular states.  
  



\subsection{Higher-order soliton molecules and their isomer structures}
To show the existence of higher-order soliton molecules  (especially a triplet SM and a quadruplet SM) and their related isomer structures, we deduce the corresponding BSS solutions from the $(\bar{N}+\bar{M})$-soliton solution of the system (\ref{manakov}). The isomer structures are soliton molecular structures that have the same analytical formula but different temporal arrangements of soliton atoms. The temporal arrangement of vector solitons can be done by tuning the free parameters of the constituent atoms, so that the total energy of each isomer structure is different. Note that the number of soliton atoms in such isomer structures remains constant. For example, by assuming the degeneracy condition: $v_1=v_2=v_1'=v_{mol}=2k_{1I}$ $(k_{1I}=k_{2I}=l_{1I}$), where $v_n=2k_{nI}$, $n=1,2$, $v_1'=2l_{1I}$, in the $(\bar{N}+\bar{M})=(2+1)$-soliton solution, we deduce the BSS solution corresponding to a triplet SM. The exact form of this BSS solution is obtained as $q_j=g^{(j)}/f$, $j=1,2$, where
\begin{eqnarray}
g^{(j)}=\begin{pmatrix}
A_{11} &A_{12} &a_{11}&1&0&0&e^{\eta_1}\\
A_{21} &A_{22} &a_{21}&0&1&0&e^{\eta_2}\\
\tilde{a}_{11} &\tilde{a}_{12} &\hat{A}_{11}&0&0&1&e^{\xi_1}\\
-1&0&0&B_{11}&B_{12}&b_{11}&0\\
0&-1&0&B_{21}&B_{22}&b_{21}&0\\
0&0&-1&\tilde{b}_{11}&\tilde{b}_{12}&\hat{B}_{11}&0\\
0&0&0&-\alpha_1^{(j)}&-\alpha_2^{(j)}&-\beta_1^{(j)}&0
\end{pmatrix},~f=\begin{pmatrix}
A_{11} &A_{12} &a_{11}&1&0&0\\
A_{21} &A_{22} &a_{21}&0&1&0\\
\tilde{a}_{11} &\tilde{a}_{12} &\hat{A}_{11}&0&0&1\\
-1&0&0&B_{11}&B_{12}&b_{11}\\
0&-1&0&B_{21}&B_{22}&b_{21}\\
0&0&-1&\tilde{b}_{11}&\tilde{b}_{12}&\hat{B}_{11}
\end{pmatrix}.~~
\end{eqnarray} 
In the above 
\bea
&&A_{11}=\frac{\text{exp}2\eta_{1R}}{2k_{1R}}, ~A_{12}=\frac{\text{exp}(\eta_{1R}+\eta_{2R}+i(\eta_{1I}-\eta_{2I}))}{k_{1R}+k_{2R}}, ~A_{22}=\frac{\text{exp}2\eta_{2R}}{2k_{2R}},\nonumber\\
&&A_{21}=\frac{\text{exp}(\eta_{1R}+\eta_{2R}-i(\eta_{1I}-\eta_{2I}))}{k_{1R}+k_{2R}}, ~a_{11}=\frac{\text{exp}(\eta_{1R}+\xi_{1R}+i(\eta_{1I}-\xi_{1I}))}{k_{1R}+l_{1R}}, \nonumber\\
&&a_{21}=\frac{\text{exp}(\eta_{2R}+\xi_{1R}+i(\eta_{2I}-\xi_{1I}))}{k_{2R}+l_{1R}},  ~\tilde{a}_{11}=\frac{\text{exp}(\eta_{1R}+\xi_{1R}-i(\eta_{1I}-\xi_{1I}))}{k_{1R}+l_{1R}}, \nonumber\\
&&\hat{A}_{11}=\frac{\text{exp}2\xi_{1R}}{2l_{1R}}, ~\tilde{a}_{12}=\frac{\text{exp}(\eta_{2R}+\xi_{1R}-i(\eta_{2I}-\xi_{1I}))}{k_{2R}+l_{1R}},~B_{11}=\frac{\psi_1^{\dagger}\sigma\psi_{1}}{2k_{1R}},\nonumber\eea\bea
&&B_{12}=\frac{\psi_1^{\dagger}\sigma\psi_{2}}{k_{1R}+k_{2R}}, ~B_{21}=\frac{\psi_2^{\dagger}\sigma\psi_{1}}{k_{1R}+k_{2R}}, B_{22}=\frac{\psi_2^{\dagger}\sigma\psi_{2}}{2k_{2R}},~\hat{B}_{11}=\frac{\psi_1'^{\dagger}\sigma\psi'_{1}}{2l_{1R}},\nonumber\\
&& b_{11}=\frac{\psi_1^{\dagger}\sigma\psi_{1}'}{k_{1R}+l_{1R}}, ~b_{21}=\frac{\psi_2^{\dagger}\sigma\psi_{1}'}{k_{2R}+l_{1R}},~
\tilde{b}_{11}=\frac{\psi_1'^{\dagger}\sigma\psi_{1}}{(l_{1R}+k_{1R})}, ~\tilde{b}_{12}=\frac{\psi_1'^{\dagger}\sigma\psi_{2}}{(l_{1R}+k_{2R})}.\nonumber
\eea
Here, $\psi_{j}=\begin{pmatrix}
\alpha_{j}^{(1)}\\
\alpha_{j}^{(2)}
\end{pmatrix}$, $\psi_{1}'=\begin{pmatrix}
\beta_{1}^{(1)}\\
\beta_{1}^{(2)}
\end{pmatrix}$, $\eta_{jR}=k_{jR}(t-2k_{1I}z)$, $\eta_{jI}=k_{1I}t+(k_{jR}^2-k_{1I}^2)z$, $j=1,2$, $\xi_{1R}=l_{1R}(t-2k_{1I}z)$, and $\xi_{1I}=k_{1I}t+(l_{1R}^2-k_{1I}^2)z$. The above BSS solution corresponding to a triplet SM is governed by sixteen real parameters: $\alpha_{nR}^{(j)}$, $\alpha_{nI}^{(j)}$,  $\beta_{1R}^{(j)}$, $\beta_{1I}^{(j)}$, $k_{jR}$, $n,j=1,2$, $l_{1R}$, and $k_{1I}$. This BSS solution also exhibits breathing pattern characterized by three recurrence frequencies: $\omega_{12}=k_{1R}^2-k_{2R}^2$, $\omega_{13}=k_{1R}^2-l_{1R}^2$, $\omega_{23}=k_{2R}^2-l_{1R}^2$ and the respective period of oscillations: $T_{12}=\frac{2\pi}{|\omega_{12}|}$, $T_{13}=\frac{2\pi}{|\omega_{13}|}$, $T_{23}=\frac{2\pi}{|\omega_{23}|}$. These characteristic frequencies essentially arise  because of the nonlinear interaction between the first and second soliton atoms,  first and third, and second and third soliton atoms, respectively. Apart from this, the temporal distributions $\Delta t_{21}$ (or $\Delta t_{12}$)  and $\Delta t_{32}$ (or $\Delta t_{23}$) between the three soliton atoms help to identify the isomer structures of triplet SM. 
\begin{figure*}
	\centering
	\includegraphics[width=0.3\linewidth]{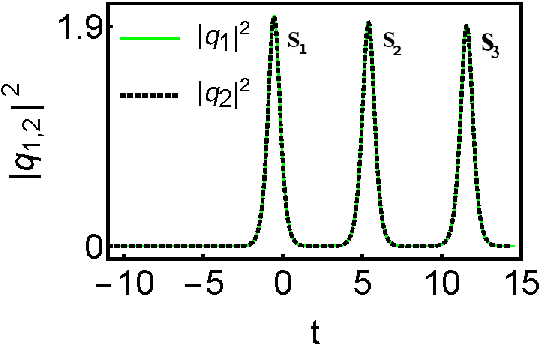}~\includegraphics[width=0.6\linewidth]{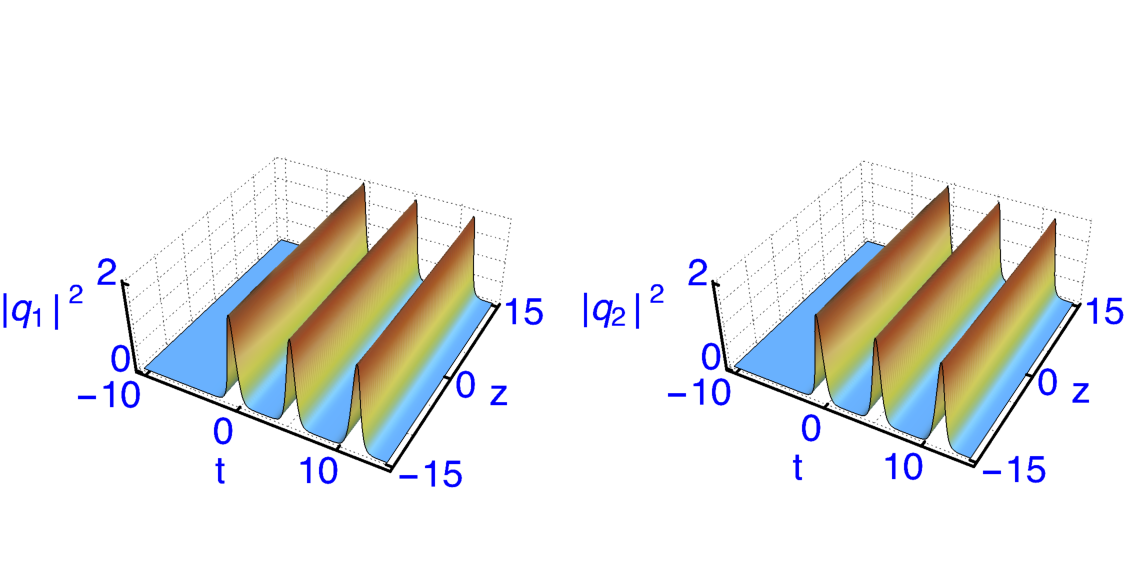}\\
	\includegraphics[width=0.3\linewidth]{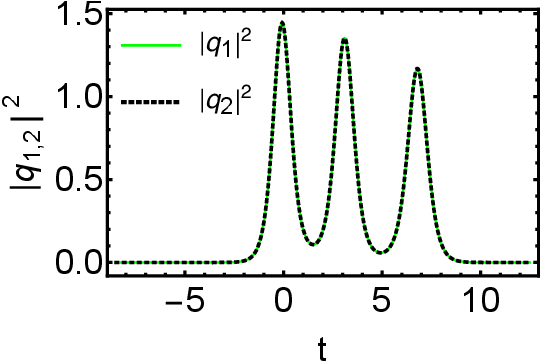}~\includegraphics[width=0.6\linewidth]{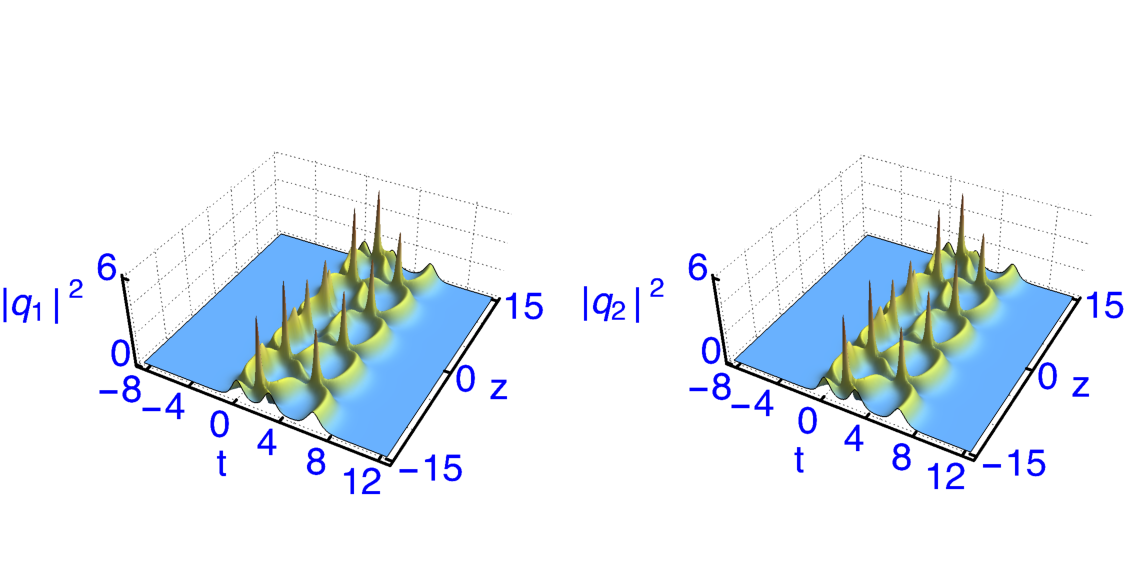}\\
	\caption{An equally spaced fully dissociated triplet soliton molecular state is demonstrated in the top row for $k_{1}=2$, $k_{2}=1.97$, $l_{1}=1.95$, $\alpha_{1}^{(j)}=1$, $\alpha_{2}^{(j)}=0.92$, $\beta_1^{(j)}=0.8$, $j=1,2$. The intensity profile of a triplet SM state at $z=0$, and the corresponding 3D-figure for all $z$ values is displayed here.      In the bottom row, an unequally spaced triplet soliton molecular structure is illustrated by fixing the values $k_{1}=2$, $k_{2}=1.7$, $l_{1}=1.4$, $\alpha_{1}^{(j)}=1$, $\alpha_{2}^{(j)}=0.9$, and $\beta_1^{(j)}=0.8$, $j=1,2$. }
	\label{fig8}
\end{figure*}

Based on the symmetry of the temporal distribution, we are able to isolate two kinds of optical isomer structures of triplet molecule in the coupled conservative fiber system (\ref{manakov}). If the temporal separations $\Delta t_{21}$ and $\Delta t_{32}$ are equal (that is, $\Delta t_{21}=\Delta t_{32}$) between the constituents then the corresponding triplet state is called as equally spaced triplet molecular state. An unequally spaced triplet molecular state emerges when the temporal separations  between the soliton atoms are not equal, that is, $\Delta t_{21}\neq \Delta t_{32}$. A typical equally spaced isomer structure of the triplet SM is illustrated in the top panel of Fig. \ref{fig8}, where the soliton atoms do not experience any force of attraction and repulsion during the propagation leading to zero oscillations. To confirm the equal temporal distribution, we have calculated the difference between $\Delta t_{32}(=11.5703-5.5569=6.0134)$ and $\Delta t_{21}(=5.3869-(-0.05541)=5.941)$. This value is equal to  $\Delta t_{32}-\Delta t_{21}=0.0724$. It implies that the three solitons are located in the $t$-direction with almost equal temporal separations. This isomer structure is formed by the symmetric intensity distribution between the modes and it belongs to the family of dissociated molecule. When the temporal distribution violates the symmetry condition, so that $\Delta t_{21}\neq \Delta t_{32}$, then the triplet SM admits unequally spaced isomer structure. Such an isomer structure is displayed in the bottom panel of Fig. \ref{fig8}. To validate this, we have calculated the temporal separation difference $\Delta t_{32}-\Delta t_{21}=3.6963-3.1727=0.5236$, where $\Delta t_{32}=6.79551-3.0992=3.6963$ and $\Delta t_{21}=3.0992-(-0.0735)=3.1727$. This value confirms that the constituents of triplet state are distributed unequally along the temporal direction as it is already confirmed from the bottom panel in Fig. \ref{fig8}. However, this isomer structure is associated with synthesized molecular state since the breathing pattern appears due to the strong interactions of three soliton atoms.
\begin{figure*}
	\centering
	\includegraphics[width=0.3\linewidth]{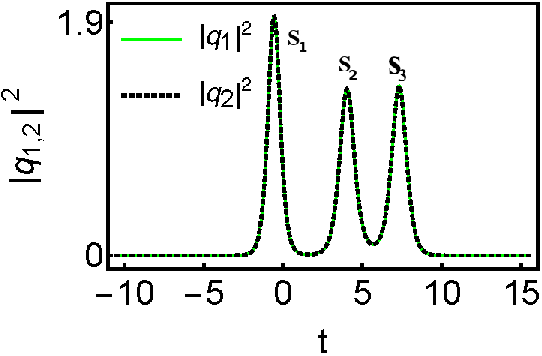}~\includegraphics[width=0.6\linewidth]{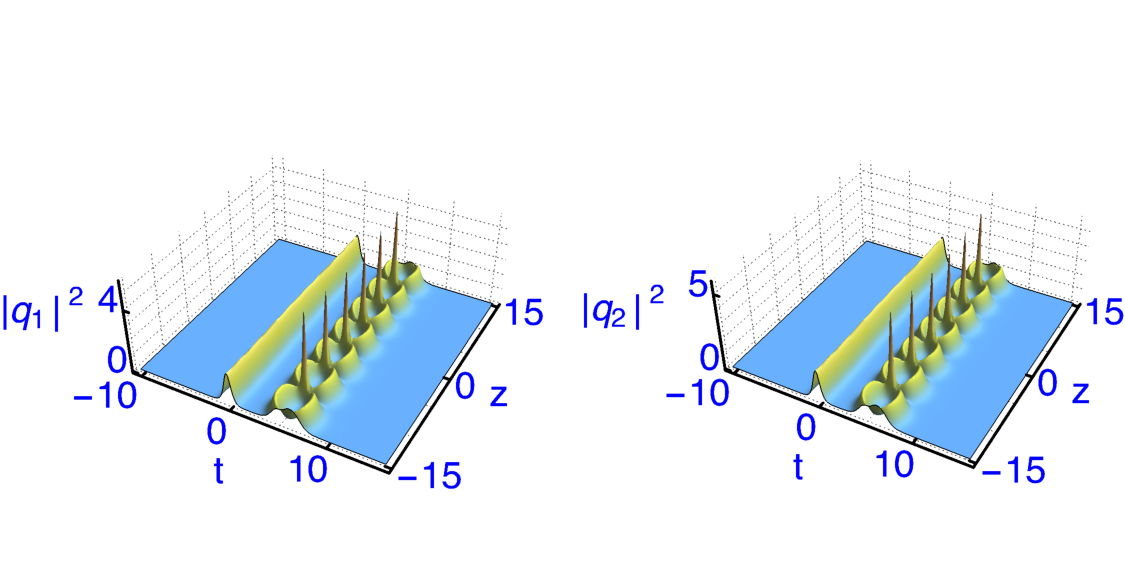}\\
		\includegraphics[width=0.3\linewidth]{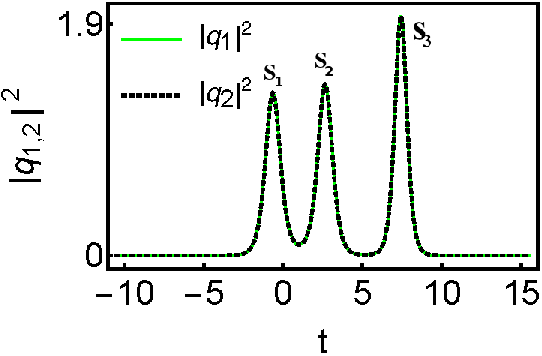}~\includegraphics[width=0.6\linewidth]{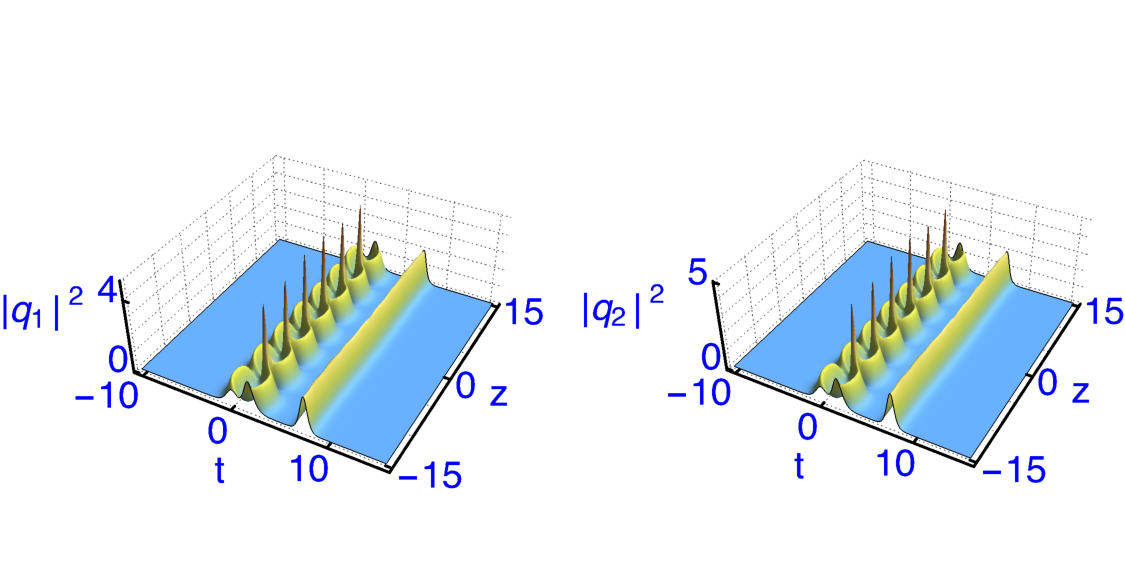}
	\caption{The intensity profiles corresponding to a triplet molecular structure, made of three soliton atoms, are displayed here by tuning the temporal separation between the constituents.  The figures presented in the top row represent a triplet molecular state in which one of the soliton atom is located separately near to $t=0$. The other two soliton atoms form a doublet SM state in both the modes. To demonstrate this we fix the parameter values as $k_{1}=2$, $k_{2}=1.92$, $l_{1}=1.5$, $\alpha_{1}^{(j)}=\alpha_{2}^{(j)}=1$, $\beta_1^{(j)}=0.8$, $j=1,2$.  The figures in bottom row show a different configuration, where the singlet state is located at a positive temporal coordinate. We obtain these figures when the parameter values are fixed as $k_{1}=1.5$, $k_{2}=1.92$, $l_{1}=2$, $\alpha_{1}^{(j)}=0.9$, $\alpha_{2}^{(j)}=1$, and $\beta_1^{(j)}=1$, $j=1,2$. }
	\label{fig9}
\end{figure*} 
 
We also came across a triplet molecular state composed of a doublet state and a singlet state by placing them with an appropriate temporal separation. This mixed isomer structure can be formed in two ways as it is demonstrated in Fig. \ref{fig9}. By allowing any one of the solitons from the set $\{N\}$ bind with a vector soliton from the set $\{M\}$ (top panel in Fig. \ref{fig9}) or the two solitons from the same set $\{N\}$ or set $\{M\}$ (bottom panel in Fig. \ref{fig9}) form a doublet SM. These two configurations show that the constituents of doublet SM exhibit breathing pattern in the overlapping region due the periodic attraction and repulsion. In the first configuration, the breathing pattern is described by the oscillation frequency, $\omega_{23}=|k_{2R}^2-l_{1R}^2|$, and the corresponding period of oscillation is $T_{23}=\frac{2\pi}{|k_{2R}^2-l_{1R}^2|}=\frac{2\pi}{1.4364}$. On the other hand, in the second configuration, the period of oscillation is governed by the frequency, $\omega_{12}=|k_{1R}^2-k_{2R}^2|$, and the corresponding period of oscillation is $T_{12}=\frac{2\pi}{k_{1R}^2-k_{2R}^2}=\frac{2\pi}{|-1.4364|}$.       
\begin{figure*}
	\centering
	\includegraphics[width=0.3\linewidth]{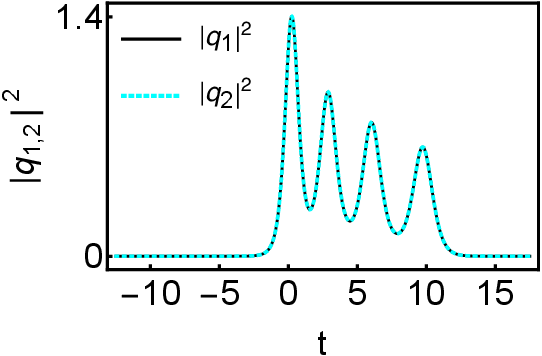}~\includegraphics[width=0.6\linewidth]{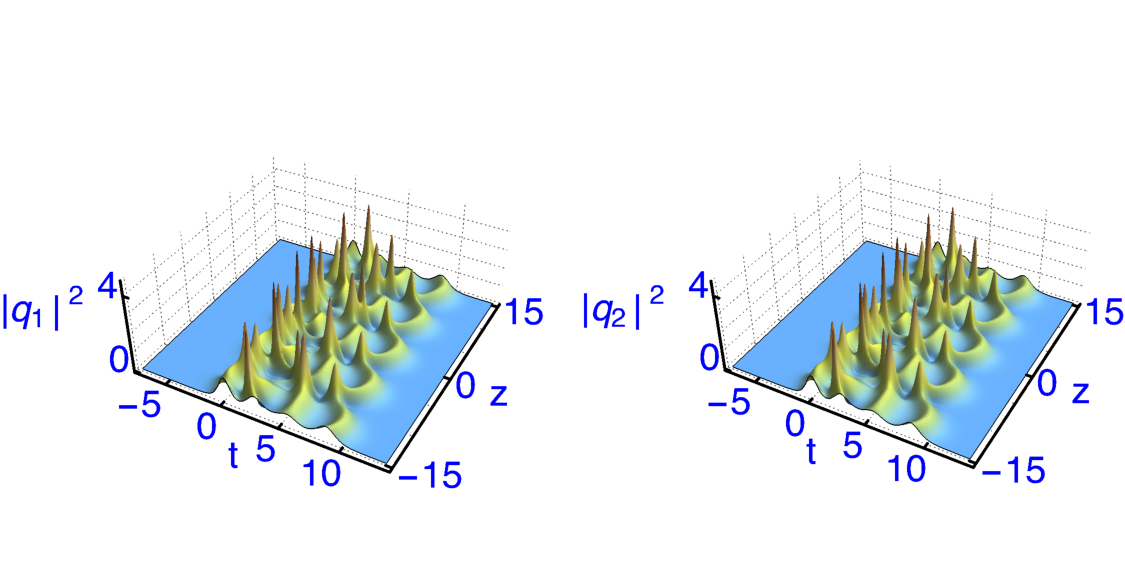}\\
	\includegraphics[width=0.3\linewidth]{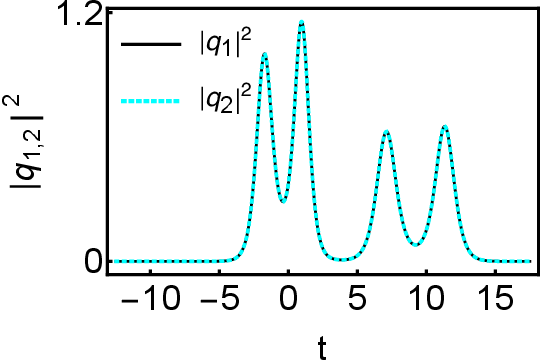}~\includegraphics[width=0.6\linewidth]{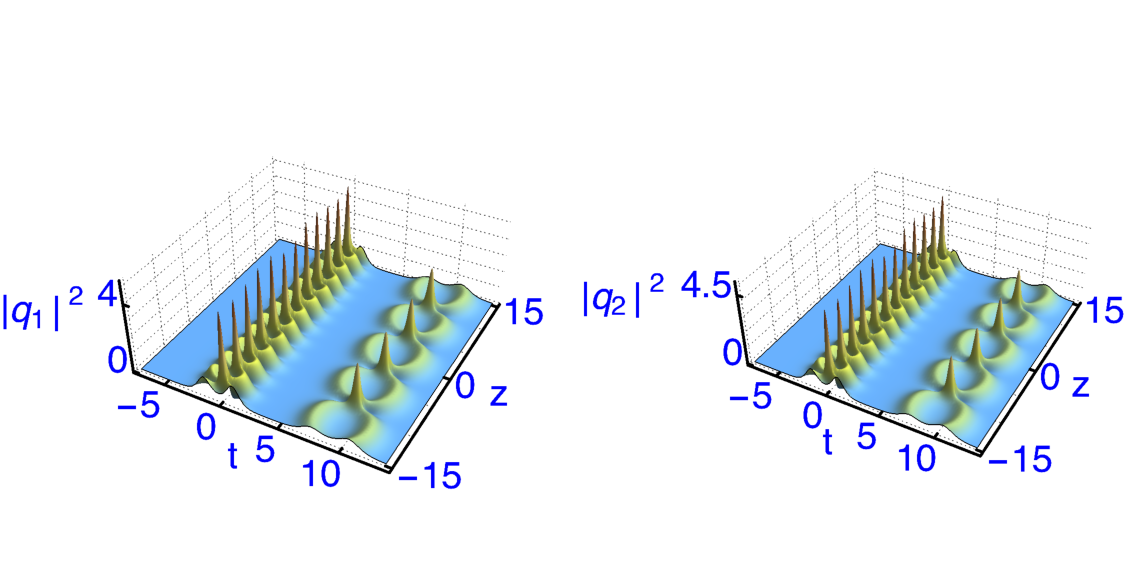}\\
	\includegraphics[width=0.3\linewidth]{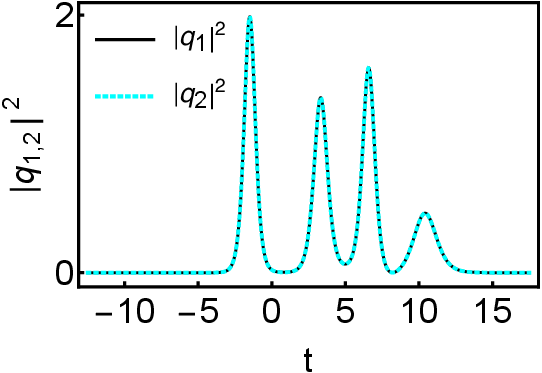}~\includegraphics[width=0.6\linewidth]{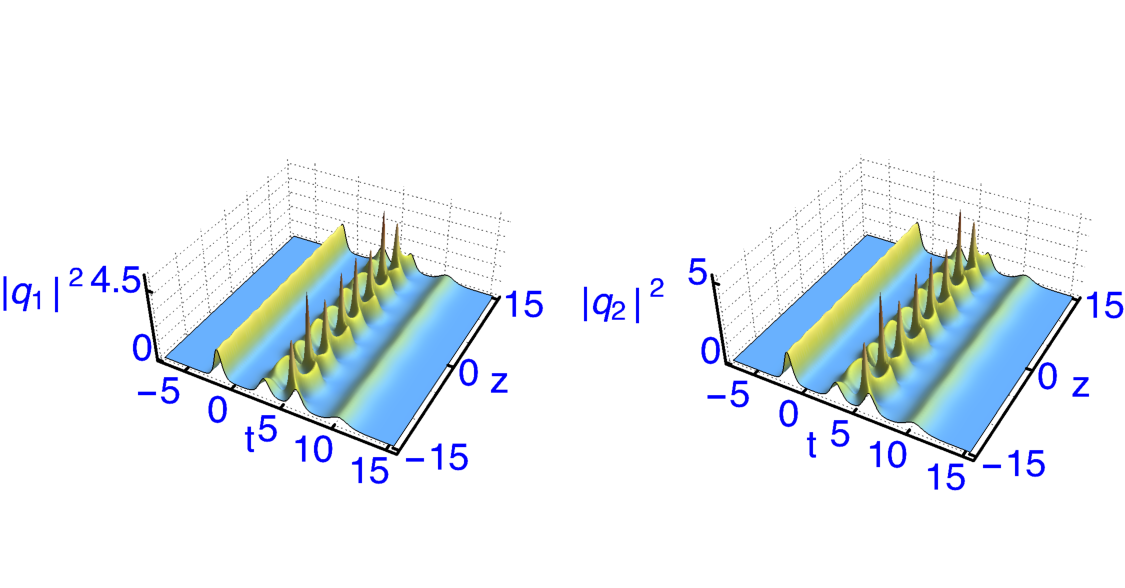}
	\caption{An unequally spaced quadruplet soliton molecular state, which is created by four overlapping solitons, is demonstrated in the top row for the values:  $k_{1}=2$, $k_{2}=1.7$, $l_{1}=1.4$, $l_{2}=1$, $\alpha_{1}^{(j)}=\alpha_{2}^{(j)}=1$, $\beta_1^{(j)}=1.2$, $\beta_2^{(j)}=1$, $j=1,2$. In the middle row, we display two distinct doublet molecular states, which are separated by finite temporal separation, for $k_{1}=2$, $k_{2}=1.25$, $l_{1}=1.4$, $l_{2}=0.99$, $\alpha_{1}^{(j)}=1$, $\alpha_{2}^{(j)}=0.95$, $\beta_1^{(j)}=1.1$, $\beta_2^{(j)}=1$, $j=1,2$. A quadruplet state, made of a doublet state and two temporally dissociated soliton atoms on either side of doublet SM, is depicted in the bottom row  while fixing the wave parameter values as $k_{1}=2$, $k_{2}=1.9$, $l_{1}=1.4$, $l_2=0.9$, $\alpha_{1}^{(j)}=1$, $\alpha_{2}^{(j)}=0.9$, $\beta_1^{(j)}=1.2$, and $\beta_2^{(j)}=1$, $j=1,2$. }
	\label{fig10}
\end{figure*}

If we increase the number of soliton atoms in a triplet state by one, then it yields an interesting quadruplet molecular state. Addition of one soliton provides an extra degree of freedom ($\Delta t_{34}$), along with the already existing two temporal distribution $\Delta t_{12}$ and $\Delta t_{23}$, to characterize the quadruplet SM. This additional freedom of temporal distribution supports four kinds of optical isomer structures, namely unequally spaced quadruplet, equally spaced quadruplet, $3+1$-quadruplet (or $1+3$-quadruplet), and $2+2$ quadruplet. To form the unequally spaced quadruplet soliton molecular state, we choose the parameter values in such a way that the temporal distributions violate the condition: $\Delta t_{12}=\Delta t_{23}=\Delta t_{34}$. This isomer structure is depicted in Fig. \ref{fig10} in the top-row, where all the soliton atoms strongly interact with each other and exhibit breathing interference pattern. Then, one can also visualize the equally spaced quadruplet isomer structure when the constituents are distributed temporally with equal separation. We do not display this possibility for brevity. Then, if the temporal distribution condition: $\Delta t_{12}=\Delta t_{23}\neq \Delta t_{34}$ (or $\Delta t_{12}\neq \Delta t_{23}=\Delta t_{34}$) is satisfied, one can have $3+1$-quadruplet (or $1+3$-quadruplet) isomer structure, which is also not demonstrated here for brevity. Further, to bring out the $2+2$ quadruplet isomer, we consider two distinct doublet states, each made by a pair of two solitons, along with temporal distribution condition $\Delta t_{12}=\Delta t_{34}$. This kind of $2+2$-quadruplet isomer structure is illustrated in the middle row of Fig. \ref{fig10}. In addition to these isomer structures, we also display an additional isomer structure in the bottom-row of Fig. \ref{fig10}, in which a doublet soliton molecule is synthesized in between two dissociated singlet states. By continuously adding a large number of soliton atoms, one can construct a macromolecular structure involving inter- and intra molecular bond. In this way, one can construct a macro molecular structure, involving intra- and inter-molecular bonds. If we continuously increase the soliton atoms in the molecular structures one may not be able to avoid turbulence or random behaviour of dynamical states created by the interactions of soliton atoms as indicated in Fig. \ref{fig10}. The unequally spaced quadruplet soliton molecular state described in Fig.  \ref{fig10} can be considered as an early signature of creation of vector soliton integrable turbulence. However, to understand this most intriguing phenomenon and related statistics one has to use for example the concept of soliton gas \cite{zakharov1,g-el}, which has been developed recently and it is one of the active areas research in the past few years in soliton theory. However, it needs a separate study similar to the vector integrable turbulence recently reported in Ref. \cite{sun-turbulence}.
\section{Collision properties of vector soliton molecular states}
To verify the robustness of the soliton molecules in the Manakov system (\ref{manakov}), it is necessary to analyse their stability properties under different environmental conditions. However, in the present conservative fiber system (\ref{manakov}), soliton-soliton interaction (or soliton molecule-soliton interaction) is the main cause for possible instability of the SMs. Therefore, it is very important to study the collision between a SM and vector solitons by treating the latter  as a strong perturbation to SM. Besides this physical situation, it is also crucial to analyse the interaction between SMs of the same or different kinds. To examine these physical situations, we first consider the doublet soliton molecular state from the set $\{\bar{N}\}$ and allow it to interact with one or two solitons from the set $\{\bar{M}\}$. Then, we consider the collision of two distinct doublet soliton molecular states as a second physical situation. We  investigate below, each of these situations with appropriate asymptotic analysis. In addition, we also numerically examine the stability of soliton molecules by introducing weak and strong random noise as perturbations in the subsequent Section V.
 \subsection{Collision between fundamental molecular state and vector soliton}
We start with analysing the collision scenario between a fundamental SM and a single vector soliton and examine their asymptotic expressions at the limit $z\rightarrow \pm\infty$. In order to do this, we impose the asymptotic nature of the wave variables of the doublet SM ($\eta_{jR}=k_{jR}(t-2k_{1I}z)$, $j=1,2$) and the  fundamental soliton  ($\xi_{1R}=l_{1R}(t-2l_{1I}z)$) in the $(\bar{N}+\bar{M})=(2+1)$-soliton solution (\ref{2}) with Eqs. (\ref{2a}) and (\ref{2b}) and deduce their corresponding asymptotic forms. To study the head-on collision between a doublet SM and a vector soliton, we consider the velocity condition: $k_{1I}>l_{1I}$. The latter choice leads to following asymptotic behaviour of $\eta_{jR}$, $j=1,2$, and $\xi_{1R}$. For the doublet SM:  $\eta_{1R},~\eta_{2R} \simeq 0$, $\xi_{1R}\rightarrow \mp \infty$ as $z\rightarrow\mp\infty$ and for the soliton: $\xi_{1R}\simeq 0$,  $\eta_{1R},~\eta_{2R}\rightarrow\pm \infty$ as $z\rightarrow \mp\infty$. Consideration of these results yields the following asymptotic expressions for the doublet SM and a fundamental vector soliton.\\   
(a) Before collision: $z\rightarrow -\infty$\\
{\bf Doublet SM}: $\eta_{1R},\eta_{2R}\simeq 0$, $\xi_{1R}\rightarrow -\infty$ \\
The asymptotic expression of the fundamental molecular state before collision is deduced from the $(2+1)$-soliton solution by setting $\bar{N}=2$, $\bar{M}=1$ in Eq. (\ref{2}) with Eqs. (\ref{2a}) and (\ref{2b}). The resultant asymptotic form is given by
\bes\begin{eqnarray}
&&q_j(z,t)=\frac{1}{D^-}\big(e^{i\eta_{2I}}a_{1j}^{-}\cosh(\eta_{1R}+\psi_1^{j-})+e^{i\eta_{1I}}a_{2j}^{-}\cosh(\eta_{2R}+\psi_2^{j-})\big),~j=1,2,\label{15a}\\
&&D^-=a_{3}^{-}\cosh(\eta_{1R}+\eta_{2R}+\psi_3^{-})+a_{4}^{-}\cosh(\eta_{1R}-\eta_{2R}+\psi_4^{-})+a_{5}^{-}\big[\cosh\psi_5^{-}\cos\theta\nonumber\\
&&\hspace{1.2cm}+i\sinh\psi_5^{1-}\sin\theta\big],\nonumber
\end{eqnarray}
where \begin{eqnarray}
&&a_{1j}^-=\frac{(k_1-k_2)^{1/2}[\alpha_1^{(j)}B_{12}-\alpha_2^{(j)}B_{11}]^{1/2}}{(k_1+k_1^*)^{1/2}(k_2+k_1^*)^{1/2}},~a_{2j}^-=\frac{(k_1-k_2)^{1/2}[\alpha_1^{(j)}B_{22}-\alpha_2^{(j)}B_{21}]^{1/2}}{(k_1+k_2^*)^{1/2}(k_2+k_2^*)^{1/2}},\nonumber\\
&&a_3^-=\frac{|k_1-k_2|[B_{11}B_{22}-B_{12}B_{21}]^{1/2}}{(k_1+k_1^*)^{1/2}(k_2+k_2^*)^{1/2}|k_1+k_2^*|},~a_4^-=\frac{(B_{11}B_{22})^{1/2}}{(k_1+k_1^*)^{1/2}(k_2+k_2^*)^{1/2}},\nonumber\\
&&a_5^-=\frac{(B_{12}B_{21})^{1/2}}{|k_1+k_2^*|},~\theta=\eta_{1I}-\eta_{2I}=(k_{1R}^2-k_{2R}^2)z.\label{15c}
\end{eqnarray}
The phase terms appearing in Eq. (\ref{15a}) are identified as 
\begin{eqnarray}
&&\psi_1^{j-}=\frac{1}{2}\ln\frac{(k_1-k_2)[\alpha_1^{(j)}B_{12}-\alpha_2^{(j)}B_{11}]}{(k_1+k_1^*)(k_2+k_1^*)}, ~\psi_2^{j-}=\frac{1}{2}\ln\frac{(k_1-k_2)[\alpha_1^{(j)}B_{22}-\alpha_2^{(j)}B_{21}]}{(k_1+k_2^*)(k_2+k_2^*)},\nonumber\\
&&\psi_3^-=\frac{1}{2}\ln\frac{|k_1-k_2|^2[B_{11}B_{22}-B_{12}B_{21}]}{(k_1+k_1^*)(k_2+k_2^*)|k_1+k_2^*|^2},~\psi_4^-=\frac{1}{2}\ln\frac{B_{11}(k_2+k_2^*)}{B_{22}(k_1+k_1^*)},\nonumber\\
&&\psi_5^-=\frac{1}{2}\ln\frac{B_{12}(k_1^*+k_2)}{B_{21}(k_1+k_2^*)}.\label{15c}
\end{eqnarray}\ees
In the above, a negative sign ($-$) appearing in the superscript represents before collision. 
\\
{\bf Soliton}: $\xi_{1R}\simeq 0$,~$\eta_{1R},\eta_{2R}\rightarrow +\infty$\\
In this limit, the asymptotic form of the fundamental soliton  is obtained as follows:

\bes
\begin{eqnarray}
	q_j(z,t)=l_{1R}A_{j}^{-}e^{i(\xi_{1I}+\Theta^{-})}\sech(\xi_{1R}+\Psi^-).\label{16a}
\end{eqnarray}
Here, 
\begin{eqnarray}
&&A_j^-=\frac{\rho_1}{(l_1+l_1^*)^{1/2}[B_{11}B_{22}-B_{12}B_{21}]^{1/2}\rho_2^{1/2}}, \nonumber\\
&&e^{i\Theta^-}=\frac{(k_1-l_1)^{1/2}(k_2-l_1)^{1/2}(k_1+l_1^*)^{1/2}(k_2+l_1^*)^{1/2}}{(k_1^*-l_1^*)^{1/2}(k_2^*-l_1^*)^{1/2}(k_1^*+l_1)^{1/2}(k_2^*+l_1)^{1/2}},\nonumber\\
&&\Psi^-=\frac{1}{2}\ln\frac{|k_1-l_1|^2|k_2-l_1|^2\rho_2}{[B_{11}B_{22}-B_{12}B_{21}](l_1+l_1^*)|k_1+l_1^*|^2|k_2+l_1^*|^2},\nonumber\\
&&\rho_1=[\alpha_1^{(j)}(B_{12}b_{21}-b_{11}B_{22})+\alpha_2^{(j)}(b_{11}B_{21}-b_{21}B_{11})+\beta_1^{(j)}(B_{11}B_{22}-B_{12}B_{21})],\nonumber\\
&&\rho_2=[B_{11}(B_{22}\hat{B}_{11}-b_{21}\tilde{b}_{12})+B_{12}(b_{21}\tilde{b}_{11}-B_{21}\hat{B}_{11})+b_{11}(B_{21}\tilde{b}_{12}-B_{22}\tilde{b}_{11})].\label{16b}
\end{eqnarray}\ees
(a) After collision: $z\rightarrow +\infty$\\
{\bf Doublet SM}: $\eta_{1R},\eta_{2R}\simeq 0$, $\xi_{1R}\rightarrow +\infty$ \\
The asymptotic expression of the fundamental molecular state after collision is deduced as follows:
\bes\begin{eqnarray}
&&q_j(z,t)=\frac{1}{D^+}\big(e^{i\eta_{2I}}a_{1j}^{+}\cosh(\eta_{1R}+\psi_1^{j+})+e^{i\eta_{1I}}a_{2j}^{+}\cosh(\eta_{2R}+\psi_2^{j+})\big),~j=1,2,\label{17a}\\
&&D^+=a_{3}^{+}\cosh(\eta_{1R}+\eta_{2R}+\psi_3^{+})+a_{4}^{+}\cosh(\eta_{1R}-\eta_{2R}+\psi_4^{+})+a_{5}^{+}\big[\cosh\psi_5^{+}\cos\theta_1\nonumber\\
&&\hspace{1.2cm}+i\sinh\psi_5^{1+}\sin\theta_1\big],\nonumber
\end{eqnarray}
where 
\bea
&&a_{1j}^+=\frac{(k_1-k_2)^{1/2}(k_1-l_1)|k_2-l_1|[\alpha_1^{(j)}\hat{B}_{11}-\beta_1^{(j)}\tilde{b}_{11}]^{1/2}\rho_3^{1/2}}{(k_1+k_2^*)^{1/2}(k_2+k_2^*)^{1/2}|k_2+l_1^*|(k_1+l_1^*)(l_1+l_1^*)},\nonumber\\
&&\rho_3=[\alpha_1^{(j)}(B_{22}\hat{B}_{11}-b_{21}\tilde{b}_{12})+\alpha_2^{(j)}(b_{21}\tilde{b}_{11}-B_{21}\hat{B}_{11})+\beta_1^{(j)}(B_{21}\tilde{b}_{12}-B_{22}\tilde{B}_{11})],\nonumber\\
&&a_{2j}^+=\frac{(k_1-k_2)^{1/2}(k_2-l_1)|k_1-l_1|[\alpha_2^{(j)}\hat{B}_{11}-\beta_1^{(j)}\tilde{b}_{12}]^{1/2}\rho_4^{1/2}}{(k_1+k_1^*)^{1/2}(k_1^*+k_2)^{1/2}|k_1+l_1^*|(k_2+l_1^*)(l_1+l_1^*)},\nonumber\\
&&\rho_4=[\alpha_1^{(j)}(B_{12}\hat{B}_{11}-b_{11}\tilde{b}_{12})+\alpha_2^{(j)}(b_{11}\tilde{b}_{11}-B_{11}\hat{B}_{11})+\beta_1^{(j)}(B_{11}\tilde{b}_{12}-B_{12}\tilde{B}_{11})],\nonumber\\
&&a_3^+=\frac{\hat{B}_{11}^{1/2}|k_1-k_2||k_1-l_1||k_2-l_1|\rho_2^{1/2}}{(k_1+k_1^*)^{1/2}(k_2+k_2^*)^{1/2}(l_1+l_1^*)^{1/2}|k_1+k_2^*||k_1+l_1^*||k_2+l_1^*|},\nonumber\\
&&a_4^+=\frac{|k_1-l_1||k_2-l_1|[B_{11}\hat{B}_{11}-b_{11}\tilde{b}_{11}]^{1/2}[B_{22}\hat{B}_{11}-b_{21}\tilde{b}_{12}]^{1/2}}{(k_1+k_1^*)^{1/2}(l_1+l_1^*)(k_2+k_2^*)^{1/2}|k_1+l_1^*||k_2+l_1^*|},\nonumber\eea\bea
&&a_5^+=\frac{|k_1-l_1||k_2-l_1|[B_{21}\tilde{b}_{11}-b_{21}]\tilde{b}_{11}]^{1/2}[B_{12}\hat{B}_{11}-b_{11}\tilde{b}_{12}]^{1/2}}{|k_1+k_2^*||k_1+l_1^*||k_2+l_1^*|(l_1+l_1^*)}.\label{17b}
\eea
The phase terms appearing in Eq. (\ref{17a}) are given below:
\begin{eqnarray}
&&\psi_1^{j+}=\frac{1}{2}\ln\frac{(k_1-k_2)|k_2-l_1|^2\rho_3}{(k_1+k_2^*)(k_2+k_2^*)|k_2+l_1|^2[\alpha_1^{(j)}\hat{B}_{11}-\beta_1^{(j)}\tilde{b}_{11}]},\nonumber\\
&&\psi_2^{j+}=\frac{1}{2}\ln\frac{(k_1-k_2)|k_1-l_1|^2\rho_4}{(k_1^*+k_2)(k_1+k_1^*)|k_1+l_1^*|^2[\alpha_2^{(j)}\hat{B}_{11}-\beta_1^{(j)}\tilde{b}_{12}]},\nonumber\\
&&\psi_3^+=\frac{1}{2}\ln\frac{|k_1-k_2|^2|k_1-l_1|^2|k_2-l_1|^2(l_1+l_1^*)\rho_2}{(k_1^*+k_1)(k_2+k_2^*)|k_1+k_2^*|^2|k_1+l_1^*|^2|k_2+l_1^*|^2},\nonumber\\
&&\psi_4^+=\frac{1}{2}\ln\frac{|k_1-l_1|^2(k_2+k_2^*)|k_2+l_1^*|^2[B_{11}\hat{B}_{11}-b_{11}\tilde{b}_{11}]}{(k_1+k_1^*)|k_1+l_1^*|^2|k_2-l_1|^2[B_{22}\hat{B}_{11}-b_{21}\tilde{b}_{12}]},\nonumber\\
&&\psi_5^+=\frac{1}{2}\ln\frac{(k_1-l_1)(k_2^*-l_1^*)(k_1^*+k_2)(k_2+l_1^*)(k_1^*+l_1)[B_{21}\hat{B}_{11}-b_{21}\tilde{b}_{11}]}{(k_1^*-l_1^*)(k_2-l_1)(k_1+k_2^*)(k_2^*+l_1)(k_1+l_1^*)[B_{12}\hat{B}_{11}-b_{11}\tilde{b}_{12}]}.\label{17c}
\end{eqnarray}
\ees
In the above, a positive sign ($+$) appearing in the superscript denotes after collision. \\
{\bf Soliton}: $\xi_{1R}\simeq 0$,~$\eta_{1R},\eta_{2R}\rightarrow -\infty$\\
In this limit, the following asymptotic form of the fundamental vector soliton is given. It reads 
\bes 
\begin{eqnarray}
q_j(z,t)=l_{1R}A_{j}^{+}e^{i\xi_{1I}}\sech(\xi_{1R}+\Psi^+),\label{18}
\end{eqnarray}
where the polarization constant $A_j^+$ and the phase constant $\Psi^+$ are 
\begin{eqnarray}
A_j^+=\frac{\beta_1^{(j)}}{[|\beta_1^{(1)}|^2+|\beta_1^{(2)}|^2]^{1/2}},~\Psi^+=\frac{1}{2}\ln\frac{|\beta_1^{(1)}|^2+|\beta_1^{(2)}|^2]^{1/2}}{(l_1+l_1^*)^2}. 
\end{eqnarray}
\ees
The constants, $B_{ij}$'s, $b_{11}$, $b_{21}$, $\tilde{b}_{11}$, $\tilde{b}_{12}$, and $\hat{B}_{11}$ appeared in the asymptotic expressions are given below:
\begin{eqnarray}
&&B_{ij}=\sum_{n=1}^2\frac{\alpha_i^{(n)}\alpha_j^{(n)*}}{k_i^*+k_j},~i,j=1,2,~b_{11}=\frac{\beta_1^{(1)}\alpha_1^{(1)*}+\beta_1^{(2)}\alpha_1^{(2)*}}{k_1^*+l_1},~b_{21}=\frac{\beta_1^{(1)}\alpha_2^{(1)*}+\beta_1^{(2)}\alpha_2^{(2)*}}{k_2^*+l_1},\nonumber\\
&&\tilde{b}_{11}=\frac{\alpha_1^{(1)}\beta_1^{(1)*}+\alpha_1^{(2)}\beta_1^{(2)*}}{l_1^*+k_1},~\tilde{b}_{12}=\frac{\alpha_2^{(1)}\beta_1^{(1)*}+\alpha_2^{(2)}\beta_1^{(2)*}}{l_1^*+k_2},~\hat{B}_{11}=\frac{|\beta_1^{(1)}|^2+|\beta_1^{(2)}|^2}{(l_1+l_1^*)}.\label{19}
\end{eqnarray}
We note that, to obtain the exact asymptotic forms of the doublet SM before and after collision one has to consider $k_{1I}=k_{2I}$ in Eqs. (\ref{15a})-(\ref{15c}) and (\ref{17a})-(\ref{17c}), respectively. 
\begin{figure*}
	\centering
	\includegraphics[width=0.8\linewidth]{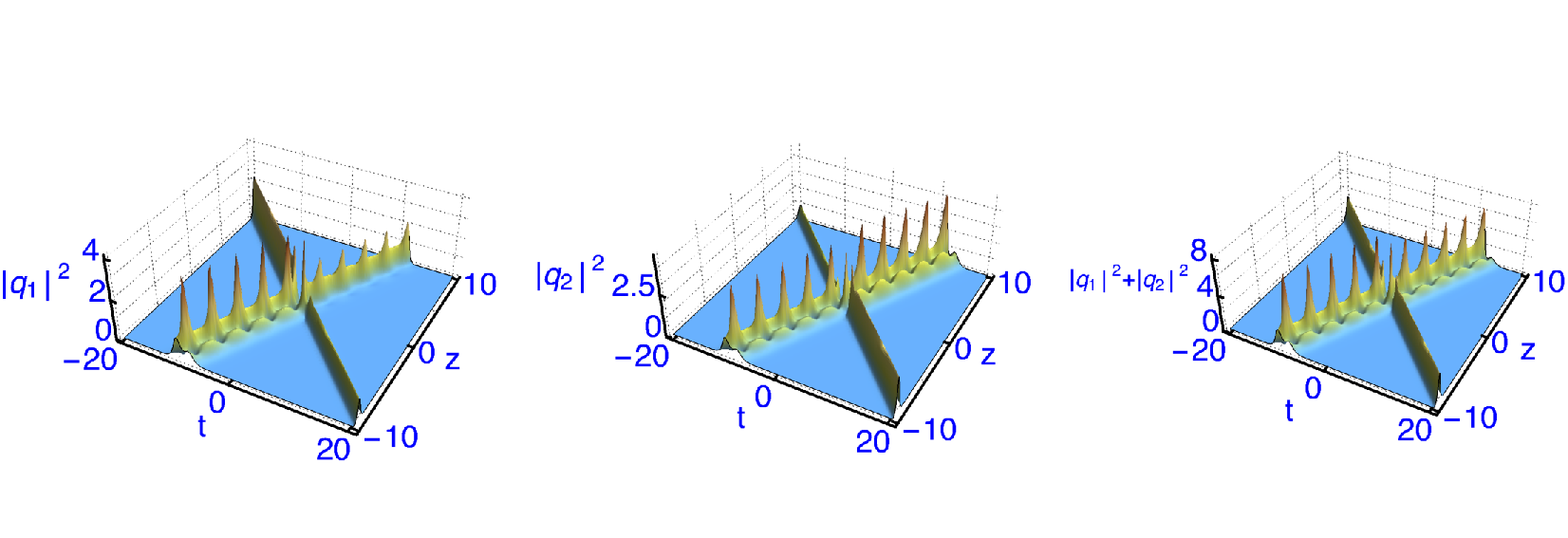}\\\includegraphics[width=0.8\linewidth]{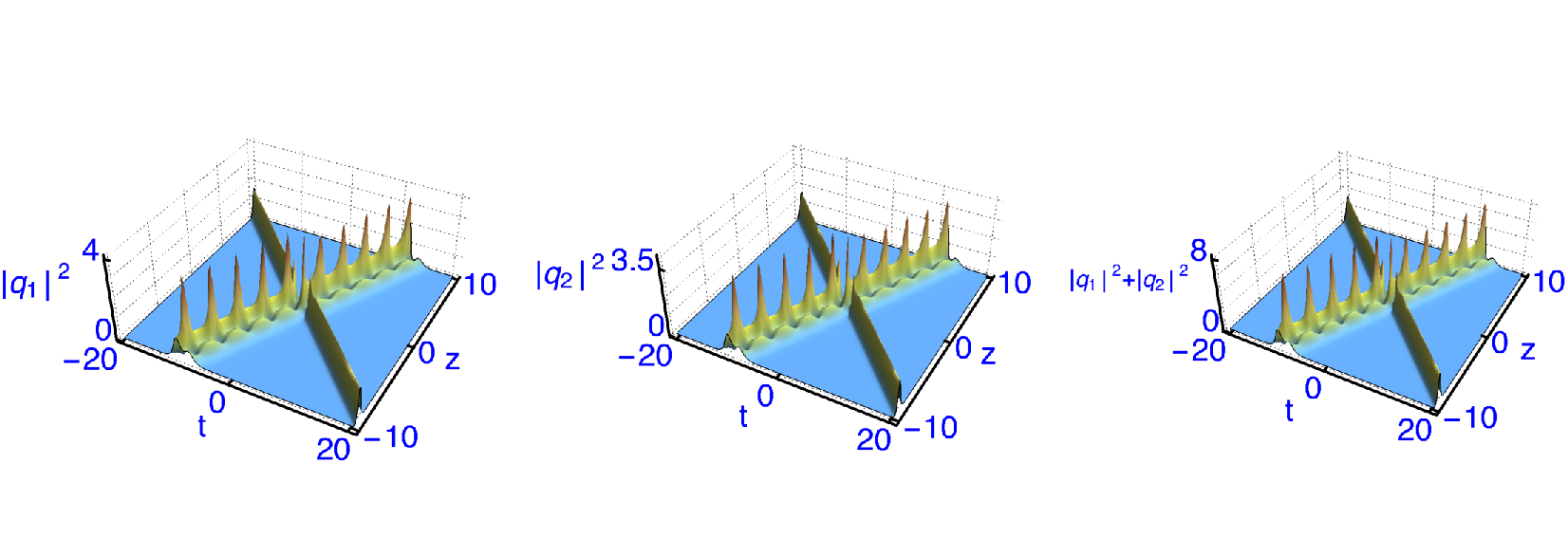}
	\caption{ In the top row an energy sharing collision between soliton molecule and a vector soliton is displayed for  $k_1=2+0.5i$, $k_{2}=0.99+0.5i$,  $l_{1}=1.8-i$, $\alpha_{1}^{(j)}=\alpha_{2}^{(j)}=1$, $j=1,2$, $\beta_1^{(1)}=1.5$, and $\beta_1^{(2)}=1$. In the bottom row panels we display the corresponding elastic collision between doublet SM and a vector soliton for the same parameter values except $\beta_1^{(1)}=1.5$ and $\beta_1^{(2)}=1.5$. }
	\label{fig10a}
\end{figure*}
\subsubsection{Energy sharing collision of doublet molecule with soliton}
The above asymptotic analysis clearly displays that the structures of both the doublet SM and the fundamental vector soliton are not preserved during the collision even though the individual energies of each of the soliton is conserved as shown below. It is because of mutual energy sharing nature of both the SM and the vector soliton. This can be confirmed from the variations in their asymptotic expressions. That is, the quantities, $a_{1j}^-$, $a_{2j}^-$, $a_{3}^-$, $a_{4}^-$, $a_{5}^-$, $\psi_1^{j-}$, $\psi_2^{j-}$, $\psi_3^-$, $\psi_4^-$, and $\psi_5^-$, $j=1,2,$ of SM  before collision get changed to $a_{1j}^+$, $a_{2j}^+$, $a_{3}^+$, $a_{4}^+$, $a_{5}^+$, $\psi_1^{j+}$, $\psi_2^{j+}$, $\psi_3^+$, $\psi_4^+$, and $\psi_5^+$, $j=1,2$, after collision. Along with these changes, the initial amplitude of the vector soliton, $l_{1R}A_j^-$, $j=1,2$,  also gets changed to $l_{1R}A_j^+$, $j=1,2$. As a consequence of these  mutual changes, the energy sharing takes place during the collision process in between the doublet SM and a vector soliton. We illustrate such a collision scenario in the top row of Fig. \ref{fig10a}, where a doublet SM is well separated initially from the vector bright soliton. From this figure, one can observe that in the first mode $q_1$, the vector soliton gains energy from the doublet SM after the collision. As a consequence, its intensity is enhanced and subsequently the energy of the one of the constituents of SM is completely suppressed in the first mode.  However, in the second mode $q_2$, the vector soliton loses its energy to the doublet SM. Consequently, this lost energy of soliton gets distributed among the constituents of SM  and subsequently their intensities are enhanced in the $q_2$ mode after collision. To characterize the amount of energy gained or lost during this collision process, we calculate the transition intensity of the fundamental soliton corresponding to the two modes, which is given by 
\begin{equation}
|T_s^j|^2=\frac{|\beta_1^{(j)}|^2(l_1+l_1^*)|\rho_2||B_{11}B_{22}-B_{12}B_{21}|}{|\beta_1^{(1)}|^2+|\beta_1^{(2)}|^2|\rho_1|^2},~j=1,2,\label{20}
\end{equation}
where $\rho_j$'s and $B_{ij}$'s, $i,j=1,2$, are defined in Eqs. (\ref{16b}) and (\ref{19}), respectively.  However, the total energy is always conserved in both the modes separately ($\int_{-\infty}^{+\infty}|q_j|^2dz=\text{const}$, $j=1,2$) as well as in all the modes ($\int_{-\infty}^{+\infty}(|q_1|^2+|q_2|^2)dz=\text{const}$, $j=1,2$). 

In addition to the above collision process, one can also capture the standard elastic collision between the doublet SM and a vector soliton, when the polarization constants, $\alpha_1^{(j)}$, $\alpha_2^{(j)}$, $\beta_1^{(j)}$, $j=1,2$, follow the ratio: $\frac{\alpha_1^{(1)}}{\alpha_2^{(1)}}=\frac{\alpha_1^{(2)}}{\alpha_2^{(2)}}=\frac{\beta_1^{(1)}}{\beta_1^{(2)}}$. We demonstrate this standard elastic collision in the bottom row of Fig. \ref{fig10a}, for the choice $\frac{\alpha_1^{(1)}}{\alpha_2^{(1)}}=\frac{\alpha_1^{(2)}}{\alpha_2^{(2)}}=\frac{\beta_1^{(1)}}{\beta_1^{(2)}}=1$.  During this collision scenario, the vector soliton acquires a phase shift when it undergoes two consecutive collisions with the constituents of the doublet SM. The phase shift suffered by the soliton is calculated as
\begin{equation}
\Delta \Psi=\frac{1}{2l_{1R}}\ln\frac{[|\beta_1^{(1)}|^2+|\beta_1^{(2)}|^2][B_{11}B_{22}-B_{12}B_{21}]|k_1+l_1^*|^2|k_2+l_1^*|^2}{|k_1-l_1|^2|k_2-l_1|^2(l_1+l_1^*)\rho_2},
\end{equation}
where  $\rho_2$ and and $B_{ij}$'s, $i,j=1,2$ are given in Eq. (\ref{16b}) and (\ref{19}), respectively. 
\subsection{Interaction between the fundamental molecular state and two vector solitons}
Now, we study the interaction dynamics between the fundamental soliton molecule and two vector bright solitons by analysing their asymptotic behaviours at the limits $z\rightarrow \pm \infty$. To understand the asymptotic nature of the doublet SM induced by the individual solitons we deduce their corresponding asymptotic expressions by substituting the asymptotic nature of the wave variables, $\eta_{jR}=k_{jR}(t-2k_{1I}z)$ and $\xi_{jR}=l_{jR}(t-2l_{jI}z)$, $j=1,2$, in the $(\bar{N}+\bar{M})=(2+2)$-soliton solution (\ref{2}) with Eqs. (\ref{2a}) and (\ref{2b}). To visualize the behaviour of these variables in the limit $z\rightarrow \pm \infty$, we consider the choice, $k_{1I}<l_{1I}<l_{2I}$, $k_{jR},l_{jR}>0$, $j=1,2$. Consideration of this parametric choice helps us to investigate the head-on collision of two vector bright solitons with a stationary doublet SM. The asymptotic behaviour of the variables $\eta_{jR}$ of SM and the wave variables $\xi_{jR}$ of two individual solitons are obtained as (i) Doublet SM:  $\eta_{1R},~\eta_{2R} \simeq 0$, $\xi_{1R}, ~\xi_{2R}\rightarrow \pm \infty$ as $z\rightarrow\mp\infty$, (ii) Soliton 1 ($S_1$): $\xi_{1R}\simeq 0$, $\eta_{1R}, ~\eta_{2R}\rightarrow \mp\infty$, $\xi_{2R}\rightarrow\pm \infty$ as $z\rightarrow\mp\infty$, and (ii) Soliton 2 ($S_2$): $\xi_{2R}\simeq 0$, $\eta_{1R},~\eta_{2R}\rightarrow\mp\infty$, $\xi_{1R}\rightarrow\mp \infty$ as $z\rightarrow\mp\infty$. These results lead to the following asymptotic forms of the individual  solitons and a doublet SM. 
\\\\
(a) Before collision: $z\rightarrow -\infty$\\
{\bf Doublet SM}: $\eta_{1R},\eta_{2R}\simeq 0$, $\xi_{1R},\xi_{2R}\rightarrow +\infty$ \\
The asymptotic expression describing the fundamental molecular state before collision is deduced from the $(\bar{N}+\bar{M})$-soliton solution (\ref{2}) by setting $\bar{N}=\bar{M}=2$. The resultant asymptotic form is
\bes\bea
&&q_j(z,t)=\frac{1}{D_1^-}\big(e^{i\eta_{2I}}c_{1j}^{-}\cosh(\eta_{1R}+\phi_1^{j-})+e^{i\eta_{1I}}c_{2j}^{-}\cosh(\eta_{2R}+\phi_2^{j-})\big),~j=1,2,\label{22a}\\
&&D_1^-=c_{3}^{-}\cosh(\eta_{1R}+\eta_{2R}+\phi_3^{-})+c_{4}^{-}\cosh(\eta_{1R}-\eta_{2R}+\phi_4^{-})+c_{5}^{-}\big[\cosh\phi_5^{-}\cos\vartheta_1\nonumber\\
&&\hspace{1.2cm}+i\sinh\phi_5^{1-}\sin\vartheta_1\big],\nonumber
\eea
where 
\begin{eqnarray}
&&c_{1,j}^-=e^{i\Omega_1}(k_2+k_2^*)^{1/2}(k_1+k_2^*)^{1/2}\nu_{1,j}\gamma_{1,j},~c_{2,j}^-=e^{i\Omega_2}(k_1+k_1^*)^{1/2}(k_1^*+k_2)^{1/2}\nu_{2,j}\gamma_{2,j}\nonumber,\\
&&c_{3}^-=(k_1^*-k_2^*)^{1/2}(\hat{B}_{11}\hat{B}_{22}-\hat{B}_{12}\hat{B}_{21})^{1/2}\tau_{1}^{1/2},~c_4^-=\frac{|k_1+k_2^*|\tau_2^{1/2}\tau_{3}^{1/2}}{(k_1-k_2)^{1/2}},~k,j=1,2, \nonumber\\
&&c_5^-=\frac{(k_1+k_1^*)^{1/2}(k_2+k_2^*)^{1/2}\tau_4^{1/2}\tau_5^{1/2}}{(k_1-k_2)^{1/2}},~e^{i\Omega_1}=\frac{(k_2-l_1)^{1/2}(k_2-l_2)^{1/2}(k_2^*+l_1)^{1/2}(k_2^*+l_2)^{1/2}}{(k_2^*-l_1^*)^{1/2}(k_2^*-l_2^*)^{1/2}(k_2+l_1^*)^{1/2}(k_2+l_2^*)^{1/2}},\nonumber\\&&e^{i\Omega_2}=\frac{(k_1-l_1)^{1/2}(k_1-l_2)^{1/2}(k_1^*+l_2)^{1/2}}{(k_1^*-l_1^*)^{1/2}(k_1^*-l_2^*)^{1/2}(k_1+l_2^*)^{1/2}},~\vartheta_1=\eta_{1I}-\eta_{2I}=(k_{1R}^2-k_{2R}^2)z.\label{22b}
\end{eqnarray} 
The expressions corresponding to the constants $\nu_{1,j}$, $\nu_{2,j}$, $\gamma_{1,j}$, $\gamma_{2,j}$, $\tau_1$, $\tau_3$, and $\tau_5$ are defined in Appendix A.   Then, the phase terms appearing in the above asymptotic form of doublet SM are given below: 
\bea
&&\phi_1^{j-}=\frac{1}{2}\ln\frac{(k_1-k_2)|k_1-l_1|^2|k_1-l_2|^2\nu_{1,j}}{(k_1+k_1^*)(k_1^*+k_2)|k_1+l_1^*|^2|k_1+l_2^*|^2\gamma_{1,j}}, \nonumber\\
&&\phi_2^{j-}=\frac{1}{2}\ln\frac{(k_1-k_2)|k_2-l_1|^2|k_2-l_2|^2\nu_{2,j}}{(k_2+k_2^*)(k_1+k_2^*)|k_2+l_1^*|^2|k_2+l_2^*|^2\gamma_{2,j}}, \nonumber\\
&&\phi_3^-=\frac{1}{2}\ln\frac{|k_1-k_2|^2|k_1-l_1|^2|k_2-l_1|^2|k_2-l_2|^2|k_1-l_2|^2\tau_1}{(k_1+k_1^*)(k_2+k_2^*)|k_1+k_2^*|^2|k_1+l_1^*|^2|k_2+l_1^*|^2|k_2+l_2^*|^2|k_1+l_2^*|^2(\hat{B}_{11}\hat{B}_{22}-\hat{B}_{12}\hat{B}_{21})}, \nonumber\eea\bea
&&\phi_4^-=\frac{1}{2}\ln\frac{|k_1-l_1|^2|k_1-l_2|^2|k_2+l_1^*|^2|k_2+l_2^*|^2(k_2+k_2^*)\tau_2}{(k_1+k_1^*)|k_1+l_1^*|^2|k_1+l_2^*|^2|k_2-l_1|^2|k_2-l_2|^2\tau_3},\nonumber\\
&&\phi_5^-=\frac{1}{2}\ln\frac{(k_1-l_1)(k_1-l_2)(k_2^*-l_1^*)(k_2^*-l_2^*)(k_1^*+k_2)(k_1^*+l_1)(k_1^*+l_2)(k_2+l_1^*)(k_2+l_2^*)\tau_4}{(k_1^*-l_1^*)(k_1^*-l_2^*)(k_2-l_1)(k_2-l_2)(k_1+k_2^*)(k_1+l_1^*)(k_1+l_2^*)(k_2^*+l_1)(k_2^*+l_2)\tau_5}.\nonumber
\end{eqnarray}\ees
Here the negative sign in the superscript indicates before collision.\\
{\bf Soliton 1}: $\xi_{1R}\simeq 0$,~$\eta_{1R},\eta_{2R}\rightarrow -\infty$, ~$\xi_{2R}\rightarrow +\infty$\\
In this limit, the asymptotic form of soliton 1 is given by 
\begin{subequations}
\begin{eqnarray}
	q_j(z,t)=l_{1R}A_{j}^{1-}e^{i(\xi_{1I}+\theta^{1-})}\sech(\xi_{1R}+\Phi_1^-).\label{23a}
\end{eqnarray}
Here, the unit phase $e^{i\theta^{1-}}$, polarization constant $A_j^{1-}$ and the central position $\Phi_1^-$ associated with the soliton $1$ before collision are given below: 
  \begin{eqnarray}
&&e^{i\theta^{1-}}=\frac{(l_1-l_2)^{1/2}(l_1^*+l_2)^{1/2}}{(l_1^*-l_2^*)^{1/2}(l_1+l_2^*)^{1/2}}, ~A_j^{1-}=\frac{\beta_1^{(j)}(1-\frac{\beta_2^{(j)}\hat{B}_{21}}{\beta_1^{(j)}\hat{B}_{22}})^{1/2}}{\sqrt{|\beta_1^{(1)}|^2+|\beta_1^{(2)}|^2}(1-\frac{\hat{B}_{12}\hat{B}_{21}}{\hat{B}_{11}\hat{B}_{22}})^{1/2}}, \label{23b}\\
&&\Phi_1^-=\frac{1}{2}\ln\frac{|l_1-l_2|^2\hat{B}_{11}(1-\frac{\hat{B}_{12}\hat{B}_{21}}{\hat{B}_{11}\hat{B}_{22}})}{(l_1+l_1^*)|l_1+l_2^*|^2},  ~j=1,2.\label{23c}
\end{eqnarray} \end{subequations}
\\
{\bf Soliton 2}: $\xi_{2R}\simeq 0$,~$\eta_{1R},\eta_{2R}\rightarrow -\infty$, ~$\xi_{1R}\rightarrow -\infty$\\
The asymptotic expression for soliton 2 is obtained from the ($2+2$)-soliton solution as
\begin{subequations}
\begin{eqnarray}
	&&q_j(z,t)=l_{2R}A_{j}^{2-}e^{i\xi_{2I}}\sech(\xi_{2R}+\Phi_2^-),\label{24a}
\end{eqnarray}\\
where the polarization constant $A_j^{2-}$ and the central position $\Phi_2^-$ associated with the soliton $2$ are given below:
\begin{eqnarray}
A_j^{2-}=\frac{\beta_2^{(j)}}{\sqrt{|\beta_2^{(1)}|^2+|\beta_2^{(2)}|^2}},~ j=1,2, ~\Phi_2^-=\frac{1}{2}\ln\frac{\sqrt{|\beta_2^{(1)}|^2+|\beta_2^{(2)}|^2}}{(l_2+l_2^*)^2}. \label{24b}
\end{eqnarray} 
\end{subequations} In the latter, the superscript $2-$ denotes the soliton 2 before collision. \\
(b) After collision: $z\rightarrow +\infty$\\
{\bf Doublet SM}: $\eta_{1R},\eta_{2R}\simeq 0$, $\xi_{1R},\xi_{2R}\rightarrow -\infty$  

In this asymptotic limit, the analytical form describing the doublet SM after collision is deduced from the $(2+2)$-soliton solution as 
\bes\bea
&&q_j(z,t)=\frac{1}{D_1^+}\big(e^{i\eta_{2I}}c_{1j}^{+}\cosh(\eta_{1R}+\phi_1^{j+})+e^{i\eta_{1I}}c_{2j}^{+}\cosh(\eta_{2R}+\phi_2^{j+})\big),~j=1,2,\label{25a}\\
&&D_1^+=c_{3}^{+}\cosh(\eta_{1R}+\eta_{2R}+\phi_3^{+})+c_{4}^{+}\cosh(\eta_{1R}-\eta_{2R}+\phi_4^{+})+c_{5}^{+}\big[\cosh\phi_5^{+}\cos\vartheta_1\nonumber\\
&&\hspace{1.2cm}+i\sinh\phi_5^{+}\sin\vartheta_1\big],\nonumber
\eea
where
\begin{eqnarray}
&&c_{1j}^+=\frac{\alpha_2^{(j)}(k_2+k_2^*)^{1/2}(k_1+k_2^*)^{1/2}}{B_{22}^{1/2}}\big(\frac{\alpha_1^{(j)}B_{12}}{\alpha_2^{(j)}B_{11}}-1\big)^{1/2}, ~c_{3}^+=(1-\frac{B_{12}B_{21}}{B_{11}B_{22}})^{1/2}(k_1^*-k_2^*)^{1/2}, \nonumber\\
&&c_{5}^+=\frac{|B_{12}|(k_1+k_1^*)^{1/2}(k_2+k_2^*)^{1/2}}{\sqrt{B_{11}B_{22}}(k_1-k_2)^{1/2}},~c_{2j}^+=\frac{\alpha_1^{(j)}(k_1+k_1^*)^{1/2}(k_1^*+k_2)^{1/2}}{B_{11}^{1/2}}\big(1-\frac{\alpha_2^{(j)}B_{21}}{\alpha_1^{(j)}B_{22}}\big)^{1/2},\nonumber\\
&&c_{4}^+=\frac{|k_1+k_2^*|}{(k_1-k_2)^{1/2}}.\label{25b}
\end{eqnarray}
The phase terms which appear in Eq. (\ref{25a}) are calculated and their forms are given below:   
\begin{eqnarray}
&& \phi_1^{j+}=\frac{1}{2}\ln\frac{(k_1-k_2)[\alpha_1^{(j)}B_{12}-\alpha_2^{(j)}B_{11}]}{\alpha_2^{(j)}(k_1+k_1^*)(k_1^*+k_2)},~~ \phi_2^{j+}=\frac{1}{2}\ln\frac{(k_1-k_2)[\alpha_1^{(j)}B_{22}-\alpha_2^{(j)}B_{21}]}{\alpha_1^{(j)}(k_2+k_2^*)(k_1+k_2^*)},\nonumber\\
&& \phi_3^{+}=\frac{1}{2}\ln\frac{|k_1-k_2|^2(B_{11}B_{22}-B_{12}B_{21})}{(k_1+k_1^*)(k_2+k_2^*)|k_1+k_2^*|^2}, ~~\phi_4^{+}=\frac{1}{2}
\ln\frac{(k_2+k_2^*)B_{11}}{(k_1+k_1^*)B_{22}},\nonumber\\
&& \phi_5^{+}=\frac{1}{2}\ln\frac{(k_2+k_1^*)B_{21}}{(k_1+k_2^*)B_{12}}.\label{25c}
\end{eqnarray}\ees 
Here the positive sign in the superscript represents after collision.\\
{\bf Soliton 1}: $\xi_{1R}\simeq 0$,~$\eta_{1R},\eta_{2R}\rightarrow +\infty$, ~$\xi_{2R}\rightarrow -\infty$\\
In this after collision limit, the asymptotic form of soliton 1 is of the form
\bes
\begin{eqnarray}
	&&q_j(z,t)=l_{1R}A_{j}^{1+}e^{i(\xi_{1I}+\theta^{1+})}\sech(\xi_{1R}+\Phi_1^+).\label{26a}
\end{eqnarray}
Here, the the unit phase $e^{i\theta^{1+}}$, polarization constant $A_j^{1+}$ and the central position $\Phi_1^+$ associated with the soliton $1$ after collision are derived and their forms read as \begin{eqnarray}
&&e^{i\theta^{1+}}=\frac{(k_1-l_1)^{1/2}(k_1+l_1^*)^{1/2}(k_2-l_1)^{1/2}(k_2+l_1^*)^{1/2}}{(k_1^*-l_1^*)^{1/2}(k_1^*+l_1)^{1/2}(k_2^*-l_1^*)^{1/2}(k_2^*+l_1)^{1/2}}, \nonumber\\
&&A_j^{1+}=\frac{\beta_1^{(j)}\gamma_{3,j}}{\tau_6\sqrt{|\beta_1^{(1)}|^2+|\beta_1^{(2)}|^2}[B_{11}B_{22}-B_{12}B_{21}]^{1/2}},\nonumber\\
&&\Phi_1^+=\frac{1}{2}\ln\frac{|k_1-l_2|^2|k_2-l_1|^2\hat{B}_{11}\tau_6}{|k_1+l_1^*|^2|k_2+l_1^*|^2(B_{11}B_{22}-B_{12}B_{21})},\nonumber\\
&&\gamma_{3,j}=\big[\frac{\alpha_1^{(j)}}{\beta_1^{(j)}}(B_{12}b_{21}-b_{11}B_{22})+\frac{\alpha_2^{(j)}}{\beta_1^{(j)}}(B_{21}b_{11}-b_{21}B_{11})+(B_{11}B_{22}-B_{12}B_{21})\big], ~j=1,2, \nonumber\\
&&\tau_6=\big(B_{11}(B_{22}-\frac{b_{11}\tilde{b}_{12}}{\hat{B}_{11}})+\frac{B_{12}}{\hat{B}_{11}}(\tilde{b}_{11}b_{21}-B_{21}\hat{B}_{11})+\frac{b_{21}}{\hat{B}_{11}}(\tilde{b}_{12}B_{21}-B_{22}\tilde{b}_{11})\big)^{1/2}.\label{26b}
\end{eqnarray} \ees
In the latter, the superscript $1+$ denotes the soliton 1 after collision. 
\\
{\bf Soliton 2}: $\xi_{2R}\simeq 0$,~$\eta_{1R},\eta_{2R}\rightarrow +\infty$, ~$\xi_{1R}\rightarrow +\infty$\\
The asymptotic expression associated with soliton 2 reads as
\bes
\begin{eqnarray}
	&&q_j(z,t)=l_{2R}A_{j}^{2+}e^{i(\xi_{2I}+\theta^{2+})}\sech(\xi_{2R}+\Phi_2^+),\label{27a}
\end{eqnarray}\\
where the the unit phase $e^{i\theta^{2+}}$, polarization constant $A_j^{2+}$ and the central position $\Phi_2^+$ associated with the soliton $2$ are obtained and their forms are given below:
\bea
&&e^{i\theta^{2+}}=\frac{(k_1-l_2)^{1/2}(k_1+l_2^*)^{1/2}(k_2-l_2)^{1/2}(l_1-l_2)^{1/2}(k_2+l_2^*)^{1/2}(l_1+l_2^*)^{1/2}}{(k_1^*-l_2^*)^{1/2}(k_1^*+l_2)^{1/2}(k_2^*-l_2^*)^{1/2}(l_1^*-l_2^*)^{1/2}(k_2^*+l_2)^{1/2}(l_1^*+l_2)^{1/2}}, \nonumber\\
&&A_j^{2+}=\frac{\beta_2^{(j)}\nu_{3,j}}{(l_2+l_2^*)^{1/2}\hat{B}_{11}^{1/2}\tau_6\tau_{1}^{1/2}}, ~j=1,2, \nonumber\\
&&\Phi_2^+=\frac{1}{2}\log\frac{|k_1-l_2|^2|k_2-l_2|^2|l_1-l_2|^2\tau_{1}}{|k_1+l_2^*|^2|k_2+l_2^*|^2|l_1+l_2^*|^2\hat{B}_{11}^{1/2}\tau_6},\nonumber\\
&&\nu_{3,j}=\frac{1}{\beta_2^{(j)}}\bigg((b_{22}\hat{B}_{11}-b_{21}\hat{B}_{12})(\alpha_1^{(j)}B_{12}-\alpha_2^{(j)}B_{11})+(B_{22}\hat{B}_{12}-b_{22}\tilde{
b}_{12})\nonumber\\
&&~~~~~~~~\times (\alpha_1^{(j)}b_{11}-\beta_1^{(j)}B_{11})+(b_{21}\tilde{b}_{12}-B_{22}\hat{B}_{11})(\alpha_1^{(j)}b_{12}-\beta_2^{(j)}B_{11})\nonumber\\
&&\hspace{1.2cm}+(b_{22}\tilde{b}_{11}-B_{21}\hat{B}_{12})(\alpha_2^{(j)}b_{11}-\beta_1^{(j)}B_{12})
+(B_{21}\hat{B}_{11}-b_{21}\tilde{b}_{11})(\alpha_2^{(j)}b_{12}-\beta_2^{(j)}B_{12})\nonumber\\
&&\hspace{1.2cm}+(B_{22}\tilde{b}_{11}-B_{21}\tilde{b}_{12})(\beta_1^{(j)}b_{12}-\beta_2^{(j)}b_{11})\bigg). \label{27b}
\eea
\ees
Here, the superscript $2+$ denotes the soliton 2 after collision. We note here that one has to impose the condition $k_{1I}=k_{2I}$ in the expressions (\ref{22a})-(\ref{22b}) and (\ref{25a})-(\ref{25c}) to obtain the exact asymptotic forms of doublet SM before and after collision.   
\subsubsection{Energy sharing collision between a doublet SM and two vector solitons}
\begin{figure*}
	\centering
	\includegraphics[width=0.9\linewidth]{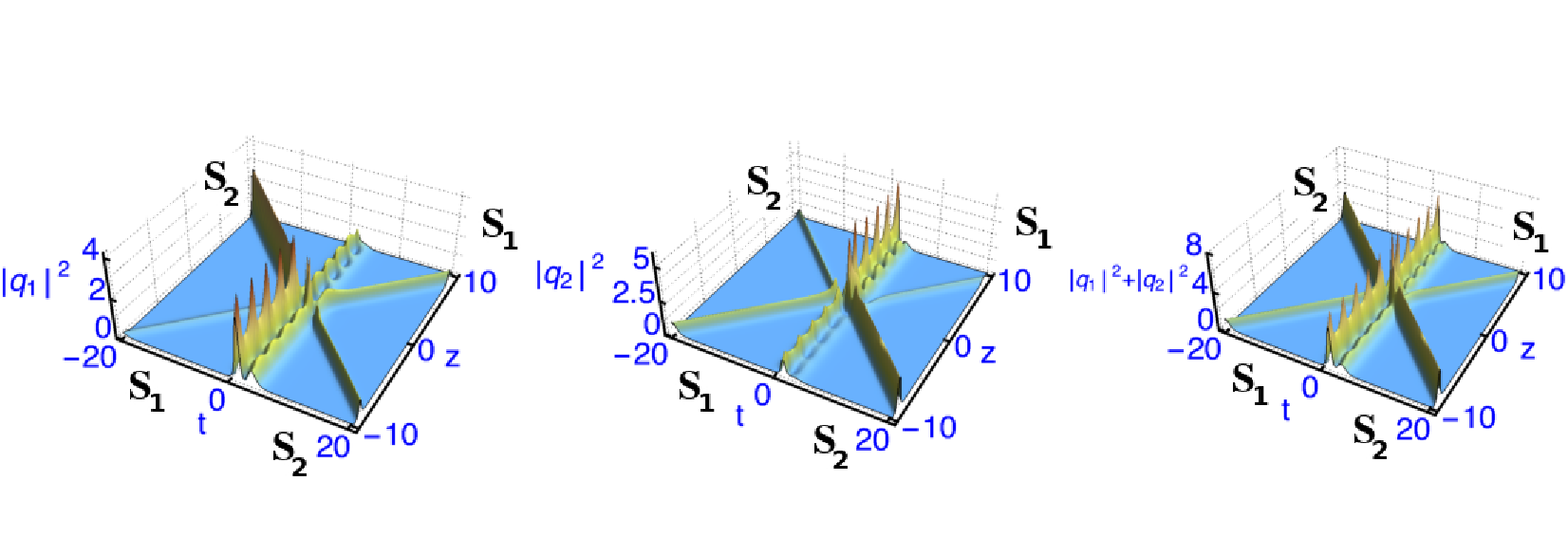}\\
	\includegraphics[width=0.9\linewidth]{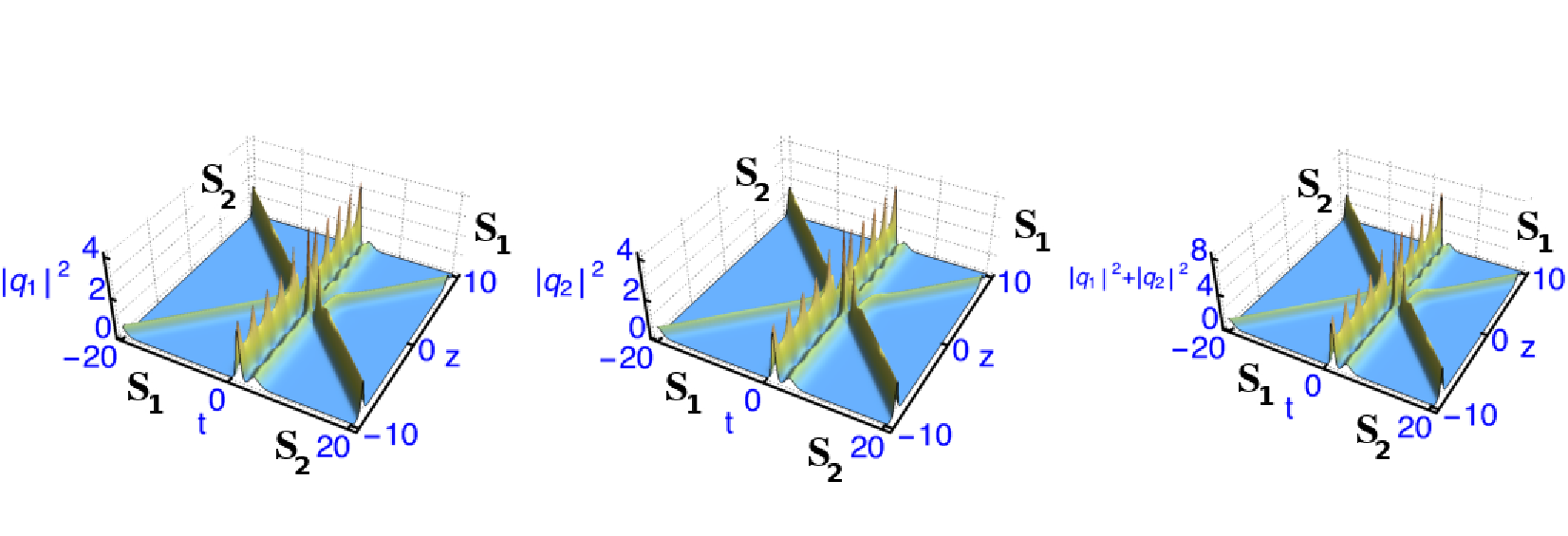}\\
	\caption{Top row panels demonstrate the energy sharing collision between a fundamental soliton molecule and two individual solitons. Here, we fix the parameter values as $k_{1}=2$, $k_{2}=0.99$,  $l_{1}=1.8-i$, $l_{2}=1+i$, $\alpha_{1}^{(j)}=\alpha_{2}^{(j)}=1$, $\beta_1^{(1)}=2$, $\beta_2^{(2)}=2+i$, and $\beta_2^{(1)}=\beta_1^{(2)}=1$. The corresponding elastic collision is displayed in the bottom row panels when we consider $k_{1}=2$, $k_{2}=0.99$,  $l_{1}=1.8-i$, $l_{2}=1+i$, $\alpha_{1}^{(j)}=\alpha_{2}^{(j)}=1$,  and  $\beta_1^{(j)}=\beta_2^{(j)}=2$, $j=1,2$.}
	\label{fig11}
\end{figure*}

The asymptotic analysis which we performed above clearly indicates that the fundamental soliton molecule undergoes energy sharing collision due to the interaction with two basic vector solitons $S_1$ and $S_2$. Because of this interaction, the quantities, $c_{1j}^-$, $c_{2j}^-$, $c_3^-$, $c_4^-$, $c_5^-$, $\phi_1^{j-}$, $\phi_2^{j-}$, $\phi_3^-$, $\phi_4^-$, and $\phi_5^-$, $j=1,2$, related to the doublet SM before collision do not preserve their forms and they get changed to $c_{1j}^+$, $c_{2j}^+$, $c_3^+$, $c_4^+$, $c_5^+$, $\phi_1^{j+}$, $\phi_2^{j+}$, $\phi_3^+$, $\phi_4^+$, and $\phi_5^+$, $j=1,2$. These changes occur essentially when the two solitons lose their original identities during the collision process. That is their initial amplitudes and central positions change from ($l_{1R}A_{j}^{1-},~\Phi_1^-$) and ($l_{2R}A_{j}^{2-},~\Phi_2^-$) to ($l_{1R}A_{j}^{1+},~\Phi_1^+$) and ($l_{2R}A_{j}^{2+},~\Phi_2^+$), $j=1,2$, respectively. It implies that there is a definite energy sharing occurring in between the doublet molecular state and the two vector solitons of the two modes. As an example, we demonstrate such an energy sharing collision among the modes in the top-row of Fig. \ref{fig11}, in which two oppositely moving  solitons collide with a stationary doublet molecular state. From this figure, we identify that the doublet molecule loses its energy to the solitons $S_1$ and $S_2$ in the first mode ($q_1$) whereas in the second mode ($q_2$) an opposite kind of energy sharing occurs. That is, soliton molecule gains energy from the two solitons. The amount of energy transferred to the molecule or gained from the molecule can be identified by calculating the transition intensities of solitons. To characterize the changes in the intensity of each of the solitons in the two modes, we derive the following expressions for the transition intensities from the asymptotic amplitude expressions. By doing so, we find the set of  transition intensities corresponding to the two components of soliton $1$ as 
 \bes\begin{eqnarray}
&&|T_j^1|^2=\frac{|A_j^{1+}|^2}{|A_j^{1-}|^2}=\frac{|1-\frac{\hat{B}_{12}\hat{B}_{21}}{\hat{B}_{11}\hat{B}_{22}}||\gamma_{3,j}|^2}{|B_{11}B_{22}-B_{12}B_{21}||\tau_6||1-\frac{\beta_2^{(j)}\hat{B}_{21}}{\beta_1^{(j)}\hat{B}_{22}}|},~j=1,2.\label{28a}
\end{eqnarray}
Then, the set of transition intensities corresponding to the two modes of soliton $2$ is obtained as 
\begin{eqnarray}
&&|T_j^2|^2=\frac{|A_j^{2+}|^2}{|A_j^{2-}|^2}=\frac{|\nu_{3,j}|^2}{|\hat{B}_{11}||\tau_6||\tau_1|},~j=1,2.\label{28b}
\end{eqnarray}
\ees
The above expressions again confirm that the molecular state undergoes energy sharing with its interacting partners in both the modes. During the collision process, the energy of the fundamental molecule in the individual modes is conserved as per the following conservation equations, that is, 
\bes\begin{eqnarray}
&&\Delta E_{mol}^1=(|A_1^{1+}|^2-|A_1^{1-}|^2)+(|A_1^{2+}|^2-|A_1^{2-}|^2),\label{29a}\\
&&\Delta E_{mol}^2=(|A_2^{1+}|^2-|A_2^{1-}|^2)+(|A_2^{2+}|^2-|A_2^{2-}|^2).\label{29b}
\end{eqnarray}\ees
The energy conservation of SM can also be visualized from the transition intensities of solitons. That is, in mode $1$, the change in energy of the molecule is given by $\Delta E_{mol}^1=|T_1^{2}|^2-|T_1^{1}|^2=\frac{|A_1^{2+}|^2}{|A_1^{2-}|^2}-\frac{|A_1^{1+}|^2}{|A_1^{1-}|^2}$ and in mode $2$ the molecular energy is,  $\Delta E_{mol}^2=|T_2^{2}|^2-|T_2^{1}|^2=\frac{|A_2^{2+}|^2}{|A_2^{2-}|^2}-\frac{|A_2^{1+}|^2}{|A_2^{1-}|^2}$. These conservation relations imply that the intensity or energy of each of the modes is separately conserved through $\int_{-\infty}^{+\infty}|q_j|^2dz=\text{const}$, $j=1,2$. In addition to this, the total energies of solitons and molecule are also conserved in all the modes, as it is illustrated in the third top-row panel of Fig. \ref{fig11}, as per the total energy conservation, $\int_{-\infty}^{+\infty}(|q_1|^2+|q_2|^2)dz=\text{const}$, $j=1,2$.     

Besides the above, another important observation that we noticed from the asymptotic expressions (\ref{22a}) and (\ref{25a}) as well as from the top-row figures of Figs. \ref{fig11} is the preservation of periodic nature of SM. The recurrence frequency: $\omega_{12}=(k_{1R}^2-k_{2R}^2)$ and the period of oscillation: $T_{12}=\frac{2\pi}{k_{1R}^2-k_{2R}^2}$ are preserved throughout the collision process. Therefore, the above analysis clearly shows that the doublet molecular state does not dissociate after collision with two individual solitons either. Thus, it survives against soliton collisions except for the change in intensity. To the best of our knowledge, this interesting collision scenario has not been reported before in the literature.  

However, one can also suppress the energy sharing property of the soliton molecule by choosing the values of the soliton parameters $\alpha_1^{(j)}$,  $\alpha_2^{(j)}$, $\beta_1^{(j)}$, and $\beta_2^{(j)}$, $j=1,2$, in such a way that they obey the relations, $\frac{\alpha_1^{(1)}}{\alpha_2^{(1)}}=\frac{\alpha_1^{(2)}}{\alpha_2^{(2)}}=\frac{\beta_1^{(1)}}{\beta_2^{(1)}}=\frac{\beta_1^{(2)}}{\beta_2^{(2)}}$. Whenever this condition is satisfied, the soliton molecule always undergoes elastic collision with its interacting partner solitons in the present conservative Kerr medium. Such collision scenario is illustrated in the bottom-panel of Figs. \ref{fig11}. From this figure, one can also notice that the shape of the molecule as well as the shapes of two solitons remain unchanged. In this circumstance, the transition intensities $|T_j^l|^2$ (Eqs. (\ref{28a}) and (\ref{28b})), $l,j=1,2$,  become unimodular and the change in energy $\Delta E_{mol}^l$ (Eqs. (\ref{29a}) and (\ref{29b})), $l=1,2$, vanishes ($\Delta E_{mol}^l=0$). This again confirms that the molecule does not gain or lose energy during the elastic collision and also its breathing nature is preserved.    

\begin{figure*}
	\centering
	\includegraphics[width=0.9\linewidth]{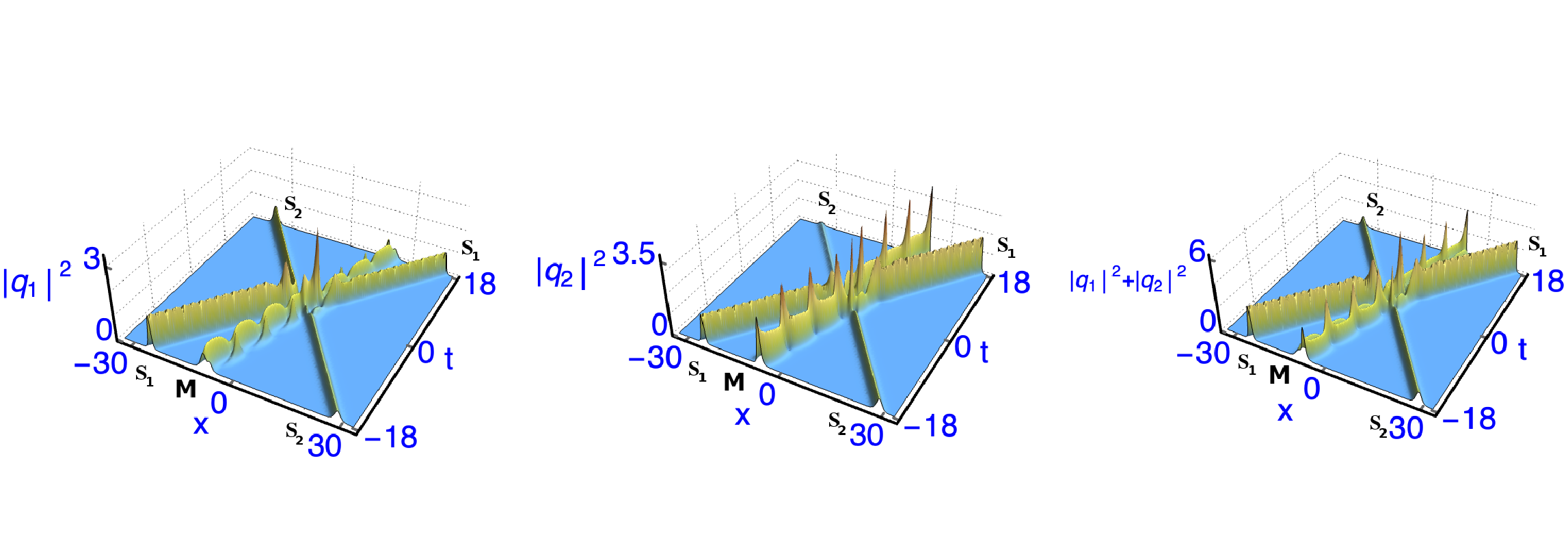}
	\caption{The energy sharing collision between a fundamental soliton molecule and two individual solitons. We fix the parameter values as $k_{1}=1.5+0.3i$, $k_{2}=1.1+0.3i$,  $l_{1}=1.8-0.7i$, $l_{2}=1+0.7i$, $\alpha_{1}^{(j)}=\alpha_{2}^{(j)}=1$, $\beta_1^{(1)}=0.7$, $\beta_2^{(2)}=1+i$, and $\beta_2^{(1)}=\beta_1^{(2)}=0.7$. }
	\label{fig12}
\end{figure*}

Further, in Fig. \ref{fig12}, we also elucidate another interesting energy sharing collision scenario between the doublet SM and two vector bright solitons. From this figure, one can observe that the two vector bright solitons exhibit energy sharing collision in the standard way. That is, in the first mode, the energy of soliton $S_1$ gets suppressed after collision with a doublet SM and the energy of second soliton $S_2$ is enhanced. The reverse scenario occurs in the second mode $q_2$. However, during this collision process, the structure of doublet SM changes in the following way: In the mode $q_1$, the intensity of the right soliton atom in the SM is enhanced after collision while that of intensity of the left soliton is suppressed. In the mode $q_2$, the opposite process takes place. That is, the intensity of the right soliton is suppressed whereas the energy of  the left bound pulse is enhanced after collision. This collision scenario is entirely different from the one that is demonstrated in the top panel of Fig. \ref{fig11}. It is because, in Fig. \ref{fig12}, the energy sharing occurs in between the constituents of SM when they collide with two solitons, whereas in Fig. \ref{fig11} the total energy of SM is completely shared (or gained) to the colliding partner solitons.

\subsection{Collision between two fundamental molecular states}
We now intend to analyse the head-on collision between two fundamental soliton molecular states, which propagate in opposite directions with distinct molecular velocities $v_{mol}=2k_{1I}$ and $v'_{mol}=2l_{1I}$. To understand this collision scenario, we assume the parametric choice $k_{jR},~l_{jR}>0$ and $k_{1I}<l_{1I}$, $j=1,2$, while performing the asymptotic analysis. Under this choice, the fundamental soliton molecular groups (say $M_1$ and $M_2$) are well separated and are located along the line, $t_1=2k_{1I}z$ and $t_2=2l_{1I}z$, respectively. In the limits $z\rightarrow\pm\infty$, their asymptotic forms are deduced from the $(\bar{N}+\bar{M})$-soliton solution (\ref{2}), with $\bar{N}=\bar{M}=2$, by incorporating the asymptotic nature of $\eta_{jR}$'s and $\xi_{jR}$'s. The asymptotic nature of these wave variables are identified as  (i) Doublet SM 1 ($M_1$): $\eta_{1R},~\eta_{2R}\simeq 0$, $\xi_{1R},~\xi_{2R}\rightarrow\pm \infty$ as $z\rightarrow\mp\infty$ and (ii) Doublet SM 2 ($M_2$): $\xi_{1R},~\xi_{2R}\simeq 0$, $\eta_{1R},~\eta_{2R}\rightarrow\mp \infty$ as $z\rightarrow\mp\infty$. The following asymptotic forms of the individual molecules are brought out from the analysis. \\
(a) Before collision: $z\rightarrow -\infty$\\
{\bf Doublet SM 1}: $\eta_{1R},~\eta_{2R}\simeq 0$, $\xi_{1R},~\xi_{2R}\rightarrow +\infty$ \\
In this asymptotic limit, the analytical expressions of the fundamental molecule $1$  are deduced from the $(\bar{N}+\bar{M})$-soliton solution by fixing $\bar{N}=\bar{M}=2$. However, the resultant forms exactly match with the expressions (\ref{22a})-(\ref{22b}), which appear in Sec. III A ,during the analysis of collision between doublet SM and two solitons. Therefore, to avoid the repetition we omit the associated mathematical details. \\  
{\bf Doublet SM 2}: $\xi_{1R},~\xi_{2R}\simeq 0$, $\eta_{1R},~\eta_{2R}\rightarrow -\infty$ \\
The forms of $q_1$ and $q_2$ related to the fundamental SM $2$ are obtained as
\bes
\bea
&&q_j(z,t)=\frac{1}{D_2^-}\big(e^{i\xi_{2I}}d_{1j}^{-}\cosh(\xi_{1R}+\varphi_1^{j-})+e^{i\xi_{1I}}d_{2j}^{-}\cosh(\xi_{2R}+\varphi_2^{j-})\big),~j=1,2,\label{30a}\\
&&D_2^-=d_{3}^{-}\cosh(\xi_{1R}+\xi_{2R}+\varphi_3^{-})+d_{4}^{-}\cosh(\xi_{1R}-\xi_{2R}+\varphi_4^{-})+d_{5}^{-}\big[\cosh\varphi_5^{-}\cos\vartheta_2\nonumber\\
&&\hspace{1.2cm}+i\sinh\varphi_5^{-}\sin\vartheta_2\big],\nonumber
\eea
where 
\bea
&&d_{1j}^-=\frac{\beta_2^{(j)}(l_2+l_2^*)^{1/2}(l_1+l_2^*)^{1/2}}{\hat{B}_{22}^{1/2}}\big(\frac{\beta_1^{(j)}\hat{B}_{12}}{\beta_2^{(j)}\hat{B}_{11}}-1\big)^{1/2}, ~\vartheta_2=\xi_{1I}-\xi_{2I}=(l_{1R}^2-l_{2R}^2)z,\nonumber\\
&&d_{2j}^-=\frac{\beta_1^{(j)}(l_1+l_1^*)^{1/2}(l_1^*+l_2)^{1/2}}{\hat{B}_{11}^{1/2}}\big(1-\frac{\beta_2^{(j)}\hat{B}_{21}}{\beta_1^{(j)}\hat{B}_{22}}\big)^{1/2},~ d_{4}^-=\frac{|l_1+l_2^*|}{(l_1-l_2)^{1/2}}, \nonumber\eea\bea
&&d_{3}^-=(1-\frac{\hat{B}_{12}\hat{B}_{21}}{\hat{B}_{11}\hat{B}_{22}})^{1/2}(l_1^*-l_2^*)^{1/2}, ~d_{5}^-=\frac{\sqrt{\hat{B}_{12}\hat{B}_{21}}(l_1+l_1^*)^{1/2}(l_2+l_2^*)^{1/2}}{\sqrt{\hat{B}_{11}\hat{B}_{22}}(l_1-l_2)^{1/2}}.\label{30b}
\eea
The phase terms related to Eq. (\ref{30a}) are appended below:  
\bea
&&\varphi_1^{j-}=\frac{1}{2}\log\frac{(l_1-l_2)[\beta_1^{(j)}\hat{B}_{12}-\beta_2^{(j)}\hat{B}_{11}]}{\beta_2^{(j)}(l_1+l_1^*)(l_1^*+l_2)}, ~\varphi_2^{j-}=\frac{1}{2}\log\frac{(l_1-l_2)[\beta_1^{(j)}\hat{B}_{22}-\beta_2^{(j)}\hat{B}_{21}]}{\beta_1^{(j)}(l_1+l_2^*)(l_2^*+l_2)}, \nonumber\\
&&\varphi_3^{-}=\frac{1}{2}\log\frac{|l_1-l_2|^2(\hat{B}_{11}\hat{B}_{22}-\hat{B}_{12}\hat{B}_{21})}{(l_1+l_1^*)(l_2+l_2^*)|l_1+l_2^*|^2}, ~\varphi_4^{-}=\frac{1}{2}
\log\frac{(l_2+l_2^*)\hat{B}_{11}}{(l_1+l_1^*)\hat{B}_{22}},\nonumber\\
&& \varphi_5^{-}=\frac{1}{2}\log\frac{(l_2+l_1^*)\hat{B}_{21}}{(l_1+l_2^*)\hat{B}_{12}}. \label{30c}
\eea
\ees
(b) After collision: $z\rightarrow +\infty$\\
{\bf Doublet SM 1}: $\eta_{1R},~\eta_{2R}\simeq 0$, $\xi_{1R},~\xi_{2R}\rightarrow -\infty$ \\
In this asymptotic limit also, the final forms of doublet molecular state  exactly coincide with the asymptotic expressions (\ref{25a})-(\ref{25c}). Therefore, we avoid this mathematical detail for brevity. \\
{\bf Doublet SM 2}: $\xi_{1R},~\xi_{2R}\simeq 0$, $\eta_{1R},~\eta_{2R}\rightarrow +\infty$ \\
In this limit, the asymptotic expressions associated with the doublet molecular state 2 are obtained from the $(2+2)$-soliton solution. They are given below:
\bes\bea
&&q_j(z,t)=\frac{1}{D_2^+}\big(e^{i\xi_{2I}}d_{1j}^{+}\cosh(\xi_{1R}+\varphi_1^{j+})+e^{i\xi_{1I}}d_{2j}^{+}\cosh(\xi_{2R}+\varphi_2^{j+})\big),~j=1,2,\label{31a}\\
&&D_2^+=d_{3}^{+}\cosh(\xi_{1R}+\xi_{2R}+\varphi_3^{+})+d_{4}^{+}\cosh(\xi_{1R}-\xi_{2R}+\varphi_4^{+})+d_{5}^{+}\big[\cosh\varphi_5^{+}\cos(\xi_{1I}-\xi_{2I})\nonumber\\
&&\hspace{1.2cm}+i\sinh\varphi_5^{+}\sin(\xi_{1I}-\xi_{2I})\big],\nonumber
\eea
where
\begin{eqnarray}
&&d_{1,j}^+=e^{i\Omega_1'}(l_2+l_2^*)^{1/2}(l_1+l_2^*)^{1/2}\mu_{1,j}\chi_{1,j},~d_{2,j}^+=e^{i\Omega_2'}(l_1+l_1^*)^{1/2}(l_1^*+l_2)^{1/2}\mu_{2,j}\chi_{2,j}\nonumber,\\
&&d_{3}^+=(l_1^*-l_2^*)^{1/2}(B_{11}B_{22}-B_{12}B_{21})^{1/2}\tau_{1}^{1/2},~d_4^+=\frac{|l_1+l_2^*|\chi_3^{1/2}\chi_{4}^{1/2}}{(l_1-l_2)^{1/2}},~k,j=1,2, \nonumber\\
&&d_5^+=\frac{(l_1+l_1^*)^{1/2}(l_2+l_2^*)^{1/2}\chi_5^{1/2}\chi_6^{1/2}}{(l_1-l_2)^{1/2}},~e^{i\Omega_1'}=\frac{(k_1-l_2)^{1/2}(k_2-l_2)^{1/2}(k_1+l_2^*)^{1/2}(k_2+l_2^*)^{1/2}}{(k_1^*-l_2^*)^{1/2}(k_2^*-l_2^*)^{1/2}(k_1^*+l_2)^{1/2}(k_2^*+l_2)^{1/2}},\nonumber\\&&e^{i\Omega_2'}=\frac{(k_1-l_1)^{1/2}(k_2-l_1)^{1/2}(k_1+l_1^*)^{1/2}(k_2+l_1^*)^{1/2}}{(k_1^*-l_2^*)^{1/2}(k_2^*-l_2^*)^{1/2}(k_1^*+l_2)^{1/2}(k_2^*+l_2)^{1/2}},~\vartheta_1=\xi_{1I}-\xi_{2I}=(l_{1R}^2-l_{2R}^2)z.\label{31b}
\end{eqnarray}
The phase terms corresponding to the asymptotic expressions are calculated and they are listed below:
\bea
\varphi_1^{j+}&=&\frac{1}{2}\log\frac{|k_1-l_1|^2|k_2-l_1|^2(l_1-l_2)\mu_{1,j}}{|k_1+l_1^*|^2|k_2+l_1^*|^2(l_1+l_1^*)(l_1^*+l_2)\chi_{1,j}},\nonumber\\
\varphi_5^+ &=&\frac{1}{2}\log\frac{(k_1-l_1)(k_2-l_1)(k_1^*-l_2^*)(k_2^*-l_2^*)(k_1+l_1^*)(k_1^*+l_2)(k_2+l_1^*)(k_2^*+l_2)(l_1^*+l_2)\chi_5}{(k_1^*-l_1^*)(k_2^*-l_1^*)(k_1-l_2)(k_2-l_2)(k_1^*+l_1)(k_1+l_2^*)(k_2^*+l_1)(k_2+l_2^*)(l_1+l_2^*)\chi_6},\nonumber\\
\varphi_3^+ &=&\frac{1}{2}\log\frac{|k_1-l_1|^2|k_2-l_1|^2|k_1-l_2|^2|k_2-l_2|^2|l_1-l_2|^2\tau_1}{|k_1+l_1^*|^2|k_2+l_1^*|^2(l_1+l_1^*)|k_1+l_2^*|^2|k_2+l_2^*|^2(l_2+l_2^*)(B_{11}B_{22}-B_{12}B_{21})},\nonumber\\
\varphi_2^{j+}&=&\frac{1}{2}\log\frac{|k_1-l_2|^2|k_2-l_2|^2(l_1-l_2)\mu_{2,j}}{|k_1+l_2^*|^2|k_2+l_2^*|^2(l_2+l_2^*)(l_1+l_2^*)\chi_{2,j}},\nonumber\\
\varphi_4^+ &=&\frac{1}{2}\log\frac{|k_1-l_1|^2|k_2-l_1|^2(l_2+l_2^*)|k_1+l_2^*|^2|k_2+l_2^*|^2\chi_3}{|k_1+l_1^*|^2|k_2+l_1^*|^2(l_1+l_1^*)|k_1-l_2|^2|k_2-l_2|^2\chi_4}.\label{31c}
\eea \ees
In this physical situation also, one has to impose the velocity resonance conditions $k_{1I}=k_{2I}$, $l_{1I}=l_{2I}$, in the expressions  (\ref{30a})-(\ref{30c}) and  (\ref{31a})-(\ref{31c}), along with Eqs. (\ref{22a})-(\ref{22b})  and (\ref{25a})-(\ref{25c}), to obtain the exact asymptotic expressions of doublet SMs before and after collision.  
\begin{figure*}
	\centering
	\includegraphics[width=0.9\linewidth]{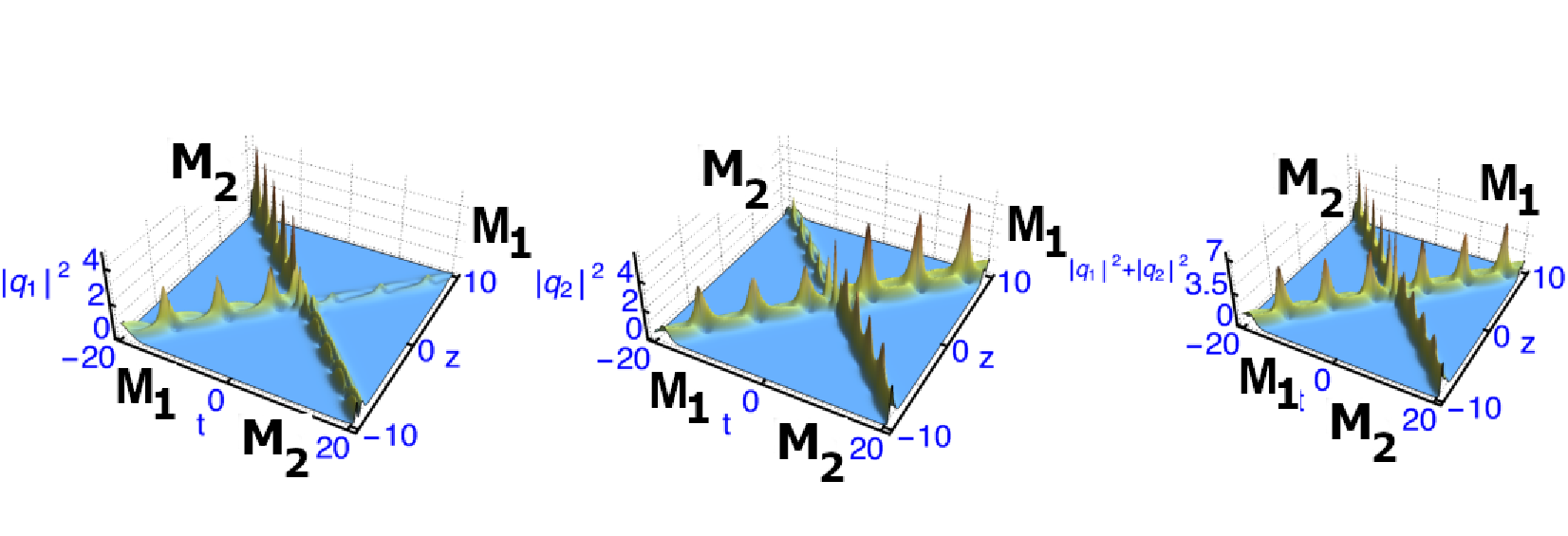}\\
	\includegraphics[width=0.9\linewidth]{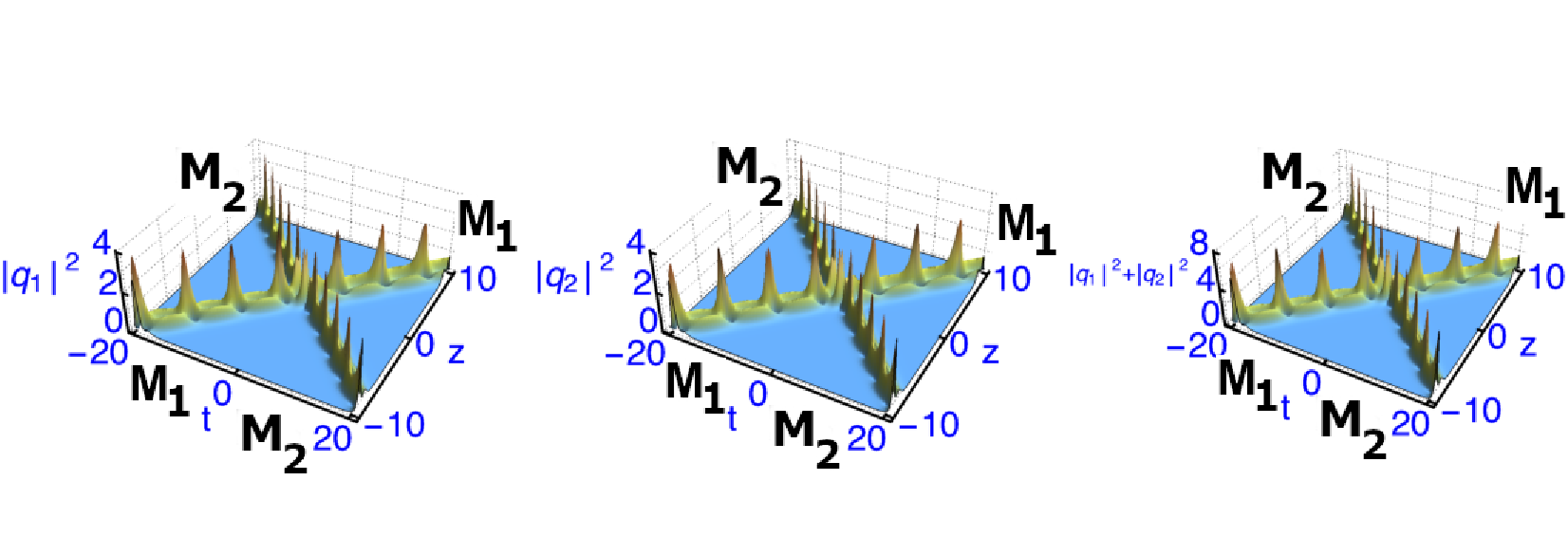}
	\caption{The shape-changing  collision between two distinct molecular groups, each made by a pair of soliton atoms which co-propagate  with degenerate velocities is demonstrated in the top panel. The parameter values are fixed as $k_{1}=2-i$, $k_{2}=1-i$,  $l_{1}=1.7+i$, $l_{2}=0.9+i$, $\alpha_{1}^{(1)}=1+i$, $\alpha_{2}^{(1)}=\alpha_{1}^{(2)}=\alpha_{2}^{(2)}=1$, $\beta_1^{(1)}=2$, and  $\beta_2^{(1)}=2-i$, $\beta_1^{(2)}=\beta_2^{(2)}=2$. In the bottom row, we display an elastic collision between the two oppositely moving soliton molecular groups by fixing the parameter values as $k_{1}=2-i$, $k_{2}=0.9-i$, $l_{1}=1.8+i$, $l_{2}=1+i$, $\alpha_{1}^{(j)}=\alpha_{2}^{(j)}=1$,  and  $\beta_1^{(j)}=\beta_2^{(j)}=2$, $j=1,2$.  }
	\label{fig13}
\end{figure*}
\subsubsection{Energy sharing collision between doublet molecular states}
The asymptotic expressions (\ref{22a})-(\ref{22b}), (\ref{25a})-(\ref{25c}), (\ref{30a})-(\ref{30c}), and  (\ref{31a})-(\ref{31c}) of doublet molecular states indicate that an energy sharing collision does occur in this physical situation. We substantiate this from the variations of all the quantities related to molecular states before and after collision. For instance, during the collision, the quantities of doublet SM $1$ before collision: $c_{1j}^-$, $c_{2j}^-$, $c_3^-$, $c_4^-$, $c_5^-$, $\phi_1^{j-}$, $\phi_2^{j-}$, $\phi_3^-$, $\phi_4^-$, and $\phi_5^-$, $j=1,2$, get varied to  $c_{1j}^+$, $c_{2j}^+$, $c_3^+$, $c_4^+$, $c_5^+$, $\phi_1^{j+}$, $\phi_2^{j+}$, $\phi_3^+$, $\phi_4^+$, and $\phi_5^+$, $j=1,2$. Similarly, the quantities of doublet SM $2$, $d_{1j}^-$, $d_{2j}^-$, $d_3^-$, $d_4^-$, $d_5^-$, $\varphi_1^{j-}$, $\varphi_2^{j-}$, $\varphi_3^-$, $\varphi_4^-$, and $\varphi_5^-$, $j=1,2$, before collision  get changed to $d_{1j}^+$, $d_{2j}^+$, $d_3^+$, $d_4^+$, $d_5^+$, $\varphi_1^{j+}$, $\varphi_2^{j+}$, $\varphi_3^+$, $\varphi_4^+$, and $\varphi_5^+$, $j=1,2$. This ensures that two fundamental molecular structures experience energy sharing collision when they interact with each other. Such an interesting collision scenario is demonstrated in the top-row of Fig. \ref{fig12}, where two doublet molecules move in  opposite directions. From this figure, we observe that energy of the molecule $M_1$ is suppressed in the first mode $q_1$ while it interacts with molecule $M_2$. To hold the energy conservation in the individual modes, as specified by $\int_{-\infty}^{+\infty}|q_j|^2dz=\text{const}$, $j=1,2$,  the energy of the molecule $M_2$ is enhanced in the same mode $q_1$ by gaining energy from the molecule $M_1$. As far as the second mode $q_2$ is concerned an opposite kind of energy sharing occurs during the collision. That is, the energy of the molecule $M_1$ is enhanced in the second mode by gaining energy from the molecule $M_2$. As a consequence of this, the energy of the molecule $M_2$ is suppressed in the mode $q_2$. Therefore, in this case, shape changing occurs in between the molecular states mainly because of energy redistribution among the modes. Apart from this, the total energy of each of the molecules is also conserved as dictated by the Hamiltonian: $\int_{-\infty}^{+\infty}(|q_1|^2+|q_2|^2)dz=\text{const}$, $j=1,2$ (third figure in top-row of Fig. \ref{fig13}).    

Furthermore, the molecular oscillation frequencies: $\omega_{12}=(k_{1R}^2-k_{2R}^2)$ and $\omega_{34}=(l_{1R}^2-l_{2R}^2)$ and the period of oscillations: $T_{12}=\frac{2\pi}{|k_{1R}^2-k_{2R}^2|}$ and $T_{34}=\frac{2\pi}{|l_{1R}^2-l_{2R}^2|}$, related to breathing patterns of both SMs are preserved throughout the collision process. This implies that molecules always maintain their periodic nature  even after undergoing collision with each other.  However, one can bring out the shape preserving nature of the soliton molecules by choosing the    values of the parameters to satisfy the condition: $\frac{\alpha_1^{(1)}}{\alpha_2^{(1)}}=\frac{\alpha_1^{(2)}}{\alpha_2^{(2)}}=\frac{\beta_1^{(1)}}{\beta_2^{(1)}}=\frac{\beta_1^{(2)}}{\beta_2^{(2)}}$. Under this situation, the energy sharing property of the fundamental molecules is suppressed. Thus, they undergo an elastic collision, which is displayed in the bottom-row of Fig. \ref{fig13}.
\section{Numerical analysis: Stability of vector soliton molecules} 
In this section, we also check the stability of vector soliton molecules numerically. To carry out the numerical analysis, we have adopted the SSF  method to verify the robustness of the obtained vector soliton molecules by considering a random noise as perturbation. In this analysis, we consider the initial conditions of the form
 \begin{equation}
  	q_{j}(t)=[1+\zeta_j f(t)]q_j(0,t), ~j=1,2. \label{31}
  \end{equation}   
In the above, $q_j(0,t)$'s are the initial profiles of the basic molecule in the two modes, which are obtained from the fundamental bound soliton solution (\ref{9a})-(\ref{9d}), at $z=0$, of the Manakov system (\ref{manakov}). Then,  $\zeta_j$'s are the amplitudes of the random perturbation for each mode and $f(t)$ is the random noise function, which is generated in the present case by uniformly distributing the random values in the interval $[-1,1]$. Further, we consider the step sizes as $\Delta t=0.1$, $\Delta z=0.002$ (to test the stability of strongly bound SM), and $\Delta z=0.003$ (to test the stability of weakly bound SM) for the dimensionless time and propagation distance respectively, in the numerical analysis. The computational domain for $t$ is fixed as $[-20,20]$ and a free boundary condition is assumed for the propagation distance $z$, so that we can locate the breaking point at which the  fundamental bound soliton state breaks into two oppositely moving individual solitons. We note that in Eq. (\ref{31}), $\zeta_j$'s can be treated as either identical ($\zeta_1=\zeta_2=\zeta$) or distinct ($\zeta_1\neq \zeta_2$) \cite{sun-stability} so that one is able to supply identical or non-identical percentage of perturbations to both the modes to study the stability of doublet SM. However, in the present analysis, we have assumed identical perturbations in all the modes.  

\begin{figure}
	\centering\includegraphics[width=0.8\linewidth]{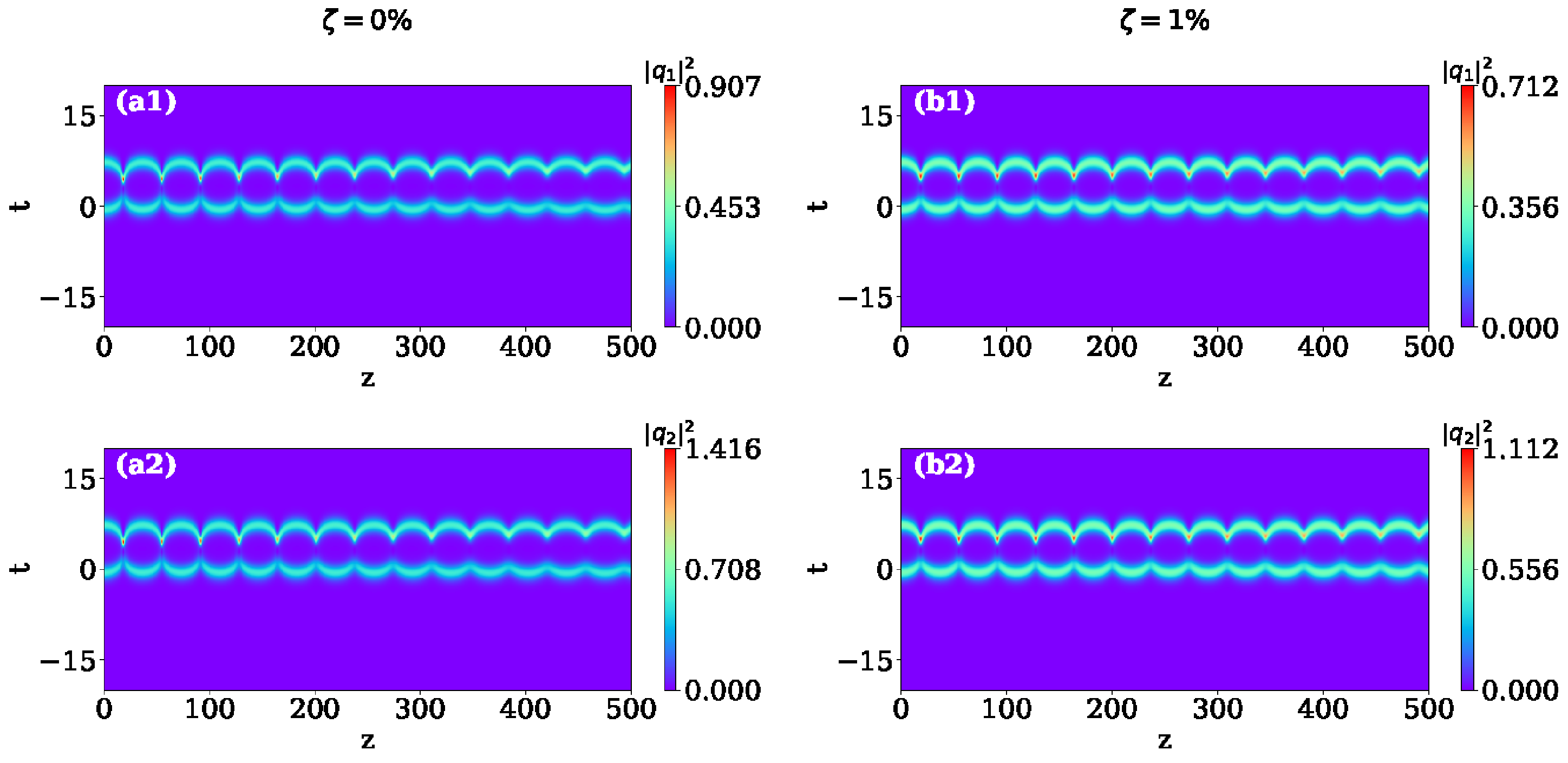} 
	\caption{The propagation of weakly bound fundamental soliton molecular structure is demonstrated in Figs. (a1) and (a2), where the bound soliton state propagates without any breaking. In Figs. (b1) and (b2), we demonstrate the evolution of basic bound soliton state with  $1\%$ of random perturbation. }
	\label{fig15}
\end{figure}

We start the analysis with a doublet SM, composed of two weakly interacting soliton atoms,  which is subjected to zero perturbation ($\zeta=0$). In this situation, the initial conditions are incorporated from Eqs. (\ref{9a})-(\ref{9d}) with $k_1=1$, $l_1=0.91$, $\alpha_1^{(1)}=0.8$, $\alpha_1^{(2)}=1$, $\alpha_2^{(1)}=1.2$, and $\alpha_1^{(1)}=1.5$. An evolution of a typical unperturbed doublet SM is demonstrated in Figs. \ref{fig15}(a1) and \ref{fig15}(a2). Two weakly interacting soliton atoms in this fundamental SM move along the propagation direction and they exhibit weak periodic oscillations due to a weak force between them. The entire soliton molecular structure remains stable and propagates without breaking. This result agrees exactly with the analytical results.
\begin{figure}
	\centering\includegraphics[width=0.8\linewidth]{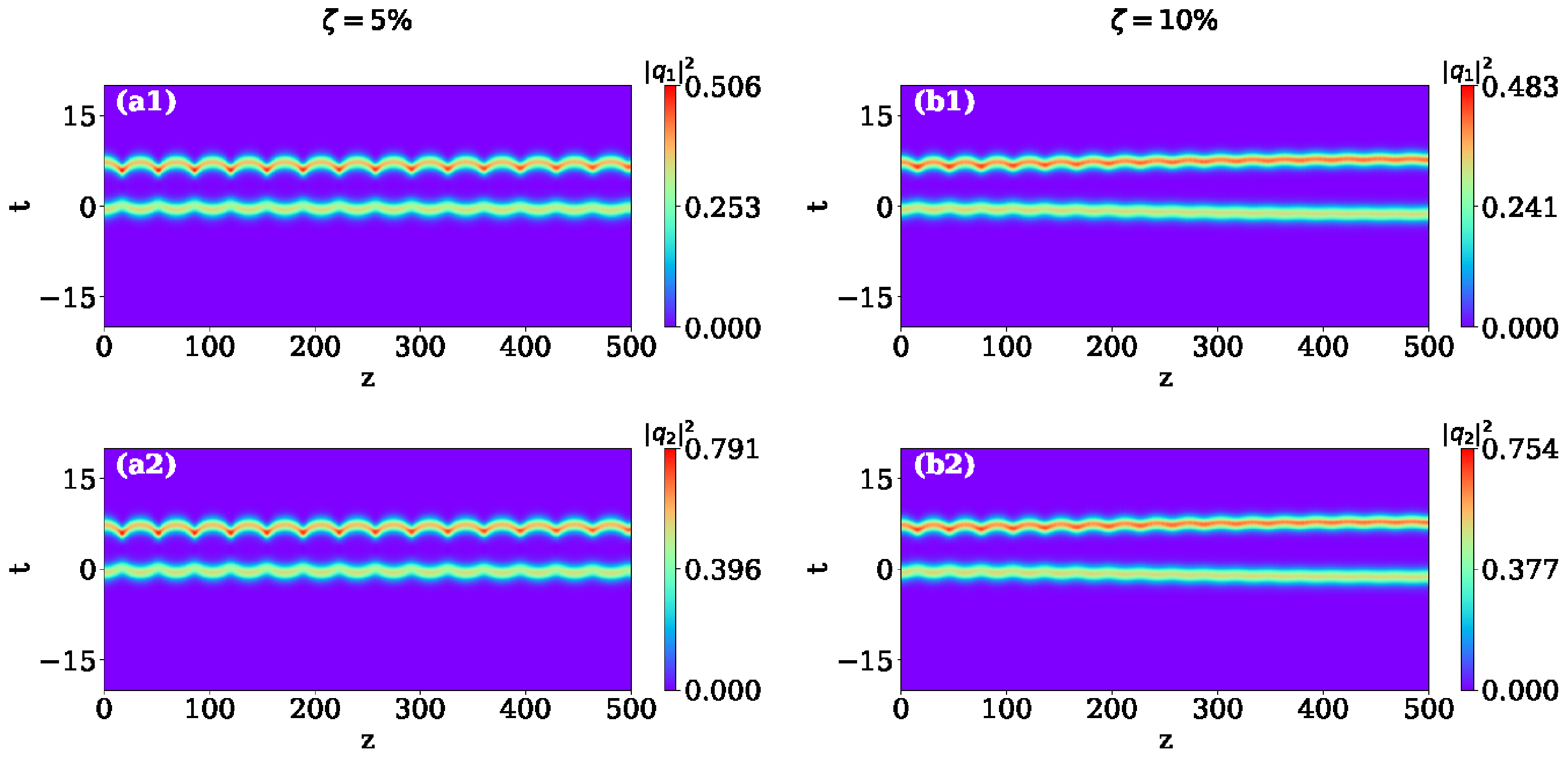} 
	\caption{The propagation of weakly bound fundamental soliton molecular structure is demonstrated in Figs. (a1) and (a2) for $5\%$ of perturbation, in which the bound state propagates without breaking. In Figs. (b1) and (b2), the evolution of basic SM is illustrated for $10\%$ of random perturbation. }
	\label{fig16}
\end{figure}

In addition to the above case, we have also studied the stability of the fundamental SM with $1\%$ ($\zeta=0.01$) of random noise as weak perturbation. Such a situation is displayed in  Figs. \ref{fig15}(b1) and \ref{fig15}(b2), where two weakly interacting solitons are able to propagate without any breaking. In this case also, the fundamental soliton molecular structure remains stable. Besides these, we have also checked the stability of SM for strong perturbations. For example, we depict the stability plots in Figs. \ref{fig16}(a1)-\ref{fig16}(a2) and \ref{fig16}(b1)-\ref{fig16}(b2) for $5\%$ ($\zeta=0.05$) and $10\%$ ($\zeta=0.1$) of perturbations, respectively. These percentage of perturbations can be considered as strong perturbations to the molecule. Figures \ref{fig16}(a1)-\ref{fig16}(a2) shows that the soliton molecular structure gets slightly disturbed for $5\%$ of perturbation but it still propagates without any distortion. However, the weakly bound molecular state gets distorted at $z> 200$ (Figs. \ref{fig16}(b1) and \ref{fig16}(b2)). Therefore, the weakly bound soliton molecular states can withstand for weak perturbations whereas for strong perturbations it covers appreciable distance. 

\begin{figure}
	\centering\includegraphics[width=0.8\linewidth]{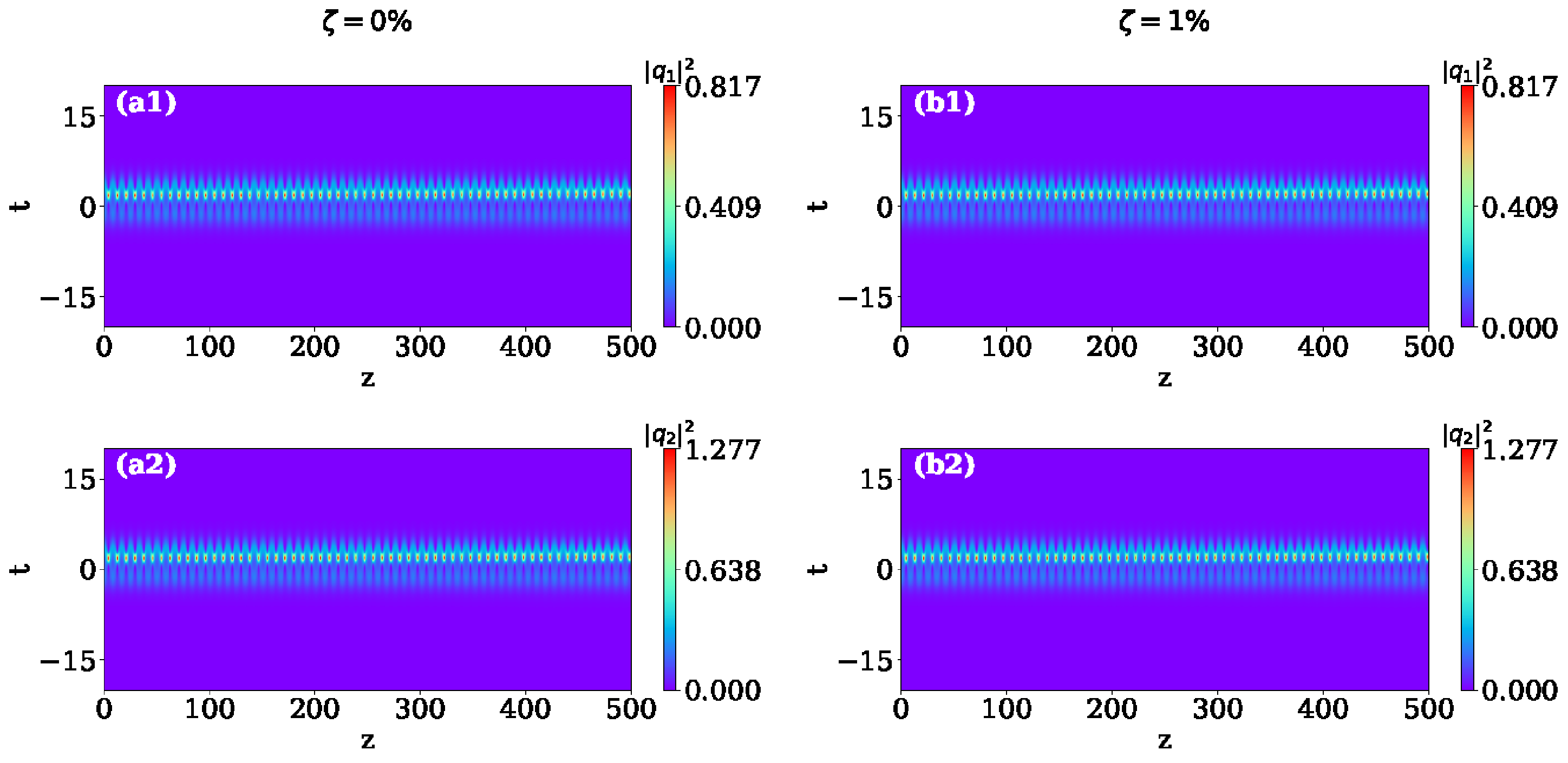} 
	\caption{The propagation of closely packed bound soliton molecular structure is demonstrated in  (a1) and (a2) for zero perturbation. In  (b1) and (b2), we illustrate the propagation of bound soliton state with  $1\%$ of random perturbation. }
	\label{fig17}
\end{figure}

\begin{figure}
\centering\includegraphics[width=0.8\linewidth]{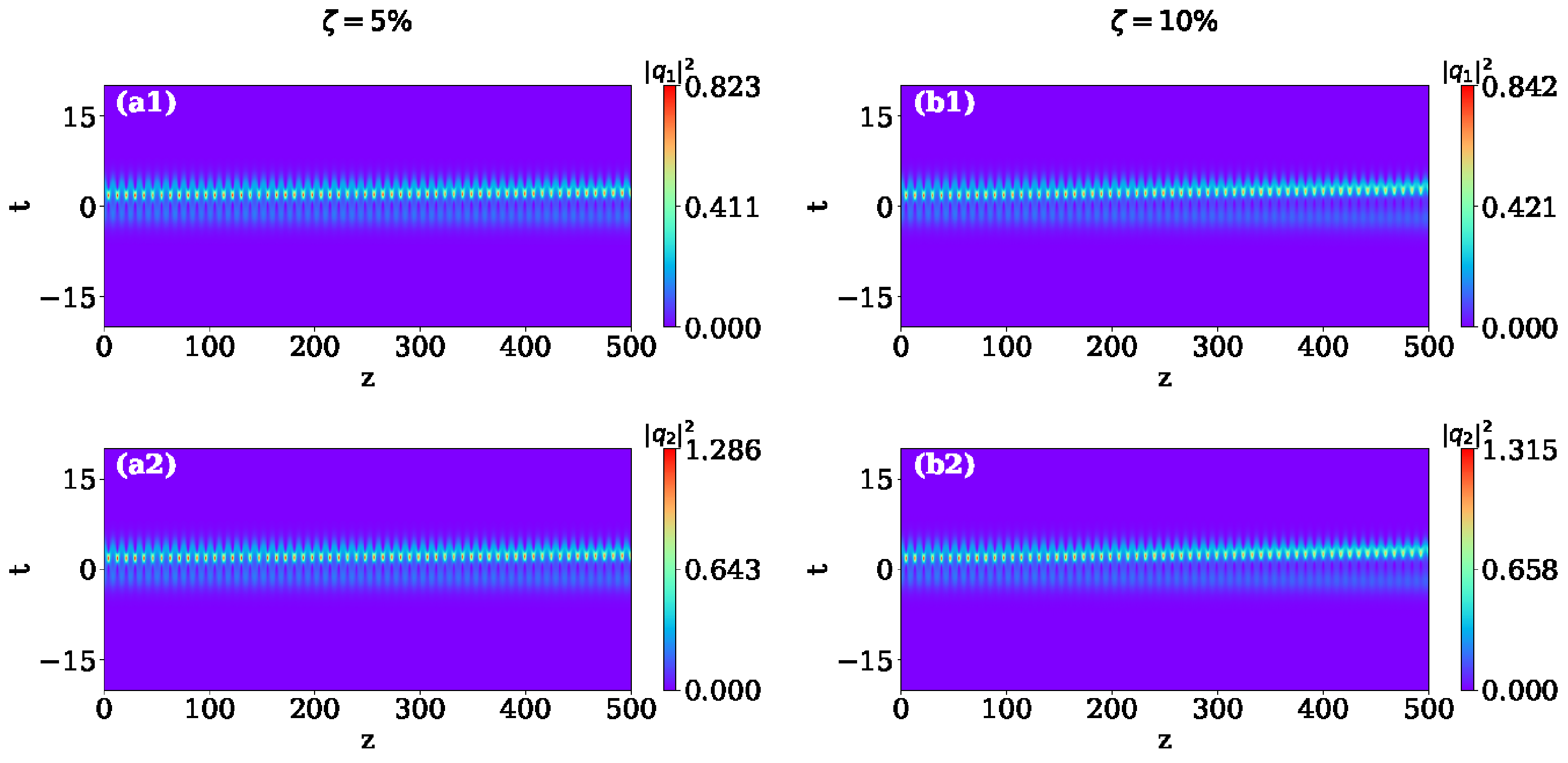} 
\caption{The propagation of two bound solitons with minimum separation is demonstrated in (a1) and (a2) for $5\%$ random perturbation. In (b1) and (b2), we illustrate the propagation of the same bound soliton state with  $10\%$ of random perturbation.}
\label{fig18}
\end{figure}

Next, we demonstrate the stability property of a strongly bound fundamental soliton molecule, in which two soliton atoms strongly interact with each other. Due to the strong interaction between the two soliton atoms, the periodic oscillations appears rapidly in the structure of doublet SM. In order to set such strongly bound SM as initial conditions, we consider the parameter values, $k_1=1$, $l_1=0.6$, $\alpha_1^{(1)}=0.8$, $\alpha_1^{(2)}=1$, $\alpha_2^{(1)}=1.2$, and $\alpha_1^{(1)}=1.5$, in the bound soliton solution (\ref{9a})-(\ref{9d}). As we have illustrated in Figs. \ref{fig17}(a1) and \ref{fig17}(a2), we initially allow the doublet SM to propagate along the $z$-direction without any perturbation. These figures show that the doublet soliton molecular structure propagates in a stable manner like the synthesized basic molecular structure predicted by the exact BSS solution. This strongly bound soliton molecular structure again shows stability when we include weak perturbation ($1\%$) in the numerical analysis. This result is displayed in Figs. \ref{fig17}(b1) and \ref{fig17}(b2). Further, the SM is subjected to the strong perturbations, such as for $5\%$ and $10\%$. The results are depicted in Figs. \ref{fig18}(a1)-\ref{fig18}(a2) and  \ref{fig18}(b1)- \ref{fig18}(b2), respectively. It is evident that the strongly bound molecular structure remains stable for  $5\%$ of random perturbation. However, it starts to respond for $10\%$ of perturbation and the bound soliton state breaking begins to occur after covering the propagation distance $z\approx 400$. From this analysis, one can confirm that the densely packed soliton molecules (that is, tightly bound molecular structures made by soliton atoms with minimum temporal separation) can propagate much longer distance than the weekly packed soliton molecules. Therefore, the soliton atoms with minimal separation can form a stable vector soliton molecules.  


\section{Conclusion}
We have demonstrated the existence of vector soliton molecules and their interesting isomer structures in the Manakov system. To obtain these molecular states, we first rewrite the degenerate $N$-soliton solution conveniently and obtain the ($\bar{N}+\bar{M}$)-soliton solution. Then, to get the bound soliton solutions, we have imposed three types of velocity degeneracy conditions on the obtained ($\bar{N}+\bar{M}$) multi-soliton solution. Especially, by imposing a complete degeneracy condition, we brought out a doublet SM, a triplet and quadruplet SMs. These molecular structures and their corresponding isomer structures are classified as synthesised molecular state and dissociated molecular state.  Further, we have also analysed the mechanism behind the formation of these molecular structures, especially the formation mechanism of a doublet SM in detail with respect to wave parameters. Besides these, we  have verified the robustness of the obtained vector soliton molecules by subjecting them under two distinct physical situations. In particular, we have studied the stability of a doublet SM by allowing it to interact with one and then two fundamental vector solitons as strong perturbations. Then, we also analysed the collision scenario between two distinct doublet SMs. The collision scenario revealed that the doublet soliton molecule always undergoes energy-sharing collisions with its interacting partners. We have confirmed this through an appropriate asymptotic analysis. Elastic collision can be achieved from the each of these collision scenarios by choosing the complex phase constants appropriately. We have also confirmed the stability of vector SMs through numerical analysis, which shows that a closely packed molecular structure can propagate over a larger distance than a weakly bound molecular structure. We believe that the results obtained in this paper will be useful for soliton molecule-based applications such as optical computation and multi-level encoding for communications \cite{fedor1,soliton-complex}.
\section*{Acknowledgment}
M. Lakshmanan thanks DST-SERB, INDIA for the award of a DST-SERB National Science Chair (NSC/2020/000029) position in which S. Stalin  is a Research Associate. 
\appendix
\section{}\label{A}
We present the constants that are appearing in Sec. III A while analysing the collision between doublet SM and two vector bright solitons: 
\bea
\nu_{k,j}&=&\big[(\hat{B}_{11}\hat{B}_{22}-\hat{B}_{12}\hat{B}_{21})(\alpha_1^{(j)}B_{k2}-\alpha_2^{(j)}B_{k1})+(\hat{B}_{12}\tilde{B}_{22}-\tilde{B}_{12}\hat{B}_{22})(\alpha_1^{(j)}b_{k1}-\beta_1^{(j)}B_{k1})\nonumber\\&&
+(\tilde{b}_{12}\hat{B}_{21}-\hat{B}_{11}\tilde{b}_{22})(\alpha_1^{(j)}b_{k2}-\beta_2^{(j)}B_{k1})+(\tilde{b}_{11}\hat{B}_{22}-\hat{B}_{12}\tilde{b}_{21})(\alpha_2^{(j)}b_{k1}-\beta_1^{(j)}B_{k2})\nonumber\\
&&+(\tilde{b}_{21}\hat{B}_{11}-\hat{B}_{21}\tilde{b}_{11})(\alpha_2^{(j)}b_{k2}-\beta_2^{(j)}B_{k2})+(\tilde{b}_{11}\tilde{b}_{22}-\tilde{b}_{21}\tilde{b}_{12})(\beta_1^{(j)}b_{k2}-\beta_2^{(j)}b_{k1})\big]^{1/2}\nonumber\\
\gamma_{1,j} &=&\big[\alpha_2^{(j)}(\hat{B}_{12}\hat{B}_{21}-\hat{B}_{11}\hat{B}_{22})+\beta_1^{(j)}(\tilde{b}_{12}\hat{B}_{22}-\tilde{b}_{22}\hat{B}_{12})+\beta_2^{(j)}(\hat{B}_{11}\tilde{b}_{22}-\tilde{b}_{12}\hat{B}_{21})\big]^{1/2},\nonumber\\
\gamma_{2,j} &=&\big[\alpha_1^{(j)}(\hat{B}_{12}\hat{B}_{21}-\hat{B}_{11}\hat{B}_{22})+\beta_1^{(j)}(\tilde{b}_{11}\hat{B}_{22}-\tilde{b}_{21}\hat{B}_{12})+\beta_2^{(j)}(\hat{B}_{11}\tilde{b}_{21}-\tilde{b}_{11}\hat{B}_{21})\big]^{1/2},\nonumber\\
\tau_1 &=&\bigg[B_{11}\big(B_{22}(\hat{B}_{11}\hat{B}_{22}-\hat{B}_{12}\hat{B}_{21})+b_{21}(\hat{B}_{12}\tilde{b}_{22}-\tilde{b}_{12}\hat{B}_{22})+b_{22}(\tilde{b}_{12}\hat{B}_{21}-\hat{B}_{11}\tilde{b}_{22})\big)\nonumber\\
&&+B_{12}\big(B_{21}(\hat{B}_{12}\hat{B}_{21}-\hat{B}_{11}\hat{B}_{22})+b_{21}(\tilde{b}_{11}\hat{B}_{22}-\hat{B}_{12}\tilde{b}_{21})+b_{22}(\tilde{b}_{21}\hat{B}_{11}-\hat{B}_{21}\tilde{b}_{11})\big)\nonumber\\
&&+b_{11}\big(B_{21}(\tilde{b}_{12}\hat{B}_{22}-\tilde{b}_{22}\hat{B}_{12})+B_{22}(\tilde{b}_{21}\hat{B}_{12}-\tilde{b}_{11}\hat{B}_{22})+b_{22}(\tilde{b}_{11}\tilde{b}_{22}-\tilde{b}_{12}\tilde{b}_{21})\big)\nonumber\\
&&+b_{12}\big(B_{21}(\tilde{b}_{22}\hat{B}_{11}-\tilde{b}_{12}\hat{B}_{21})+B_{22}(\tilde{b}_{11}\hat{B}_{21}-\tilde{b}_{21}\hat{B}_{11})+b_{21}(\tilde{b}_{21}\tilde{b}_{12}-\tilde{b}_{11}\tilde{b}_{22})\big)\bigg],\nonumber\\
\tau_2 &=&\big[B_{11}(\hat{B}_{12}\hat{B}_{21}-\hat{B}_{11}\hat{B}_{22})+b_{11}(\tilde{b}_{11}\hat{B}_{22}-\tilde{b}_{21}\hat{B}_{12})+b_{12}(\tilde{b}_{21}\hat{B}_{11}-\tilde{b}_{11}\hat{B}_{21})\big],\nonumber\\
\tau_3 &=&\big[B_{22}(\hat{B}_{12}\hat{B}_{21}-\hat{B}_{11}\hat{B}_{22})+b_{21}(\tilde{b}_{12}\hat{B}_{22}-\tilde{b}_{22}\hat{B}_{12})+b_{22}(\tilde{b}_{22}\hat{B}_{11}-\tilde{b}_{12}\hat{B}_{21})\big],\nonumber\\
\tau_4 &=&\big[B_{21}(\hat{B}_{12}\hat{B}_{21}-\hat{B}_{11}\hat{B}_{22})+b_{21}(\tilde{b}_{11}\hat{B}_{22}-\tilde{b}_{21}\hat{B}_{12})+b_{22}(\tilde{b}_{21}\hat{B}_{11}-\tilde{b}_{11}\hat{B}_{21})\big],\nonumber\\
\tau_5 &=&\big[B_{12}(\hat{B}_{12}\hat{B}_{21}-\hat{B}_{11}\hat{B}_{22})+b_{11}(\tilde{b}_{12}\hat{B}_{22}-\tilde{b}_{22}\hat{B}_{12})+b_{12}(\tilde{b}_{22}\hat{B}_{11}-\tilde{b}_{12}\hat{B}_{21})\big].\nonumber\\
\end{eqnarray}
Similarly, we also present the constants that are appearing in Sec. III B while analysing the collision between two doublet SMs: 
\bea
\mu_{1,j}&=&\bigg[(b_{22}\hat{B}_{11}-b_{21}\hat{B}_{12})(\alpha_1^{(j)}B_{12}-\alpha_2^{(j)}B_{11})+(B_{22}\hat{B}_{12}-b_{22}\tilde{b}_{12})(\alpha_1^{(j)}b_{11}-\beta_1^{(j)}B_{11})\nonumber\\
&&+(b_{21}\tilde{b}_{12}-B_{22}\hat{B}_{11})(\alpha_1^{(j)}b_{12}-\beta_2^{(j)}B_{11})+(b_{22}\tilde{b}_{11}-B_{21}\hat{B}_{12})(\alpha_2^{(j)}b_{11}-\beta_1^{(j)}B_{12})\nonumber\\
&&+(B_{21}\hat{B}_{11}-b_{21}\tilde{b}_{11})(\alpha_2^{(j)}b_{12}-\beta_2^{(j)}B_{12})+(B_{22}\tilde{b}_{11}-B_{21}\tilde{b}_{12})(\beta_1^{(j)}b_{12}-\beta_2^{(j)}b_{11})\bigg]^{1/2}\nonumber\eea
\bea
\chi_{1,j}&=&\big[\alpha_1^{(j)}(b_{12}B_{22}-B_{12}b_{22})+\alpha_2^{(j)}(B_{11}b_{22}-b_{12}B_{21})+\beta_2^{(j)}(B_{12}B_{21}-B_{11}B_{22})\big]^{1/2}\nonumber\\
\mu_{2,j}&=&\bigg[(b_{22}\hat{B}_{21}-b_{21}\hat{B}_{22})(\alpha_1^{(j)}B_{12}-\alpha_2^{(j)}B_{11})+(B_{22}\hat{B}_{22}-b_{22}\tilde{b}_{22})(\alpha_1^{(j)}b_{11}-\beta_1^{(j)}B_{11})\nonumber\\
&&+(b_{21}\tilde{b}_{22}-B_{22}\hat{B}_{21})(\alpha_1^{(j)}b_{12}-\beta_2^{(j)}B_{11})+(b_{22}\tilde{b}_{21}-B_{21}\hat{B}_{22})(\alpha_2^{(j)}b_{11}-\beta_1^{(j)}B_{12})\nonumber\\
&&+(B_{21}\hat{B}_{21}-b_{21}\tilde{b}_{21})(\alpha_2^{(j)}b_{12}-\beta_2^{(j)}B_{12})+(B_{22}\tilde{b}_{21}-B_{21}\tilde{b}_{22})(\beta_1^{(j)}b_{12}-\beta_2^{(j)}b_{11})\bigg]^{1/2}\nonumber\\
\chi_{2,j}&=&\big[\alpha_1^{(j)}(b_{11}B_{22}-B_{12}b_{21})+\alpha_2^{(j)}(B_{11}b_{21}-b_{11}B_{21})+\beta_1^{(j)}(B_{12}B_{21}-B_{11}B_{22})\big]^{1/2}\nonumber\\
\chi_3 &=&\big[B_{11}(b_{21}\tilde{b}_{12}-B_{22}\hat{B}_{11})+B_{12}(B_{21}\hat{B}_{11}-b_{21}\tilde{b}_{11})+b_{11}(B_{22}\tilde{b}_{11}-B_{21}\tilde{b}_{12})\big]\nonumber\\
\chi_4 &=&\big[B_{11}(b_{22}\tilde{b}_{22}-B_{22}\hat{B}_{22})+B_{12}(B_{21}\hat{B}_{22}-b_{22}\hat{b}_{21})+b_{12}(B_{22}\tilde{b}_{21}-B_{21}\tilde{b}_{22})\big]\nonumber\\
\chi_5 &=&\big[B_{11}(b_{21}\tilde{b}_{22}-B_{22}\hat{B}_{21})+B_{12}(B_{21}\hat{B}_{21}-b_{21}\tilde{b}_{21})+b_{11}(B_{22}\tilde{b}_{21}-B_{21}\tilde{b}_{22})\big]\nonumber\\
\chi_6 &=&\big[B_{11}(b_{22}\tilde{b}_{12}-B_{22}b_{22})+B_{12}(B_{21}\hat{B}_{12}-b_{22}\tilde{b}_{11})+b_{12}(B_{22}\tilde{b}_{11}-B_{21}\tilde{b}_{12})\big].
\end{eqnarray}

The constants $B_{ij}$, $b_{ij}$, $\tilde{b}_{ij}$, and $\hat{B}_{ij}$ are defined below: 
\begin{eqnarray}
&&B_{ij}=\sum_{n=1}^2\frac{\alpha_i^{(n)}\alpha_j^{(n)*}}{k_i^*+k_j},~b_{ij}=\sum_{n=1}^2\frac{\beta_i^{(n)}\alpha_j^{(n)*}}{k_i^*+l_j},\nonumber\\
&&\tilde{b}_{ij}=\sum_{n=1}^2\frac{\alpha_i^{(n)}\beta_j^{(n)*}}{l_i^*+k_j},~\hat{B}_{ij}=\sum_{n=1}^2\frac{\beta_i^{(n)}\beta_j^{(n)*}}{l_i^*+l_j},~i,j=1,2.\label{A3}
\end{eqnarray}

\end{document}